\documentclass[traditabstract,longauth]{aa}
\usepackage{amsmath}
\usepackage{txfonts}
\usepackage{ifpdf}

\ifpdf
\usepackage[pdftex]{graphicx}
\usepackage{epstopdf}

\else
\usepackage[dvips]{graphicx}
\usepackage[dvips]{hyperref}

\fi
\usepackage{url}
\usepackage{framed}

\def\setsymbol#1#2{\expandafter\def\csname #1\endcsname{#2}}
\def\getsymbol#1{\csname #1\endcsname}

\def\Planck{\textit{Planck}}

\def\all2013resultspapers{\nocite{planck2013-p01, planck2013-p02, planck2013-p02a, planck2013-p02d, planck2013-p02b, planck2013-p03, planck2013-p03c, planck2013-p03f, planck2013-p03d, planck2013-p03e, planck2013-p06b, planck2013-p06, planck2013-p03a, planck2013-pip88, planck2013-p08, planck2013-p11, planck2013-p12, planck2013-p13, planck2013-p14, planck2013-p15, planck2013-p05b, planck2013-p17, planck2013-p09, planck2013-p09a, planck2013-p20, planck2013-p19, planck2013-pipaberration, planck2013-p05, planck2013-p05a, planck2013-pip56, planck2013-p01a}}

\newbox\tablebox    \newdimen\tablewidth
\def\leaderfil{\leaders\hbox to 5pt{\hss.\hss}\hfil}
\def\endPlancktable{\tablewidth=\columnwidth 
    $$\hss\copy\tablebox\hss$$
    \vskip-\lastskip\vskip -2pt}

\def\tablenote#1 #2\par{\begingroup \parindent=0.8em
    \abovedisplayshortskip=0pt\belowdisplayshortskip=0pt
    \noindent
    $$\hss\vbox{\hsize\tablewidth \hangindent=\parindent \hangafter=1 \noindent
    \hbox to \parindent{$^#1$\hss}\strut#2\strut\par}\hss$$
    \endgroup}
\def\doubleline{\vskip 3pt\hrule \vskip 1.5pt \hrule \vskip 5pt}

\def\L2{\ifmmode L_2\else $L_2$\fi}

\def\DeltaT{\ifmmode \Delta T\else $\Delta T$\fi}
\def\deltat{\ifmmode \Delta t\else $\Delta t$\fi}
\def\fknee{\ifmmode f_{\rm knee}\else $f_{\rm knee}$\fi}
\def\Fmax{\ifmmode F_{\rm max}\else $F_{\rm max}$\fi}
\def\solar{\ifmmode{\rm M}_{\mathord\odot}\else${\rm M}_{\mathord\odot}$\fi}
\def\Msolar{\ifmmode{\rm M}_{\mathord\odot}\else${\rm M}_{\mathord\odot}$\fi}
\def\Lsolar{\ifmmode{\rm L}_{\mathord\odot}\else${\rm L}_{\mathord\odot}$\fi}

\def\inv{\ifmmode^{-1}\else$^{-1}$\fi}
\def\mo{\ifmmode^{-1}\else$^{-1}$\fi}
\def\sup#1{\ifmmode ^{\rm #1}\else $^{\rm #1}$\fi}
\def\expo#1{\ifmmode \times 10^{#1}\else $\times 10^{#1}$\fi}
\def\,{\thinspace}
\def\lsim{\mathrel{\raise .4ex\hbox{\rlap{$<$}\lower 1.2ex\hbox{$\sim$}}}}
\def\gsim{\mathrel{\raise .4ex\hbox{\rlap{$>$}\lower 1.2ex\hbox{$\sim$}}}}

\def\simprop{\mathrel{\raise .4ex\hbox{\rlap{$\propto$}\lower 1.2ex\hbox{$\sim$}}}}
\def\deg{\ifmmode^\circ\else$^\circ$\fi}
\def\pdeg{\ifmmode $\setbox0=\hbox{$^{\circ}$}\rlap{\hskip.11\wd0 .}$^{\circ}
          \else \setbox0=\hbox{$^{\circ}$}\rlap{\hskip.11\wd0 .}$^{\circ}$\fi}
\def\arcs{\ifmmode {^{\scriptstyle\prime\prime}}
          \else $^{\scriptstyle\prime\prime}$\fi}
\def\arcm{\ifmmode {^{\scriptstyle\prime}}
          \else $^{\scriptstyle\prime}$\fi}
\newdimen\sa  \newdimen\sb
\def\parcs{\sa=.07em \sb=.03em
     \ifmmode \hbox{\rlap{.}}^{\scriptstyle\prime\kern -\sb\prime}\hbox{\kern -\sa}
     \else \rlap{.}$^{\scriptstyle\prime\kern -\sb\prime}$\kern -\sa\fi}
\def\parcm{\sa=.08em \sb=.03em
     \ifmmode \hbox{\rlap{.}\kern\sa}^{\scriptstyle\prime}\hbox{\kern-\sb}
     \else \rlap{.}\kern\sa$^{\scriptstyle\prime}$\kern-\sb\fi}
\def\ra[#1 #2 #3.#4]{#1\sup{h}#2\sup{m}#3\sup{s}\llap.#4}
\def\dec[#1 #2 #3.#4]{#1\deg#2\arcm#3\arcs\llap.#4}
\def\deco[#1 #2 #3]{#1\deg#2\arcm#3\arcs}
\def\rra[#1 #2]{#1\sup{h}#2\sup{m}}

\def\dots{\relax\ifmmode \ldots\else $\ldots$\fi}
\def\WHzsr{\ifmmode $W\,Hz\mo\,sr\mo$\else W\,Hz\mo\,sr\mo\fi}
\def\mHz{\ifmmode $\,mHz$\else \,mHz\fi}
\def\GHz{\ifmmode $\,GHz$\else \,GHz\fi}
\def\mKs{\ifmmode $\,mK\,s$^{1/2}\else \,mK\,s$^{1/2}$\fi}
\def\muKs{\ifmmode \,\mu$K\,s$^{1/2}\else \,$\mu$K\,s$^{1/2}$\fi}
\def\muKRJs{\ifmmode \,\mu$K$_{\rm RJ}$\,s$^{1/2}\else \,$\mu$K$_{\rm RJ}$\,s$^{1/2}$\fi}
\def\muKHz{\ifmmode \,\mu$K\,Hz$^{-1/2}\else \,$\mu$K\,Hz$^{-1/2}$\fi}
\def\MJysr{\ifmmode \,$MJy\,sr\mo$\else \,MJy\,sr\mo\fi}
\def\MJysrmK{\ifmmode \,$MJy\,sr\mo$\,mK$_{\rm CMB}\mo\else \,MJy\,sr\mo\,mK$_{\rm CMB}\mo$\fi}
\def\microns{\ifmmode \,\mu$m$\else \,$\mu$m\fi}

\def\muK{\ifmmode \,\mu$K$\else \,$\mu$\hbox{K}\fi}
\def\microK{\ifmmode \,\mu$K$\else \,$\mu$\hbox{K}\fi}
\def\muW{\ifmmode \,\mu$W$\else \,$\mu$\hbox{W}\fi}
\def\kms{\ifmmode $\,km\,s$^{-1}\else \,km\,s$^{-1}$\fi}
\def\kmsMpc{\ifmmode $\,\kms\,Mpc\mo$\else \,\kms\,Mpc\mo\fi}
\setsymbol{LFI:center:frequency:70GHz:units}{70.3\,GHz}
\setsymbol{LFI:center:frequency:44GHz:units}{44.1\,GHz}
\setsymbol{LFI:center:frequency:30GHz:units}{28.5\,GHz}

\setsymbol{LFI:center:frequency:70GHz}{70.3}
\setsymbol{LFI:center:frequency:44GHz}{44.1}
\setsymbol{LFI:center:frequency:30GHz}{28.5}

\setsymbol{LFI:center:frequency:LFI18:Rad:M:units}{71.7\GHz}
\setsymbol{LFI:center:frequency:LFI19:Rad:M:units}{67.5\GHz}
\setsymbol{LFI:center:frequency:LFI20:Rad:M:units}{69.2\GHz}
\setsymbol{LFI:center:frequency:LFI21:Rad:M:units}{70.4\GHz}
\setsymbol{LFI:center:frequency:LFI22:Rad:M:units}{71.5\GHz}
\setsymbol{LFI:center:frequency:LFI23:Rad:M:units}{70.8\GHz}
\setsymbol{LFI:center:frequency:LFI24:Rad:M:units}{44.4\GHz}
\setsymbol{LFI:center:frequency:LFI25:Rad:M:units}{44.0\GHz}
\setsymbol{LFI:center:frequency:LFI26:Rad:M:units}{43.9\GHz}
\setsymbol{LFI:center:frequency:LFI27:Rad:M:units}{28.3\GHz}
\setsymbol{LFI:center:frequency:LFI28:Rad:M:units}{28.8\GHz}
\setsymbol{LFI:center:frequency:LFI18:Rad:S:units}{70.1\GHz}
\setsymbol{LFI:center:frequency:LFI19:Rad:S:units}{69.6\GHz}
\setsymbol{LFI:center:frequency:LFI20:Rad:S:units}{69.5\GHz}
\setsymbol{LFI:center:frequency:LFI21:Rad:S:units}{69.5\GHz}
\setsymbol{LFI:center:frequency:LFI22:Rad:S:units}{72.8\GHz}
\setsymbol{LFI:center:frequency:LFI23:Rad:S:units}{71.3\GHz}
\setsymbol{LFI:center:frequency:LFI24:Rad:S:units}{44.1\GHz}
\setsymbol{LFI:center:frequency:LFI25:Rad:S:units}{44.1\GHz}
\setsymbol{LFI:center:frequency:LFI26:Rad:S:units}{44.1\GHz}
\setsymbol{LFI:center:frequency:LFI27:Rad:S:units}{28.5\GHz}
\setsymbol{LFI:center:frequency:LFI28:Rad:S:units}{28.2\GHz}

\setsymbol{LFI:center:frequency:LFI18:Rad:M}{71.7}
\setsymbol{LFI:center:frequency:LFI19:Rad:M}{67.5}
\setsymbol{LFI:center:frequency:LFI20:Rad:M}{69.2}
\setsymbol{LFI:center:frequency:LFI21:Rad:M}{70.4}
\setsymbol{LFI:center:frequency:LFI22:Rad:M}{71.5}
\setsymbol{LFI:center:frequency:LFI23:Rad:M}{70.8}
\setsymbol{LFI:center:frequency:LFI24:Rad:M}{44.4}
\setsymbol{LFI:center:frequency:LFI25:Rad:M}{44.0}
\setsymbol{LFI:center:frequency:LFI26:Rad:M}{43.9}
\setsymbol{LFI:center:frequency:LFI27:Rad:M}{28.3}
\setsymbol{LFI:center:frequency:LFI28:Rad:M}{28.8}
\setsymbol{LFI:center:frequency:LFI18:Rad:S}{70.1}
\setsymbol{LFI:center:frequency:LFI19:Rad:S}{69.6}
\setsymbol{LFI:center:frequency:LFI20:Rad:S}{69.5}
\setsymbol{LFI:center:frequency:LFI21:Rad:S}{69.5}
\setsymbol{LFI:center:frequency:LFI22:Rad:S}{72.8}
\setsymbol{LFI:center:frequency:LFI23:Rad:S}{71.3}
\setsymbol{LFI:center:frequency:LFI24:Rad:S}{44.1}
\setsymbol{LFI:center:frequency:LFI25:Rad:S}{44.1}
\setsymbol{LFI:center:frequency:LFI26:Rad:S}{44.1}
\setsymbol{LFI:center:frequency:LFI27:Rad:S}{28.5}
\setsymbol{LFI:center:frequency:LFI28:Rad:S}{28.2}

\setsymbol{LFI:white:noise:sensitivity:70GHz:units}{134.7\muKs}
\setsymbol{LFI:white:noise:sensitivity:44GHz:units}{164.7\muKs}
\setsymbol{LFI:white:noise:sensitivity:30GHz:units}{143.4\muKs}

\setsymbol{LFI:white:noise:sensitivity:70GHz}{134.7}
\setsymbol{LFI:white:noise:sensitivity:44GHz}{164.7}
\setsymbol{LFI:white:noise:sensitivity:30GHz}{143.4}

\setsymbol{LFI:white:noise:sensitivity:LFI18:Rad:M:units}{512.0\muKs}
\setsymbol{LFI:white:noise:sensitivity:LFI19:Rad:M:units}{581.4\muKs}
\setsymbol{LFI:white:noise:sensitivity:LFI20:Rad:M:units}{590.8\muKs}
\setsymbol{LFI:white:noise:sensitivity:LFI21:Rad:M:units}{455.2\muKs}
\setsymbol{LFI:white:noise:sensitivity:LFI22:Rad:M:units}{492.0\muKs}
\setsymbol{LFI:white:noise:sensitivity:LFI23:Rad:M:units}{507.7\muKs}
\setsymbol{LFI:white:noise:sensitivity:LFI24:Rad:M:units}{462.2\muKs}
\setsymbol{LFI:white:noise:sensitivity:LFI25:Rad:M:units}{413.6\muKs}
\setsymbol{LFI:white:noise:sensitivity:LFI26:Rad:M:units}{478.6\muKs}
\setsymbol{LFI:white:noise:sensitivity:LFI27:Rad:M:units}{277.7\muKs}
\setsymbol{LFI:white:noise:sensitivity:LFI28:Rad:M:units}{312.3\muKs}
\setsymbol{LFI:white:noise:sensitivity:LFI18:Rad:S:units}{465.7\muKs}
\setsymbol{LFI:white:noise:sensitivity:LFI19:Rad:S:units}{555.6\muKs}
\setsymbol{LFI:white:noise:sensitivity:LFI20:Rad:S:units}{623.2\muKs}
\setsymbol{LFI:white:noise:sensitivity:LFI21:Rad:S:units}{564.1\muKs}
\setsymbol{LFI:white:noise:sensitivity:LFI22:Rad:S:units}{534.4\muKs}
\setsymbol{LFI:white:noise:sensitivity:LFI23:Rad:S:units}{542.4\muKs}
\setsymbol{LFI:white:noise:sensitivity:LFI24:Rad:S:units}{399.2\muKs}
\setsymbol{LFI:white:noise:sensitivity:LFI25:Rad:S:units}{392.6\muKs}
\setsymbol{LFI:white:noise:sensitivity:LFI26:Rad:S:units}{418.6\muKs}
\setsymbol{LFI:white:noise:sensitivity:LFI27:Rad:S:units}{302.9\muKs}
\setsymbol{LFI:white:noise:sensitivity:LFI28:Rad:S:units}{285.3\muKs}

\setsymbol{LFI:white:noise:sensitivity:LFI18:Rad:M}{512.0}
\setsymbol{LFI:white:noise:sensitivity:LFI19:Rad:M}{581.4}
\setsymbol{LFI:white:noise:sensitivity:LFI20:Rad:M}{590.8}
\setsymbol{LFI:white:noise:sensitivity:LFI21:Rad:M}{455.2}
\setsymbol{LFI:white:noise:sensitivity:LFI22:Rad:M}{492.0}
\setsymbol{LFI:white:noise:sensitivity:LFI23:Rad:M}{507.7}
\setsymbol{LFI:white:noise:sensitivity:LFI24:Rad:M}{462.2}
\setsymbol{LFI:white:noise:sensitivity:LFI25:Rad:M}{413.6}
\setsymbol{LFI:white:noise:sensitivity:LFI26:Rad:M}{478.6}
\setsymbol{LFI:white:noise:sensitivity:LFI27:Rad:M}{277.7}
\setsymbol{LFI:white:noise:sensitivity:LFI28:Rad:M}{312.3}
\setsymbol{LFI:white:noise:sensitivity:LFI18:Rad:S}{465.7}
\setsymbol{LFI:white:noise:sensitivity:LFI19:Rad:S}{555.6}
\setsymbol{LFI:white:noise:sensitivity:LFI20:Rad:S}{623.2}
\setsymbol{LFI:white:noise:sensitivity:LFI21:Rad:S}{564.1}
\setsymbol{LFI:white:noise:sensitivity:LFI22:Rad:S}{534.4}
\setsymbol{LFI:white:noise:sensitivity:LFI23:Rad:S}{542.4}
\setsymbol{LFI:white:noise:sensitivity:LFI24:Rad:S}{399.2}
\setsymbol{LFI:white:noise:sensitivity:LFI25:Rad:S}{392.6}
\setsymbol{LFI:white:noise:sensitivity:LFI26:Rad:S}{418.6}
\setsymbol{LFI:white:noise:sensitivity:LFI27:Rad:S}{302.9}
\setsymbol{LFI:white:noise:sensitivity:LFI28:Rad:S}{285.3}

\setsymbol{LFI:knee:frequency:70GHz:units}{29.5\mHz}
\setsymbol{LFI:knee:frequency:44GHz:units}{56.2\mHz}
\setsymbol{LFI:knee:frequency:30GHz:units}{113.7\mHz}

\setsymbol{LFI:knee:frequency:70GHz}{29.5}
\setsymbol{LFI:knee:frequency:44GHz}{56.2}
\setsymbol{LFI:knee:frequency:30GHz}{113.7}

\setsymbol{LFI:knee:frequency:LFI18:Rad:M:units}{16.3\mHz}
\setsymbol{LFI:knee:frequency:LFI19:Rad:M:units}{15.1\mHz}
\setsymbol{LFI:knee:frequency:LFI20:Rad:M:units}{18.7\mHz}
\setsymbol{LFI:knee:frequency:LFI21:Rad:M:units}{37.2\mHz}
\setsymbol{LFI:knee:frequency:LFI22:Rad:M:units}{12.7\mHz}
\setsymbol{LFI:knee:frequency:LFI23:Rad:M:units}{34.6\mHz}
\setsymbol{LFI:knee:frequency:LFI24:Rad:M:units}{46.2\mHz}
\setsymbol{LFI:knee:frequency:LFI25:Rad:M:units}{24.9\mHz}
\setsymbol{LFI:knee:frequency:LFI26:Rad:M:units}{67.6\mHz}
\setsymbol{LFI:knee:frequency:LFI27:Rad:M:units}{187.4\mHz}
\setsymbol{LFI:knee:frequency:LFI28:Rad:M:units}{122.2\mHz}
\setsymbol{LFI:knee:frequency:LFI18:Rad:S:units}{17.7\mHz}
\setsymbol{LFI:knee:frequency:LFI19:Rad:S:units}{22.0\mHz}
\setsymbol{LFI:knee:frequency:LFI20:Rad:S:units}{8.7\mHz}
\setsymbol{LFI:knee:frequency:LFI21:Rad:S:units}{25.9\mHz}
\setsymbol{LFI:knee:frequency:LFI22:Rad:S:units}{15.8\mHz}
\setsymbol{LFI:knee:frequency:LFI23:Rad:S:units}{129.8\mHz}
\setsymbol{LFI:knee:frequency:LFI24:Rad:S:units}{100.9\mHz}
\setsymbol{LFI:knee:frequency:LFI25:Rad:S:units}{38.9\mHz}
\setsymbol{LFI:knee:frequency:LFI26:Rad:S:units}{58.9\mHz}
\setsymbol{LFI:knee:frequency:LFI27:Rad:S:units}{104.4\mHz}
\setsymbol{LFI:knee:frequency:LFI28:Rad:S:units}{40.7\mHz}

\setsymbol{LFI:knee:frequency:LFI18:Rad:M}{16.3}
\setsymbol{LFI:knee:frequency:LFI19:Rad:M}{15.1}
\setsymbol{LFI:knee:frequency:LFI20:Rad:M}{18.7}
\setsymbol{LFI:knee:frequency:LFI21:Rad:M}{37.2}
\setsymbol{LFI:knee:frequency:LFI22:Rad:M}{12.7}
\setsymbol{LFI:knee:frequency:LFI23:Rad:M}{34.6}
\setsymbol{LFI:knee:frequency:LFI24:Rad:M}{46.2}
\setsymbol{LFI:knee:frequency:LFI25:Rad:M}{24.9}
\setsymbol{LFI:knee:frequency:LFI26:Rad:M}{67.6}
\setsymbol{LFI:knee:frequency:LFI27:Rad:M}{187.4}
\setsymbol{LFI:knee:frequency:LFI28:Rad:M}{122.2}
\setsymbol{LFI:knee:frequency:LFI18:Rad:S}{17.7}
\setsymbol{LFI:knee:frequency:LFI19:Rad:S}{22.0}
\setsymbol{LFI:knee:frequency:LFI20:Rad:S}{8.7}
\setsymbol{LFI:knee:frequency:LFI21:Rad:S}{25.9}
\setsymbol{LFI:knee:frequency:LFI22:Rad:S}{15.8}
\setsymbol{LFI:knee:frequency:LFI23:Rad:S}{129.8}
\setsymbol{LFI:knee:frequency:LFI24:Rad:S}{100.9}
\setsymbol{LFI:knee:frequency:LFI25:Rad:S}{38.9}
\setsymbol{LFI:knee:frequency:LFI26:Rad:S}{58.9}
\setsymbol{LFI:knee:frequency:LFI27:Rad:S}{104.4}
\setsymbol{LFI:knee:frequency:LFI28:Rad:S}{40.7}

\setsymbol{LFI:slope:70GHz:units}{$-1.03$\mHz}
\setsymbol{LFI:slope:44GHz:units}{$-0.89$\mHz}
\setsymbol{LFI:slope:30GHz:units}{$-0.87$\mHz}

\setsymbol{LFI:slope:70GHz}{$-1.03$}
\setsymbol{LFI:slope:44GHz}{$-0.89$}
\setsymbol{LFI:slope:30GHz}{$-0.87$}

\setsymbol{LFI:slope:LFI18:Rad:M:units}{$-1.04$\mHz}
\setsymbol{LFI:slope:LFI19:Rad:M:units}{$-1.09$\mHz}
\setsymbol{LFI:slope:LFI20:Rad:M:units}{$-0.69$\mHz}
\setsymbol{LFI:slope:LFI21:Rad:M:units}{$-1.56$\mHz}
\setsymbol{LFI:slope:LFI22:Rad:M:units}{$-1.01$\mHz}
\setsymbol{LFI:slope:LFI23:Rad:M:units}{$-0.96$\mHz}
\setsymbol{LFI:slope:LFI24:Rad:M:units}{$-0.83$\mHz}
\setsymbol{LFI:slope:LFI25:Rad:M:units}{$-0.91$\mHz}
\setsymbol{LFI:slope:LFI26:Rad:M:units}{$-0.95$\mHz}
\setsymbol{LFI:slope:LFI27:Rad:M:units}{$-0.87$\mHz}
\setsymbol{LFI:slope:LFI28:Rad:M:units}{$-0.88$\mHz}
\setsymbol{LFI:slope:LFI18:Rad:S:units}{$-1.15$\mHz}
\setsymbol{LFI:slope:LFI19:Rad:S:units}{$-1.00$\mHz}
\setsymbol{LFI:slope:LFI20:Rad:S:units}{$-0.95$\mHz}
\setsymbol{LFI:slope:LFI21:Rad:S:units}{$-0.92$\mHz}
\setsymbol{LFI:slope:LFI22:Rad:S:units}{$-1.01$\mHz}
\setsymbol{LFI:slope:LFI23:Rad:S:units}{$-0.95$\mHz}
\setsymbol{LFI:slope:LFI24:Rad:S:units}{$-0.73$\mHz}
\setsymbol{LFI:slope:LFI25:Rad:S:units}{$-1.16$\mHz}
\setsymbol{LFI:slope:LFI26:Rad:S:units}{$-0.79$\mHz}
\setsymbol{LFI:slope:LFI27:Rad:S:units}{$-0.82$\mHz}
\setsymbol{LFI:slope:LFI28:Rad:S:units}{$-0.91$\mHz}

\setsymbol{LFI:slope:LFI18:Rad:M}{$-1.04$}
\setsymbol{LFI:slope:LFI19:Rad:M}{$-1.09$}
\setsymbol{LFI:slope:LFI20:Rad:M}{$-0.69$}
\setsymbol{LFI:slope:LFI21:Rad:M}{$-1.56$}
\setsymbol{LFI:slope:LFI22:Rad:M}{$-1.01$}
\setsymbol{LFI:slope:LFI23:Rad:M}{$-0.96$}
\setsymbol{LFI:slope:LFI24:Rad:M}{$-0.83$}
\setsymbol{LFI:slope:LFI25:Rad:M}{$-0.91$}
\setsymbol{LFI:slope:LFI26:Rad:M}{$-0.95$}
\setsymbol{LFI:slope:LFI27:Rad:M}{$-0.87$}
\setsymbol{LFI:slope:LFI28:Rad:M}{$-0.88$}
\setsymbol{LFI:slope:LFI18:Rad:S}{$-1.15$}
\setsymbol{LFI:slope:LFI19:Rad:S}{$-1.00$}
\setsymbol{LFI:slope:LFI20:Rad:S}{$-0.95$}
\setsymbol{LFI:slope:LFI21:Rad:S}{$-0.92$}
\setsymbol{LFI:slope:LFI22:Rad:S}{$-1.01$}
\setsymbol{LFI:slope:LFI23:Rad:S}{$-0.95$}
\setsymbol{LFI:slope:LFI24:Rad:S}{$-0.73$}
\setsymbol{LFI:slope:LFI25:Rad:S}{$-1.16$}
\setsymbol{LFI:slope:LFI26:Rad:S}{$-0.79$}
\setsymbol{LFI:slope:LFI27:Rad:S}{$-0.82$}
\setsymbol{LFI:slope:LFI28:Rad:S}{$-0.91$}

\setsymbol{LFI:FWHM:70GHz:units}{13\parcm01}
\setsymbol{LFI:FWHM:44GHz:units}{27\parcm92}
\setsymbol{LFI:FWHM:30GHz:units}{32\parcm65}

\setsymbol{LFI:FWHM:70GHz}{13.01}
\setsymbol{LFI:FWHM:44GHz}{27.92}
\setsymbol{LFI:FWHM:30GHz}{32.65}

\setsymbol{LFI:FWHM:LFI18:units}{13\parcm39}
\setsymbol{LFI:FWHM:LFI19:units}{13\parcm01}
\setsymbol{LFI:FWHM:LFI20:units}{12\parcm75}
\setsymbol{LFI:FWHM:LFI21:units}{12\parcm74}
\setsymbol{LFI:FWHM:LFI22:units}{12\parcm87}
\setsymbol{LFI:FWHM:LFI23:units}{13\parcm27}
\setsymbol{LFI:FWHM:LFI24:units}{22\parcm98}
\setsymbol{LFI:FWHM:LFI25:units}{30\parcm46}
\setsymbol{LFI:FWHM:LFI26:units}{30\parcm31}
\setsymbol{LFI:FWHM:LFI27:units}{32\parcm65}
\setsymbol{LFI:FWHM:LFI28:units}{32\parcm66}

\setsymbol{LFI:FWHM:LFI18}{13.39}
\setsymbol{LFI:FWHM:LFI19}{13.01}
\setsymbol{LFI:FWHM:LFI20}{12.75}
\setsymbol{LFI:FWHM:LFI21}{12.74}
\setsymbol{LFI:FWHM:LFI22}{12.87}
\setsymbol{LFI:FWHM:LFI23}{13.27}
\setsymbol{LFI:FWHM:LFI24}{22.98}
\setsymbol{LFI:FWHM:LFI25}{30.46}
\setsymbol{LFI:FWHM:LFI26}{30.31}
\setsymbol{LFI:FWHM:LFI27}{32.65}
\setsymbol{LFI:FWHM:LFI28}{32.66}

\setsymbol{LFI:FWHM:uncertainty:LFI18:units}{0.170\arcm}
\setsymbol{LFI:FWHM:uncertainty:LFI19:units}{0.174\arcm}
\setsymbol{LFI:FWHM:uncertainty:LFI20:units}{0.170\arcm}
\setsymbol{LFI:FWHM:uncertainty:LFI21:units}{0.156\arcm}
\setsymbol{LFI:FWHM:uncertainty:LFI22:units}{0.164\arcm}
\setsymbol{LFI:FWHM:uncertainty:LFI23:units}{0.171\arcm}
\setsymbol{LFI:FWHM:uncertainty:LFI24:units}{0.652\arcm}
\setsymbol{LFI:FWHM:uncertainty:LFI25:units}{1.075\arcm}
\setsymbol{LFI:FWHM:uncertainty:LFI26:units}{1.131\arcm}
\setsymbol{LFI:FWHM:uncertainty:LFI27:units}{1.266\arcm}
\setsymbol{LFI:FWHM:uncertainty:LFI28:units}{1.287\arcm}

\setsymbol{LFI:FWHM:uncertainty:LFI18}{0.170}
\setsymbol{LFI:FWHM:uncertainty:LFI19}{0.174}
\setsymbol{LFI:FWHM:uncertainty:LFI20}{0.170}
\setsymbol{LFI:FWHM:uncertainty:LFI21}{0.156}
\setsymbol{LFI:FWHM:uncertainty:LFI22}{0.164}
\setsymbol{LFI:FWHM:uncertainty:LFI23}{0.171}
\setsymbol{LFI:FWHM:uncertainty:LFI24}{0.652}
\setsymbol{LFI:FWHM:uncertainty:LFI25}{1.075}
\setsymbol{LFI:FWHM:uncertainty:LFI26}{1.131}
\setsymbol{LFI:FWHM:uncertainty:LFI27}{1.266}
\setsymbol{LFI:FWHM:uncertainty:LFI28}{1.287}

\setsymbol{HFI:center:frequency:100GHz:units}{100\,GHz}
\setsymbol{HFI:center:frequency:143GHz:units}{143\,GHz}
\setsymbol{HFI:center:frequency:217GHz:units}{217\,GHz}
\setsymbol{HFI:center:frequency:353GHz:units}{353\,GHz}
\setsymbol{HFI:center:frequency:545GHz:units}{545\,GHz}
\setsymbol{HFI:center:frequency:857GHz:units}{857\,GHz}

\setsymbol{HFI:center:frequency:100GHz}{100}
\setsymbol{HFI:center:frequency:143GHz}{143}
\setsymbol{HFI:center:frequency:217GHz}{217}
\setsymbol{HFI:center:frequency:353GHz}{353}
\setsymbol{HFI:center:frequency:545GHz}{545}
\setsymbol{HFI:center:frequency:857GHz}{857}

\setsymbol{HFI:Ndetectors:100GHz}{8}
\setsymbol{HFI:Ndetectors:143GHz}{11}
\setsymbol{HFI:Ndetectors:217GHz}{12}
\setsymbol{HFI:Ndetectors:353GHz}{12}
\setsymbol{HFI:Ndetectors:545GHz}{3}
\setsymbol{HFI:Ndetectors:857GHz}{4}

\setsymbol{HFI:FWHM:Maps:100GHz:units}{9\parcm88}
\setsymbol{HFI:FWHM:Maps:143GHz:units}{7\parcm18}
\setsymbol{HFI:FWHM:Maps:217GHz:units}{4\parcm87}
\setsymbol{HFI:FWHM:Maps:353GHz:units}{4\parcm65}
\setsymbol{HFI:FWHM:Maps:545GHz:units}{4\parcm72}
\setsymbol{HFI:FWHM:Maps:857GHz:units}{4\parcm39}
\setsymbol{HFI:FWHM:Maps:100GHz}{9.88}
\setsymbol{HFI:FWHM:Maps:143GHz}{7.18}
\setsymbol{HFI:FWHM:Maps:217GHz}{4.87}
\setsymbol{HFI:FWHM:Maps:353GHz}{4.65}
\setsymbol{HFI:FWHM:Maps:545GHz}{4.72}
\setsymbol{HFI:FWHM:Maps:857GHz}{4.39}

\setsymbol{HFI:beam:ellipticity:Maps:100GHz}{1.15}
\setsymbol{HFI:beam:ellipticity:Maps:143GHz}{1.01}
\setsymbol{HFI:beam:ellipticity:Maps:217GHz}{1.06}
\setsymbol{HFI:beam:ellipticity:Maps:353GHz}{1.05}
\setsymbol{HFI:beam:ellipticity:Maps:545GHz}{1.14}
\setsymbol{HFI:beam:ellipticity:Maps:857GHz}{1.19}

\setsymbol{HFI:FWHM:Mars:100GHz:units}{9\parcm37}
\setsymbol{HFI:FWHM:Mars:143GHz:units}{7\parcm04}
\setsymbol{HFI:FWHM:Mars:217GHz:units}{4\parcm68}
\setsymbol{HFI:FWHM:Mars:353GHz:units}{4\parcm43}
\setsymbol{HFI:FWHM:Mars:545GHz:units}{3\parcm80}
\setsymbol{HFI:FWHM:Mars:857GHz:units}{3\parcm67}

\setsymbol{HFI:FWHM:Mars:100GHz}{9.37}
\setsymbol{HFI:FWHM:Mars:143GHz}{7.04}
\setsymbol{HFI:FWHM:Mars:217GHz}{4.68}
\setsymbol{HFI:FWHM:Mars:353GHz}{4.43}
\setsymbol{HFI:FWHM:Mars:545GHz}{3.80}
\setsymbol{HFI:FWHM:Mars:857GHz}{3.67}

\setsymbol{HFI:beam:ellipticity:Mars:100GHz}{1.18}
\setsymbol{HFI:beam:ellipticity:Mars:143GHz}{1.03}
\setsymbol{HFI:beam:ellipticity:Mars:217GHz}{1.14}
\setsymbol{HFI:beam:ellipticity:Mars:353GHz}{1.09}
\setsymbol{HFI:beam:ellipticity:Mars:545GHz}{1.25}
\setsymbol{HFI:beam:ellipticity:Mars:857GHz}{1.03}

\setsymbol{HFI:CMB:relative:calibration:100GHz}{$\lsim 1\%$}
\setsymbol{HFI:CMB:relative:calibration:143GHz}{$\lsim 1\%$}
\setsymbol{HFI:CMB:relative:calibration:217GHz}{$\lsim 1\%$}
\setsymbol{HFI:CMB:relative:calibration:353GHz}{$\lsim 1\%$}
\setsymbol{HFI:CMB:relative:calibration:545GHz}{}
\setsymbol{HFI:CMB:relative:calibration:857GHz}{}

\setsymbol{HFI:CMB:absolute:calibration:100GHz}{$\lsim 2\%$}
\setsymbol{HFI:CMB:absolute:calibration:143GHz}{$\lsim 2\%$}
\setsymbol{HFI:CMB:absolute:calibration:217GHz}{$\lsim 2\%$}
\setsymbol{HFI:CMB:absolute:calibration:353GHz}{$\lsim 2\%$}
\setsymbol{HFI:CMB:absolute:calibration:545GHz}{}
\setsymbol{HFI:CMB:absolute:calibration:857GHz}{}

\setsymbol{HFI:FIRAS:gain:calibration:accuracy:statistical:100GHz}{}
\setsymbol{HFI:FIRAS:gain:calibration:accuracy:statistical:143GHz}{}
\setsymbol{HFI:FIRAS:gain:calibration:accuracy:statistical:217GHz}{}
\setsymbol{HFI:FIRAS:gain:calibration:accuracy:statistical:353GHz}{2.5\%}
\setsymbol{HFI:FIRAS:gain:calibration:accuracy:statistical:545GHz}{1\%}
\setsymbol{HFI:FIRAS:gain:calibration:accuracy:statistical:857GHz}{0.5\%}

\setsymbol{HFI:FIRAS:gain:calibration:accuracy:systematic:100GHz}{}
\setsymbol{HFI:FIRAS:gain:calibration:accuracy:systematic:143GHz}{}
\setsymbol{HFI:FIRAS:gain:calibration:accuracy:systematic:217GHz}{}
\setsymbol{HFI:FIRAS:gain:calibration:accuracy:systematic:353GHz}{}
\setsymbol{HFI:FIRAS:gain:calibration:accuracy:systematic:545GHz}{7\%}
\setsymbol{HFI:FIRAS:gain:calibration:accuracy:systematic:857GHz}{7\%}

\setsymbol{HFI:FIRAS:zero:point:accuracy:100GHz:units}{0.8\MJysr}
\setsymbol{HFI:FIRAS:zero:point:accuracy:143GHz:units}{}
\setsymbol{HFI:FIRAS:zero:point:accuracy:217GHz:units}{}
\setsymbol{HFI:FIRAS:zero:point:accuracy:353GHz:units}{1.4\MJysr}
\setsymbol{HFI:FIRAS:zero:point:accuracy:545GHz:units}{2.2\MJysr}
\setsymbol{HFI:FIRAS:zero:point:accuracy:857GHz:units}{1.7\MJysr}

\setsymbol{HFI:FIRAS:zero:point:accuracy:100GHz}{0.8}
\setsymbol{HFI:FIRAS:zero:point:accuracy:143GHz}{}
\setsymbol{HFI:FIRAS:zero:point:accuracy:217GHz}{}
\setsymbol{HFI:FIRAS:zero:point:accuracy:353GHz}{1.4}
\setsymbol{HFI:FIRAS:zero:point:accuracy:545GHz}{2.2}
\setsymbol{HFI:FIRAS:zero:point:accuracy:857GHz}{1.7}

\setsymbol{HFI:unit:conversion:100GHz:units}{0.2415\MJysrmK}
\setsymbol{HFI:unit:conversion:143GHz:units}{0.3694\MJysrmK}
\setsymbol{HFI:unit:conversion:217GHz:units}{0.4811\MJysrmK}
\setsymbol{HFI:unit:conversion:353GHz:units}{0.2883\MJysrmK}
\setsymbol{HFI:unit:conversion:545GHz:units}{0.05826\MJysrmK}
\setsymbol{HFI:unit:conversion:857GHz:units}{0.002238\MJysrmK}

\setsymbol{HFI:unit:conversion:100GHz}{0.2415}
\setsymbol{HFI:unit:conversion:143GHz}{0.3694}
\setsymbol{HFI:unit:conversion:217GHz}{0.4811}
\setsymbol{HFI:unit:conversion:353GHz}{0.2883}
\setsymbol{HFI:unit:conversion:545GHz}{0.05826}
\setsymbol{HFI:unit:conversion:857GHz}{0.002238}

\setsymbol{HFI:colour:correction:alpha=-2:V1.01:100GHz}{0.9893}
\setsymbol{HFI:colour:correction:alpha=-2:V1.01:143GHz}{0.9759}
\setsymbol{HFI:colour:correction:alpha=-2:V1.01:217GHz}{1.0007}
\setsymbol{HFI:colour:correction:alpha=-2:V1.01:353GHz}{1.0028}
\setsymbol{HFI:colour:correction:alpha=-2:V1.01:545GHz}{1.0019}
\setsymbol{HFI:colour:correction:alpha=-2:V1.01:857GHz}{0.9889}

\setsymbol{HFI:colour:correction:alpha=0:V1.01:100GHz}{1.0008}
\setsymbol{HFI:colour:correction:alpha=0:V1.01:143GHz}{1.0148}
\setsymbol{HFI:colour:correction:alpha=0:V1.01:217GHz}{0.9909}
\setsymbol{HFI:colour:correction:alpha=0:V1.01:353GHz}{0.9888}
\setsymbol{HFI:colour:correction:alpha=0:V1.01:545GHz}{0.9878}
\setsymbol{HFI:colour:correction:alpha=0:V1.01:857GHz}{1.0014}

\providecommand{\sorthelp}[1]{}

\usepackage{natbib}
\usepackage[breaklinks,colorlinks,citecolor=blue]{hyperref}
\usepackage{longtable}
\usepackage{multirow}
\usepackage{fixltx2e}
\usepackage{breakurl}
\bibpunct{(}{)}{;}{a}{}{,} 
\usepackage{ifthen}

\begin{document}

\title{\Planck{} 2015 results. VI. LFI mapmaking}

\abstract{
This paper describes the mapmaking procedure applied to \Planck\ LFI (Low Frequency Instrument) data.
The mapmaking step takes as input the calibrated timelines and pointing information.
The main products are sky maps of $I,Q$, and $U$ Stokes components.
For the first time, we present polarization maps at LFI frequencies.
The mapmaking algorithm is based on a destriping technique, enhanced with a noise prior.
The Galactic region is masked to reduce errors arising  from bandpass mismatch and high signal gradients.
We apply horn-uniform radiometer weights to reduce effects of beam shape mismatch.
The algorithm is the same as used for the 2013 release, apart from small changes in parameter settings.
We validate the procedure through simulations. Special emphasis is put on the control of systematics,
which is particularly important for accurate polarization analysis.
We also produce low-resolution versions of the maps, and corresponding noise covariance matrices.
These serve as input in later analysis steps and parameter estimation.  
The noise covariance matrices are validated through noise Monte Carlo simulations.
The residual noise in the map products is characterized through analysis of half-ring maps,
noise covariance matrices, and simulations.

}

\authorrunning{Planck Collaboration}
\titlerunning{LFI mapmaking}

\author{\small
Planck Collaboration: P.~A.~R.~Ade\inst{82}
\and
N.~Aghanim\inst{57}
\and
M.~Ashdown\inst{66, 5}
\and
J.~Aumont\inst{57}
\and
C.~Baccigalupi\inst{81}
\and
A.~J.~Banday\inst{90, 7}
\and
R.~B.~Barreiro\inst{62}
\and
N.~Bartolo\inst{27, 63}
\and
E.~Battaner\inst{91, 92}
\and
K.~Benabed\inst{58, 89}
% ...
}
\institute{\small
APC, AstroParticule et Cosmologie, Universit\'{e} Paris Diderot, CNRS/IN2P3, CEA/lrfu, Observatoire de Paris, Sorbonne Paris Cit\'{e}, 10, rue Alice Domon et L\'{e}onie Duquet, 75205 Paris Cedex 13, France\goodbreak
\and
Aalto University Mets\"{a}hovi Radio Observatory and Dept of Radio Science and Engineering, P.O. Box 13000, FI-00076 AALTO, Finland\goodbreak
\and
African Institute for Mathematical Sciences, 6-8 Melrose Road, Muizenberg, Cape Town, South Africa\goodbreak
\and
Agenzia Spaziale Italiana Science Data Center, Via del Politecnico snc, 00133, Roma, Italy\goodbreak
\and
Astrophysics Group, Cavendish Laboratory, University of Cambridge, J J Thomson Avenue, Cambridge CB3 0HE, U.K.\goodbreak
\and
CITA, University of Toronto, 60 St. George St., Toronto, ON M5S 3H8, Canada\goodbreak
\and
CNRS, IRAP, 9 Av. colonel Roche, BP 44346, F-31028 Toulouse cedex 4, France\goodbreak
\and
CRANN, Trinity College, Dublin, Ireland\goodbreak
\and
California Institute of Technology, Pasadena, California, U.S.A.\goodbreak
\and
Centre for Theoretical Cosmology, DAMTP, University of Cambridge, Wilberforce Road, Cambridge CB3 0WA, U.K.\goodbreak
\and
Computational Cosmology Center, Lawrence Berkeley National Laboratory, Berkeley, California, U.S.A.\goodbreak
\and
Consejo Superior de Investigaciones Cient\'{\i}ficas (CSIC), Madrid, Spain\goodbreak
\and
DSM/Irfu/SPP, CEA-Saclay, F-91191 Gif-sur-Yvette Cedex, France\goodbreak
\and
DTU Space, National Space Institute, Technical University of Denmark, Elektrovej 327, DK-2800 Kgs. Lyngby, Denmark\goodbreak
\and
D\'{e}partement de Physique Th\'{e}orique, Universit\'{e} de Gen\`{e}ve, 24, Quai E. Ansermet,1211 Gen\`{e}ve 4, Switzerland\goodbreak
\and
Departamento de F\'{\i}sica, Universidad de Oviedo, Avda. Calvo Sotelo s/n, Oviedo, Spain\goodbreak
\and
Department of Astronomy and Astrophysics, University of Toronto, 50 Saint George Street, Toronto, Ontario, Canada\goodbreak
\and
Department of Astrophysics/IMAPP, Radboud University Nijmegen, P.O. Box 9010, 6500 GL Nijmegen, The Netherlands\goodbreak
\and
Department of Physics \& Astronomy, University of British Columbia, 6224 Agricultural Road, Vancouver, British Columbia, Canada\goodbreak
\and
Department of Physics and Astronomy, Dana and David Dornsife College of Letter, Arts and Sciences, University of Southern California, Los Angeles, CA 90089, U.S.A.\goodbreak
\and
Department of Physics and Astronomy, University College London, London WC1E 6BT, U.K.\goodbreak
\and
Department of Physics, Florida State University, Keen Physics Building, 77 Chieftan Way, Tallahassee, Florida, U.S.A.\goodbreak
\and
Department of Physics, Gustaf H\"{a}llstr\"{o}min katu 2a, University of Helsinki, Helsinki, Finland\goodbreak
\and
Department of Physics, Princeton University, Princeton, New Jersey, U.S.A.\goodbreak
\and
Department of Physics, University of California, Santa Barbara, California, U.S.A.\goodbreak
\and
Department of Physics, University of Illinois at Urbana-Champaign, 1110 West Green Street, Urbana, Illinois, U.S.A.\goodbreak
\and
Dipartimento di Fisica e Astronomia G. Galilei, Universit\`{a} degli Studi di Padova, via Marzolo 8, 35131 Padova, Italy\goodbreak
\and
Dipartimento di Fisica e Scienze della Terra, Universit\`{a} di Ferrara, Via Saragat 1, 44122 Ferrara, Italy\goodbreak
\and
Dipartimento di Fisica, Universit\`{a} La Sapienza, P. le A. Moro 2, Roma, Italy\goodbreak
\and
Dipartimento di Fisica, Universit\`{a} degli Studi di Milano, Via Celoria, 16, Milano, Italy\goodbreak
\and
Dipartimento di Fisica, Universit\`{a} degli Studi di Trieste, via A. Valerio 2, Trieste, Italy\goodbreak
\and
Dipartimento di Fisica, Universit\`{a} di Roma Tor Vergata, Via della Ricerca Scientifica, 1, Roma, Italy\goodbreak
\and
Dipartimento di Matematica, Universit\`{a} di Roma Tor Vergata, Via della Ricerca Scientifica, 1, Roma, Italy\goodbreak
\and
Discovery Center, Niels Bohr Institute, Blegdamsvej 17, Copenhagen, Denmark\goodbreak
\and
Dpto. Astrof\'{i}sica, Universidad de La Laguna (ULL), E-38206 La Laguna, Tenerife, Spain\goodbreak
\and
European Space Agency, ESAC, Planck Science Office, Camino bajo del Castillo, s/n, Urbanizaci\'{o}n Villafranca del Castillo, Villanueva de la Ca\~{n}ada, Madrid, Spain\goodbreak
\and
European Space Agency, ESTEC, Keplerlaan 1, 2201 AZ Noordwijk, The Netherlands\goodbreak
\and
Facolt\`{a} di Ingegneria, Universit\`{a} degli Studi e-Campus, Via Isimbardi 10, Novedrate (CO), 22060, Italy\goodbreak
\and
Gran Sasso Science Institute, INFN, viale F. Crispi 7, 67100 L'Aquila, Italy\goodbreak
\and
HGSFP and University of Heidelberg, Theoretical Physics Department, Philosophenweg 16, 69120, Heidelberg, Germany\goodbreak
\and
Haverford College Astronomy Department, 370 Lancaster Avenue, Haverford, Pennsylvania, U.S.A.\goodbreak
\and
Helsinki Institute of Physics, Gustaf H\"{a}llstr\"{o}min katu 2, University of Helsinki, Helsinki, Finland\goodbreak
\and
INAF - Osservatorio Astronomico di Padova, Vicolo dell'Osservatorio 5, Padova, Italy\goodbreak
\and
INAF - Osservatorio Astronomico di Roma, via di Frascati 33, Monte Porzio Catone, Italy\goodbreak
\and
INAF - Osservatorio Astronomico di Trieste, Via G.B. Tiepolo 11, Trieste, Italy\goodbreak
\and
INAF/IASF Bologna, Via Gobetti 101, Bologna, Italy\goodbreak
\and
INAF/IASF Milano, Via E. Bassini 15, Milano, Italy\goodbreak
\and
INFN, Sezione di Bologna, Via Irnerio 46, I-40126, Bologna, Italy\goodbreak
\and
INFN, Sezione di Roma 1, Universit\`{a} di Roma Sapienza, Piazzale Aldo Moro 2, 00185, Roma, Italy\goodbreak
\and
INFN, Sezione di Roma 2, Universit\`{a} di Roma Tor Vergata, Via della Ricerca Scientifica, 1, Roma, Italy\goodbreak
\and
INFN/National Institute for Nuclear Physics, Via Valerio 2, I-34127 Trieste, Italy\goodbreak
\and
IUCAA, Post Bag 4, Ganeshkhind, Pune University Campus, Pune 411 007, India\goodbreak
\and
Imperial College London, Astrophysics group, Blackett Laboratory, Prince Consort Road, London, SW7 2AZ, U.K.\goodbreak
\and
Infrared Processing and Analysis Center, California Institute of Technology, Pasadena, CA 91125, U.S.A.\goodbreak
\and
Institut N\'{e}el, CNRS, Universit\'{e} Joseph Fourier Grenoble I, 25 rue des Martyrs, Grenoble, France\goodbreak
\and
Institut Universitaire de France, 103, bd Saint-Michel, 75005, Paris, France\goodbreak
\and
Institut d'Astrophysique Spatiale, CNRS (UMR8617) Universit\'{e} Paris-Sud 11, B\^{a}timent 121, Orsay, France\goodbreak
\and
Institut d'Astrophysique de Paris, CNRS (UMR7095), 98 bis Boulevard Arago, F-75014, Paris, France\goodbreak
\and
Institute of Astronomy, University of Cambridge, Madingley Road, Cambridge CB3 0HA, U.K.\goodbreak
\and
Institute of Theoretical Astrophysics, University of Oslo, Blindern, Oslo, Norway\goodbreak
\and
Instituto de Astrof\'{\i}sica de Canarias, C/V\'{\i}a L\'{a}ctea s/n, La Laguna, Tenerife, Spain\goodbreak
\and
Instituto de F\'{\i}sica de Cantabria (CSIC-Universidad de Cantabria), Avda. de los Castros s/n, Santander, Spain\goodbreak
\and
Istituto Nazionale di Fisica Nucleare, Sezione di Padova, via Marzolo 8, I-35131 Padova, Italy\goodbreak
\and
Jet Propulsion Laboratory, California Institute of Technology, 4800 Oak Grove Drive, Pasadena, California, U.S.A.\goodbreak
\and
Jodrell Bank Centre for Astrophysics, Alan Turing Building, School of Physics and Astronomy, The University of Manchester, Oxford Road, Manchester, M13 9PL, U.K.\goodbreak
\and
Kavli Institute for Cosmology Cambridge, Madingley Road, Cambridge, CB3 0HA, U.K.\goodbreak
\and
LAL, Universit\'{e} Paris-Sud, CNRS/IN2P3, Orsay, France\goodbreak
\and
LAPTh, Univ. de Savoie, CNRS, B.P.110, Annecy-le-Vieux F-74941, France\goodbreak
\and
LERMA, CNRS, Observatoire de Paris, 61 Avenue de l'Observatoire, Paris, France\goodbreak
\and
Laboratoire AIM, IRFU/Service d'Astrophysique - CEA/DSM - CNRS - Universit\'{e} Paris Diderot, B\^{a}t. 709, CEA-Saclay, F-91191 Gif-sur-Yvette Cedex, France\goodbreak
\and
Laboratoire Traitement et Communication de l'Information, CNRS (UMR 5141) and T\'{e}l\'{e}com ParisTech, 46 rue Barrault F-75634 Paris Cedex 13, France\goodbreak
\and
Laboratoire de Physique Subatomique et Cosmologie, Universit\'{e} Grenoble-Alpes, CNRS/IN2P3, 53, rue des Martyrs, 38026 Grenoble Cedex, France\goodbreak
\and
Laboratoire de Physique Th\'{e}orique, Universit\'{e} Paris-Sud 11 \& CNRS, B\^{a}timent 210, 91405 Orsay, France\goodbreak
\and
Lawrence Berkeley National Laboratory, Berkeley, California, U.S.A.\goodbreak
\and
Lebedev Physical Institute of the Russian Academy of Sciences, Astro Space Centre, 84/32 Profsoyuznaya st., Moscow, GSP-7, 117997, Russia\goodbreak
\and
Max-Planck-Institut f\"{u}r Astrophysik, Karl-Schwarzschild-Str. 1, 85741 Garching, Germany\goodbreak
\and
McGill Physics, Ernest Rutherford Physics Building, McGill University, 3600 rue University, Montr\'{e}al, QC, H3A 2T8, Canada\goodbreak
\and
National University of Ireland, Department of Experimental Physics, Maynooth, Co. Kildare, Ireland\goodbreak
\and
Niels Bohr Institute, Blegdamsvej 17, Copenhagen, Denmark\goodbreak
\and
SB-ITP-LPPC, EPFL, CH-1015, Lausanne, Switzerland\goodbreak
\and
SISSA, Astrophysics Sector, via Bonomea 265, 34136, Trieste, Italy\goodbreak
\and
School of Physics and Astronomy, Cardiff University, Queens Buildings, The Parade, Cardiff, CF24 3AA, U.K.\goodbreak
\and
School of Physics and Astronomy, University of Nottingham, Nottingham NG7 2RD, U.K.\goodbreak
\and
Sorbonne Universit\'{e}-UPMC, UMR7095, Institut d'Astrophysique de Paris, 98 bis Boulevard Arago, F-75014, Paris, France\goodbreak
\and
Space Sciences Laboratory, University of California, Berkeley, California, U.S.A.\goodbreak
\and
Special Astrophysical Observatory, Russian Academy of Sciences, Nizhnij Arkhyz, Zelenchukskiy region, Karachai-Cherkessian Republic, 369167, Russia\goodbreak
\and
Sub-Department of Astrophysics, University of Oxford, Keble Road, Oxford OX1 3RH, U.K.\goodbreak
\and
Theory Division, PH-TH, CERN, CH-1211, Geneva 23, Switzerland\goodbreak
\and
UPMC Univ Paris 06, UMR7095, 98 bis Boulevard Arago, F-75014, Paris, France\goodbreak
\and
Universit\'{e} de Toulouse, UPS-OMP, IRAP, F-31028 Toulouse cedex 4, France\goodbreak
\and
University of Granada, Departamento de F\'{\i}sica Te\'{o}rica y del Cosmos, Facultad de Ciencias, Granada, Spain\goodbreak
\and
University of Granada, Instituto Carlos I de F\'{\i}sica Te\'{o}rica y Computacional, Granada, Spain\goodbreak
% ...
}


\maketitle

\section{Introduction}

This paper is one of a set associated with the 2015 release of data from the \Planck\footnote{\Planck\ (\url{http://www.esa.int/Planck}) is a project of the 
European Space Agency  (ESA) with instruments provided by two scientific 
consortia funded by ESA member states and led by Principal Investigators 
from France and Italy, telescope reflectors provided through a collaboration 
between ESA and a scientific consortium led and funded by Denmark, and 
additional contributions from NASA (USA).}
 mission \citep{planck2013-p01}.
It describes the mapmaking procedure applied to time-ordered data from the \Planck\ Low Frequency Instrument (LFI).
Mapmaking is the next step in the LFI pipeline after calibration and dipole removal. It is followed by
bandpass correction and component separation.
A description of the complete data analysis pipeline is given in \cite{planck2014-a03}.

The mapmaking procedure applied to 2013 data was discussed in \cite{planck2013-p02}.
In this work we present a thorough validation on the procedure,
and set focus on the control of systematics \citep{planck2014-a04}.  For the first time we present polarisation results.
The mapmaking approach adopted for HFI is described in \cite{planck2014-a09}.

Mapmaking takes as its input the calibrated timelines, from which the cosmological and orbital dipole signals have been removed.
Also, an estimate for the Galactic straylight  is subtracted from the timelines prior to mapmaking,
since this is difficult to correct for at map level. 
Calibration and removal of stray light are described in \cite{planck2014-a06}.

In addition to the time-ordered data (TOI), required inputs are, the radiometer pointing ($\theta,\phi,\psi$) for each sample,
information of the polarization sensitivity of each radiometer, and noise description.
Output consists of sky maps of temperature and Stokes $Q$ and $U$ polarization, 
and a description of residual noise in them.
Effects of beam shape are not corrected for at the mapmaking level.
The LFI beams and the associated window functions are described in \cite{planck2014-a05}.

The time-ordered data are contaminated by correlated $1/f$ noise.
The aim of the mapmaking step is to remove the correlated noise as accurately as possible,
while simultaneously minimising systematic errors. 
Various mapmaking algorithms in the context of \Planck\ are compared in \cite{poutanen2006,ashdown2007a, ashdown2007b,ashdown2009}.

The mapmaking step is computationally challenging, due to the huge amount of data.
For optimal removal of noise, all of the time-ordered data must be held in computer memory at the same time.
This leads to a large memory requirement (of the order of a TB for 70\,GHz).
The size of the output map is much less (of the order of 100 MB per map).
Mapmaking is thus also a data compression step.

Providing a statistical description of the residual noise present in the maps, in the form of pixel-pixel noise covariance matrices, 
is part of the mapmaking step. Covariance matrices are used in a number of different places in the data-analysis pipeline.  
These matrices are impossible to employ at the native map resolution because of resource limitations. 
At present, the low-resolution products can only be used efficiently at resolution $N_{\textrm{side}} = 16$ or lower. 
Consequently, we also produce corresponding low-resolution maps from the high-resolution maps.

This paper is organized as follows. 
In Sect. 2 we provide an overview of the mapmaking procedure,
with emphasis on the control of systematics.
Validation of the procedure with simulations is presented in Sect. 3. 
Production of low-resolution maps and related noise covariance matrices are discussed in Sect. 4.
In Sect. 5 we give an overview of the mapmaking products and characterize the residual noise in them.
A summary and conclusions are given in Sect. 6.

\section{Mapmaking procedure}
\label{sec:mapmaking}

\subsection{Madam}

LFI maps were produced by the {\tt Madam} mapmaking code, version number 3.7.4.
The algorithm is based on the destriping technique, enhanced with a noise prior.
The destriping technique as a tool for mapmaking is discussed in
\cite{maino2002}, \cite{keihanen2004}, and \cite{kurki-suonio2009}.
A noise prior allows us to extend the destriping technique to very short baseline lengths,
allowing more accurate noise removal.
The {\tt Madam}  algorithm is described in \cite{keihanen2005} and \cite{keihanen2010}.
The code is the same as used for the 2013 release,
but the parameter settings have been altered.
The 2013 mapmaking procedure was discussed in \cite{planck2013-p02}.
Here we give an overview of the method for convenience.
We focus on changes made after the 2013 release, and on aspects relating to polarization.

Consider a time-ordered data stream
\begin{equation}
\vec{y} = \vec{s}+\vec{n'}
\end{equation}
where $\vec{s}$ represents the sky signal and $\vec{n'}$ is noise.
In the destriping technique, the correlated noise component is modelled by a sequence of offsets, here called baselines. 
The same offset applies for $N$ adjacent samples in time-ordered-data.
In the conventional language $N$ is referred to as the ``baseline length,'' while ``baseline'' alone refers to the value of the offset. 

The baselines are assumed to be the result of a random process.
On top of the correlated noise, pure white noise with known variance is assumed.
Formally we can write
\begin{equation}
\vec{n'} = \tens{F}\vec{a}+\vec{n}.
\end{equation}
Vector $\vec{a}$  represents the baselines, and $\tens{F}$ is formally a matrix that spreads the baselines 
into the time-ordered data.  Vector $n$ represents the white noise component.
The noise model ignores correlated noise at frequencies that exceed the inverse of the baseline length.
The technique is suitable for $1/f$-type noise that dies out at high frequencies.
The baseline length should be chosen to be short enough to capture all frequencies below 
or comparable to the knee frequency of the $1/f$ spectrum.

The signal component is modelled as
\begin{equation}
\vec{s}= \tens{P}\vec{m} ,  \label{signalmodel}
\end{equation}
where $\vec{m}$ is a pixelized sky map, and $\tens{P}$ is a pointing matrix.
Map $\vec{m}$ consists of three components, representing temperature and $Q$ and $U$ polarization.
We can thus write for sample $i$:
\begin{equation}
s_i=T(p_i)+Q(p_i)\cos(2\psi_i)+U(p_i)\sin(2\psi_i).
\end{equation}
Here $p_i$ is the pixel hit by sample $i$, and $\psi_i$ is the rotation angle between the local meridian
and the vector that defines the direction of polarization sensitivity of the beam.

Each sample is assigned entirely to the pixel in which the beam centre falls.
Beam shape is not taken into account in this context. 
We are also ignoring cross-polarization, which for LFI is below 1\% \citep{planck2014-a04}.
Formally then, $\tens{P}$ is  a sparse matrix with three non-zero elements in each row.
The factors $\cos(2\psi_i)$ and $\sin(2\psi_i)$ are included in $\tens{P}$.
Limitations of the model and their impact on mapmaking are discussed in Sect. \ref{sec:systematics}.

In the conventional destriping technique no prior information on the baselines is used.
On the other hand, if the noise spectrum is known, one can construct a {\em noise prior}
\begin{equation}
\tens{C} _{\rm a}=\langle \vec{a} \vec{a}^\mathrm{T}\rangle.   \label{noiseprior}
\end{equation}
The noise prior provides the extra constraint that allows us to extend the destriping technique to very short baselines.
We construct the noise prior from the known noise parameters (knee frequency, white noise variance, and spectral slope),
as described in \cite{keihanen2010}.
It can be shown that the destriping solution for the map gives the maximum-likelihood map
in the limiting case where the baseline length approaches one sample \citep{keihanen2005}.

With the assumptions listed above, the baseline vector $\vec{a}$ can be solved from the linear system of equations
\begin{equation}
(\tens{F}^{\rm T} \tens{C}_{\rm w}^{-1}\tens{Z}\tens{F} +\tens{C}_{\rm a}^{-1})\vec{\hat a} 
= \tens{F}^{\rm T} \tens{C}_{\rm w}^{-1}\tens{Z} \vec y,  \label{baseeq}
\end{equation}
where
\begin{equation}
\tens{Z}=\tens{I}-\tens{P}(\tens{P}^{\rm ^{\rm T}T}\tens{C}_{\rm w}^{-1}\tens{P})^{-1}\tens{P}^{\rm T}\tens{C}_{\rm w}^{-1}.   \label{Zmatrix}
\end{equation}
Here $\tens{C}_{\rm w}$ is the white noise covariance, formally a diagonal matrix with the white noise variance on the diagonal.
The final {destriped map} is then constructed as 
\begin{equation}
  \vec{\hat m} = (\tens{P}^{\rm T}\tens{C}_{\rm w}^{-1}\tens{P})^{-1}\tens{P}^{\rm T}\tens{C}_{\rm w}^{-1}
       (\vec{y}-\tens{F}\vec{\hat a}),   \label{map-binning}
\end{equation}
where $\vec{\hat a}$ is the baseline solution from (\ref{baseeq}).
We use the hat symbol to indicate that $\vec{\hat a}$ and $\vec{\hat m}$ are estimates for the true $\vec{a}$ and $\vec{m}$.
The ``binned map" is constructed from the timeline the same way without baseline removal
\begin{equation}
  \vec{m}_{\rm b} = (\tens{P}^{\rm T}\tens{C}_{\rm w}^{-1}\tens{P})^{-1}\tens{P}^{\rm T}\tens{C}_{\rm n}^{-1} \vec{y}.    \label{eq:binned}
\end{equation}

We include the same set of radiometers and the same time range in destriping phase and in map-binning phase.
In this respect our approach is different from the one adopted for HFI \citep{planck2014-a09}.
Our approach has the benefit that single-year or single-survey maps,
or maps constructed from different radiometer sets,
have independent noise contributions.
The price to pay is that the baseline solution in the partial maps is less accurate than
in the full survey maps.

Adding a uniform offset to the time-ordered data does not have an impact on the baseline solution. 
This can be seen by inserting a unity vector $\vec{y}=\vec{1}$ to the right-hand side of Eq. (\ref{baseeq}).
A unity vector can be expressed as $\vec{1}=\tens{P}\vec{m}_1$, where $\vec{m}_1$ is an ($I,Q,U$) 
map with temperature component equal to one
and with vanishing polarization. The operation $\tens{ZP}\vec{m}_1$ yields zero, and consequently to a zero baseline solution.
Based on similar arguments, destriping cannot determine the absolute monopole of the temperature map,
since this is indistinguishable from an offset in the timeline.

We do not impose an explicit constraint on the sequence of baselines to avoid the degeneracy
between the monopole of the maps and an offset in TOI.
Instead, we make use of the fact the conjugate gradient iteration method works for singular systems as well,
under the condition that the same eigenmodes vanish on both sides of the equation. 
The raw output temperature map will in general have an arbitrary non-zero mean that depends on the input signal
and noise in a non-trivial way. 
A monopole is subtracted from each released temperature map in post-processing,
as described in \cite{planck2014-a03}.

Similar arguments do not apply to an offset difference between radiometers.
In contrast to a common offset, a differential offset cannot be presented as a map,
because it does not obey the rotation properties of the $Q,U$ maps.
A differential offset is distinguished from a map by the destriping procedure, and does not leak into polarization,
as long as there are enough crossing points where the same detector scans the same pixel in different orientations.

The destriping technique produces the final map through a procedure in which one first solves for the baselines,
and then bins the map from the data stream from which the baselines have been removed. 
This two-step procedure has a benefit over direct maximum-likelihood methods, 
because it enables us to use techniques to make the map diverge from the maximum-likelihood solution
and thus have a better control of systematics.
For instance, one can apply a mask in the destriping phase while still binning the final map to cover the whole sky,
in order to reduce signal error due to strong signal gradients.
We discuss this in more detail in Sect. \ref{sec:systematics}.

  \begin{table*}
  \begingroup
  \newdimen\tblskip \tblskip=5pt
  \caption{Frequency-specific mapmaking parameters and related information. From left: sampling frequency;
  knee frequency range;   and chosen baseline length  in seconds and as a number of samples.  
  The three rightmost columns show the FWHM range of the main beams of the radiometers of the channel,
  the chosen destriping resolution as the $N_\textrm{side}$ parameter of {\tt HEALPix}, and the average size of one pixel.}
  \label{tab:frequency_parameters}
  \nointerlineskip
  \vskip -3mm
  \footnotesize
  \setbox\tablebox=\vbox{
  \newdimen\digitwidth
  \setbox0=\hbox{\rm 0}
  \digitwidth=\wd0
  \catcode`*=\active
  \def*{\kern\digitwidth}
  \newdimen\signwidth
  \setbox0=\hbox{+}
  \signwidth=\wd0
  \catcode`!=\active
  \def!{\kern\signwidth}
  \halign{\hbox {#}\tabskip=2em&
      \hfil#\hfil& \hfil#\hfil&\hfil#\hfil& \hfil#\hfil& \hfil#\hfil&  \hfil#\hfil&
      \hfil#\hfil\tabskip=0pt\cr
  \noalign{\doubleline}
  \omit&\omit & \omit 
  &\multispan2\hfil B{\sc aseline length} \hfil &
  &\multispan2\hfil R{\sc esolution}\hfil
  \cr
  \noalign{\vskip -4pt}
  \omit&\omit &\omit &\multispan2\hrulefill &\omit&\multispan2\hrulefill \cr
Channel & $f_\mathrm{samp}$ [Hz]  & $f_{\rm kn}$ [Hz]& s & samples  & FWHM [arcmin] & $N_\textrm{side}$  & arcmin\cr
  \noalign{\vskip 3pt\hrule\vskip 5pt}
 30\,GHz & 32.508  & 0.043--0.175 & 0.246 & 8  & 33.0-33.2 & 1024 & 3.44 \cr
44\,GHz & 46.545  & 0.020--0.088  & 0.988 & 46  & 23.0-30.8 & 1024 & 3.44 \cr
70\,GHz & 78.769  & 0.006--0.059 & 1.000 & 79 & 12.8-13.5 & 1024/2048 & 3.44/1.72\cr
  \noalign{\vskip 5pt\hrule\vskip 3pt}}}
  \endPlancktable
  \endgroup
  \end{table*}

\subsubsection{Baseline length and resolution}

The baseline length is a key parameter in the  {\tt Madam} mapmaking algorithm.
The chosen baseline length is a trade-off between computational burden and optimal noise removal.
We have selected baselines that are well below the timescale given by the knee frequencies of each channel.
In the case of 44\,GHz and 70\,GHz we have chosen to use one second long baselines.
For 30\,GHz, where the knee frequencies are higher, we have used shorter baselines of 0.25\,s.
This is different from the 2013 release, where we used one second baselines for all channels.
The exact baseline lengths and channel sampling frequencies are given in Table \ref{tab:frequency_parameters}.
The LFI knee frequencies are given in Table \ref{tab:noise_parameters} along with other noise parameters.

{\tt Madam} maps follow the  {\tt HEALPix}\footnote{\url{http://healpix.sourceforge.net}} pixelization scheme, where map resolution is defined by parameter $N_\textrm{side}$.
The total number of pixels in map is $12N_\textrm{side} ^2$ \citep{gorski2005}.
Map resolution was chosen according to the beam width.
For the majority of maps we used resolution $N_\textrm{side}$=1024.
At this resolution, the average pixel size corresponds to 1/10 of the beam FWHM at 30\,GHz or 1/4 at 70\,GHz.
A subset of 70\,GHz maps was redone with resolution $N_\textrm{side}$=2048.
All computations were carried out in Galactic coordinate frame.
Pixel sizes are compared with beam widths in Table \ref{tab:noise_parameters}.

\subsubsection{Missing data}

Periods of stable pointing are regularly interrupted by manoeuvring periods, where the telescope is repointed according to the scanning strategy.
These repointing periods are discarded from mapmaking. 
In addition there are other data sections that must be discarded, for instance due to missing data, gain saturation or crossing of a bright object \citep{planck2014-a03}.
We accomplish this by formally setting the white noise variance $\tens{C}_{\rm w}$ to infinity for the flagged samples.
The procedure is different from removing the flagged data sections and appending the remaining sections end-to-end.
As a result, the flagged samples do not contribute to the output map, but the correct noise correlation is preserved across the gaps.

\subsubsection{Single-survey maps and handling of missing pixels}
\label{sec:missing_pixels}

The 2015 release includes polarization maps for individual surveys (6 months of data),
which do not cover the whole sky.  
Obviously, pixels that are not observed at all, are not included in the maps.
In addition there are pixels that are scanned by a single horn.
The polarization analysis of those pixels requires special attention.
To reliably solve the $I$, $Q$, and $U$ components for a pixel, the pixel has to be scanned by two horns with orthogonal polarization sensitivity,
or, alternatively,  twice by same horn with sufficiently different orientations of the beam.
If this is not the case, we discard the pixel from the map. 
Even these pixels, however, can be used for destriping as we now describe.

We assess the goodness of a pixel through the matrix
\begin{equation}
\tens{P}^{\rm T}\tens{C}_{\rm w}^{-1}\tens{P}  \label{pixelmatrix}.
\end{equation}
This is a block-diagonal matrix with $3\times3$ blocks along the diagonal. Each $3\times3$ block corresponds to one pixel, 
and represents the inverse of the white noise covariance within the pixel. The matrix
gives an accurate diagnosis 
of a pixel's usefulness for polarization analysis.
Pixels with insufficient polarization coverage yield a singular or badly scaled matrix.
The inverse of the pixel matrix appears in the destriping Eqs.  (\ref{baseeq}) and (\ref{Zmatrix}).
It is obvious that a singular matrix leads to a non-stable solution for the map.
However, in Eq. (\ref{baseeq}) the pixel matrix appears sandwiched between two occurrences
of the same pointing matrix, which lack the same eigenmodes. Matrix $\tens{Z}$ is thus finite even if the pixel matrix is singular.
This can also be seen from the fact that $\tens{Z}$ is a projection matrix, which means that all its eigenvalues are 1 or 0.

If the pixel matrix is singular, but not identically zero, we invert the non-singular eigenmodes.
That way we can also utilize pixels that have been observed by a single horn.
In the final map-binning phase we cannot use this same procedure. Instead, the badly determined pixels are excluded.
We use as rejection criterion the reciprocal condition ``rcond" value of the pixel matrix.
For the 2015 release we chose the conservative limit $\mathrm{rcond}>0.01$.
Statistics on the sky coverage of LFI maps is provided in Sect. 5.

\subsubsection{Single-horn and single-radiometer maps}

In order to decompose the signal into $I,Q,U$ components, it is necessary to have at least three measurements from the same pixel, 
in different polarization angles. 
In a limited region around the ecliptic poles it is possible to determine the polarization signal already from one LFI horn,
since the region is scanned in different angles by the same radiometer pair,  but for most of the sky we need observations from 
two horns (four radiometers).

We have produced temperature maps from each horn's or radiometer's data.
The same formulation as above applies, but the pointing matrix $\tens{P}$ only contains the temperature column.
Consequently, the matrix in Eq. (\ref{pixelmatrix}) becomes diagonal and is trivial to invert.

In the case of horn maps, the polarization signal largely cancels out when the two radiometers are combined,
and the map gives a reliable estimate of the temperature signal. 
The single-radiometer maps, instead, are a mixture of temperature and polarization signals,
which must be kept in mind when analysing these maps.

  \begin{table*}
  \begingroup
  \newdimen\tblskip \tblskip=5pt
  \caption{Noise parameters for LFI radiometers.
  Values for $f_{\rm knee}$, $\beta$, and $\sigma$ given here were used for the construction of the noise prior.
  We give also radiometer weights for pure noise weighting and for the horn-uniform weighting scheme. 
  Horn-uniform weighting was used for actual mapmaking.
  }
  \label{tab:noise_parameters}
  \nointerlineskip
  \vskip -3mm
  \footnotesize
 \setbox\tablebox=\vbox{
  \newdimen\digitwidth
  \setbox0=\hbox{\rm 0}
  \digitwidth=\wd0
  \catcode`*=\active
  \def*{\kern\digitwidth}
  \newdimen\signwidth
  \setbox0=\hbox{+}
  \signwidth=\wd0
  \catcode`!=\active
  \def!{\kern\signwidth}
  %
 \halign{\hbox to 18mm{#}\tabskip=2em&
      \hfil#\hfil& \hfil#\hfil&\hfil$#$\hfil& \hfil$#$\hfil& \hfil#\hfil&  \hfil#\hfil&
      \hfil#\hfil& \hfil#\hfil&  \hfil#\hfil\tabskip=0pt\cr
  \noalign{\doubleline}
  \omit&\multispan2\hfil K{\sc nee} F{\sc requency} $f_{\rm knee}$ [mHz]\hfil
  &\multispan2\hfil S{\sc lope} $\beta$\hfil
  &\multispan2\hfil W{\sc hite noise}  $\sigma$ [mK]\hfil
  &\multispan2 W{\sc eight}  $\tens{C}_{\rm{w}}^{-1}$ (norm.)\hfil 
  &H{\sc orn-uniform}\hfil 
  \cr
  \noalign{\vskip -4pt}
  \omit&\multispan2\hrulefill&\multispan2\hrulefill&\multispan2\hrulefill &\multispan2\hrulefill &\hrulefill\cr
  \omit& M&S&\rm M&\rm S&M&S&M&S&M \& S\cr
  \noalign{\vskip 3pt\hrule\vskip 5pt}
  \omit{\bf 70\,\,GHz}\hfil\cr
  \noalign{\vskip 4pt}
  \hglue 1em LFI-18 & 14.82 & 17.78  & -1.060 & -1.180  & 4.553 & 4.146 &   0.1676  & 0.2226  & 0.1945 \cr
  \hglue 1em LFI-19 & 11.72 & 13.71  & -1.207 & -1.111 & 5.144 & 4.926 &   0.1313 &  0.1577  & 0.1454 \cr
  \hglue 1em LFI-20 &  7.96 & 5.67 & -1.198 & -1.298 & 5.212 & 5.507   &0.1279 & 0.1262 &  0.1283  \cr
  \hglue 1em LFI-21 & 37.89 & 13.27 & -1.247 &-1.205 & 4.003 & 4.971  &  0.2168  & 0.1548  & 0.1811 \cr
  \hglue 1em LFI-22 & 9.68  & 14.80 & -1.424 & -1.237 & 4.356 & 4.715   & 0.1831  & 0.1721  & 0.1790 \cr
  \hglue 1em LFI-23 & 29.65  & 59.03 & -1.073 & -1.211  & 4.476 & 4.790  & 0.1734   & 0.1667 &  0.1716 \cr
  \noalign{\vskip 5pt}
  \omit{\bf 44\,\,GHzGHz}\hfil\cr
  \noalign{\vskip 4pt}
  \hglue 1em  LFI-24 & 26.78  & 88.30  & -0.942 & -0.908 & 3.159  & 2.734  &  0.3163  & 0.3421 &  0.3279 \cr
  \hglue 1em LFI-25 & 20.07  & 46.37  & -0.845 & -0.904 & 2.834  & 2.698  &   0.3931  & 0.3513 & 0.3739 \cr
  \hglue 1em LFI-26 & 64.42 & 68.19  & -0.918 & -0.758 & 3.295 & 2.887  &  0.2907  & 0.3066 &  0.2982 \cr
   \noalign{\vskip 5pt}
  \omit{\bf 30\,GHz}\hfil\cr
  \noalign{\vskip 4pt}
  \hglue 1em LFI-27   & 174.53 &  108.79 & -0.927 & -0.907 & 1.605  & 1.729 &   0.5602  & 0.4716 &  0.5167 \cr
  \hglue 1em LFI-28   & 130.08 & 43.08 & -0.931 & -0.900  & 1.812 & 1.633  &  0.4398 &  0.5284 &  0.4833 \cr
  \noalign{\vskip 5pt\hrule\vskip 3pt}}}
  \endPlancktable
  \endgroup
  \end{table*}

\subsection{Noise prior}

We use a prior $\tens{C}_{\rm a}^{-1}$ for the noise baselines  to constrain the destriping solution. 
Matrix $\tens{C}_{\rm a}$ gives the expected correlation between noise baselines,  as in Eq. (\ref{noiseprior}).
The details of the computation from a given input spectrum are given in \cite{keihanen2010}.
The prior makes it possible to use very short baselines,
and thus to model the noise accurately.

We construct the prior from the noise parameters ($\sigma$, $f_{\rm kn}$, $\beta$),
which are listed in Table \ref{tab:noise_parameters}.
The noise spectrum is given by
\begin{equation}
P(f) = \frac{\sigma^2}{f_{\rm samp}}  \left( \frac{f}{f_{\rm kn}} \right) ^\beta ,  \label{noisespec}
\end{equation}
where $f_{\rm samp}$ is the sampling frequency (see Table \ref{tab:frequency_parameters}).
Below a cut-off frequency $f_{\rm min}=$1/(3600\,s), the spectrum is assigned a constant value $P(f_{\rm min})$,
to avoid numerical problems related to the divergence at $f=0$.

On top of the correlated noise component there is assumed to be pure white noise with constant spectrum
\begin{equation}
P_\mathrm{wn}(f) = \frac{\sigma^2}{f_{\rm samp}}. 
\end{equation}
The noise prior is based on the correlated component of Eq. (\ref{noisespec}) only, since this is the component modelled by the baselines.

We assume that the noise properties do not change significantly with time,
and apply the same prior throughout the mission.
We also assume that noise is uncorrelated between radiometers,
and from one pointing period to another.
Under these assumptions the correlation matrix $\tens{C}_{\rm a}$
has a simple form and can be applied efficiently using Fourier techniques.

\subsection{Handling of systematic effects}
\label{sec:systematics}

Signal variations that are not captured by the signal model (Eq. (\ref{signalmodel})) give rise to additional uncertainty in baseline determination,
referred to as ``signal error". 
There are three main sources for signal error:  signal variations within a pixel; differences in radiometer 
frequency response (bandpass mismatch) \citep{leahy2010};
and beam shape mismatch. 
Cross-polarization in this context can be understood as part of beam mismatch.

Other sources include variable sources and gain estimation error from the 
preceding calibration step.
All these have the effect that the same sky pixel, when observed by different radiometers 
or in different orientation of the beam,
or just with the beam centre in slightly different position, yields different signals.  
{\tt Madam} interprets this as additional noise and fits baselines to it.

Note that a symmetric beam component, with equal width at all radiometers that participate in the mapmaking process, 
does not give rise to signal error, since it does not create inconsistencies between the measurements.
Instead, it the beam smoothing is just interpreted as a property of the sky,
and can be corrected for in a power spectrum analysis by applying an appropriate beam window function.

The same error sources also cause leakage from the temperature 
signal into the weaker polarization maps.
The relative importance of systematic effects, and their effect on LFI maps, is discussed in
\cite{planck2014-a04}.

The effects of beam shape mismatch and cross-polarization are not 
taken into account in the mapmaking step, but they are handled in the later data processing steps
through application of a beam window function \citep{planck2014-a05}.
Other methods have been also developed for full beam deconvolution \citep{keihanen2012}.

Ideally, optimal noise removal is achieved by destriping the data according to the maximum-likelihood
solution of Eqs. (\ref{baseeq}-\ref{Zmatrix}), with $\tens{C}_{\rm w}$ equal to the actual white noise covariance,
and using all available data.
However, this does not take into account the systematic effects,
which are particularly important for accurate polarization analysis. 

The two-step nature of the destriping procedure enables us to diverge from the ML solution to reduce systematics,
at the cost of slightly increased residual noise.
Residual noise is easier to account for, and is thus less harmful in subsequent analysis, than systematic effects.

\subsubsection{Destriping resolution}

Signal error arising from high signal gradients, and consequent variations inside a pixel,
can be reduced with an increased destriping resolution.
This leads to slightly higher residual noise, as there are more unknowns to be solved for in the same amount of data.

The destriping resolution can be chosen independently from the final map resolution.
In high-resolution mapmaking we have used the same resolution ($N_\textrm{side}$=1024 or $N_\textrm{side}$=2048) 
both for destriping and for map binning.
The choice of resolution plays a more important role in the construction of low-resolution maps (Sect. \ref{sec:lowres_products}).

\subsubsection{Destriping mask}

\begin{figure}
\centering
\includegraphics[width=8.86cm]{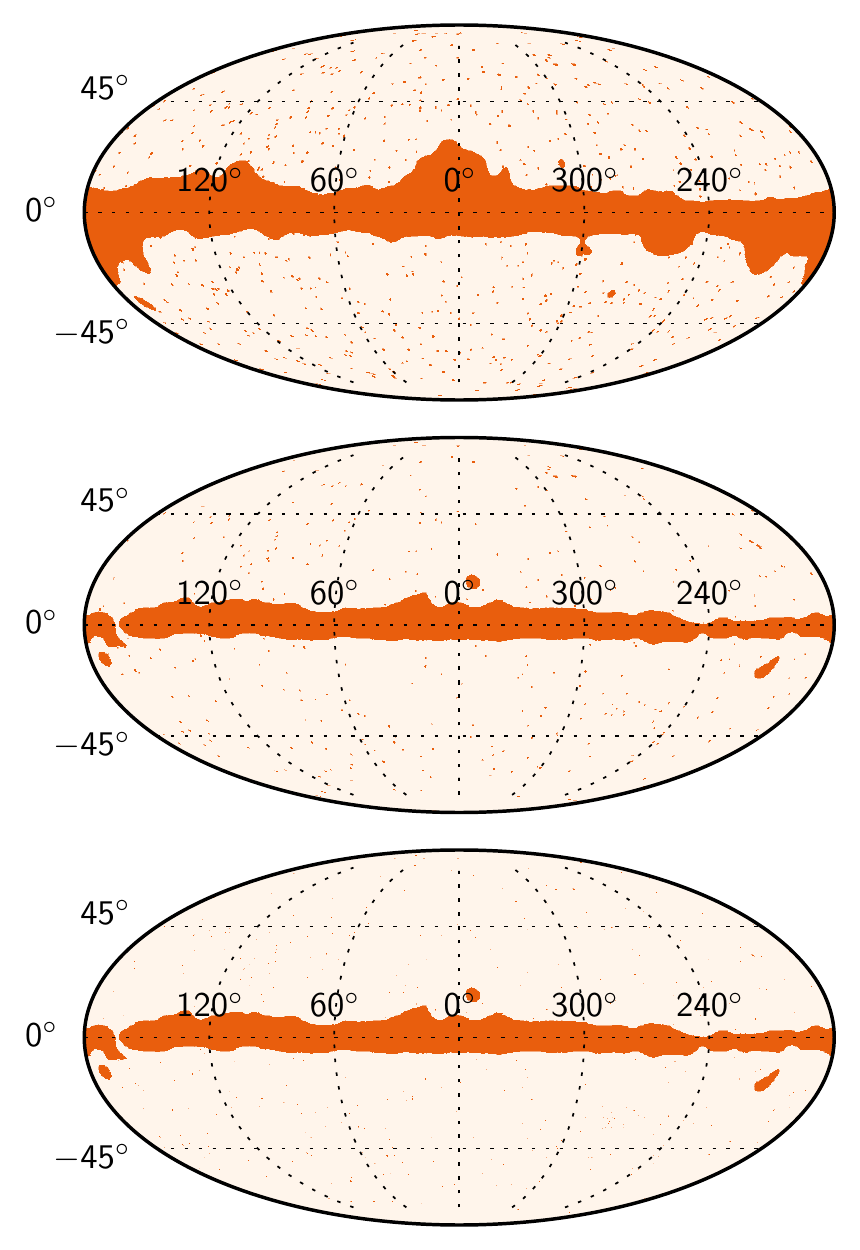}
\caption{Destriping masks for 30\,GHz (top) , 44\,GHz, and 70\,GHz (bottom).
}
\label{fig:masks}
\end{figure}

\begin{figure}
\centering
\includegraphics[width=8.64cm]{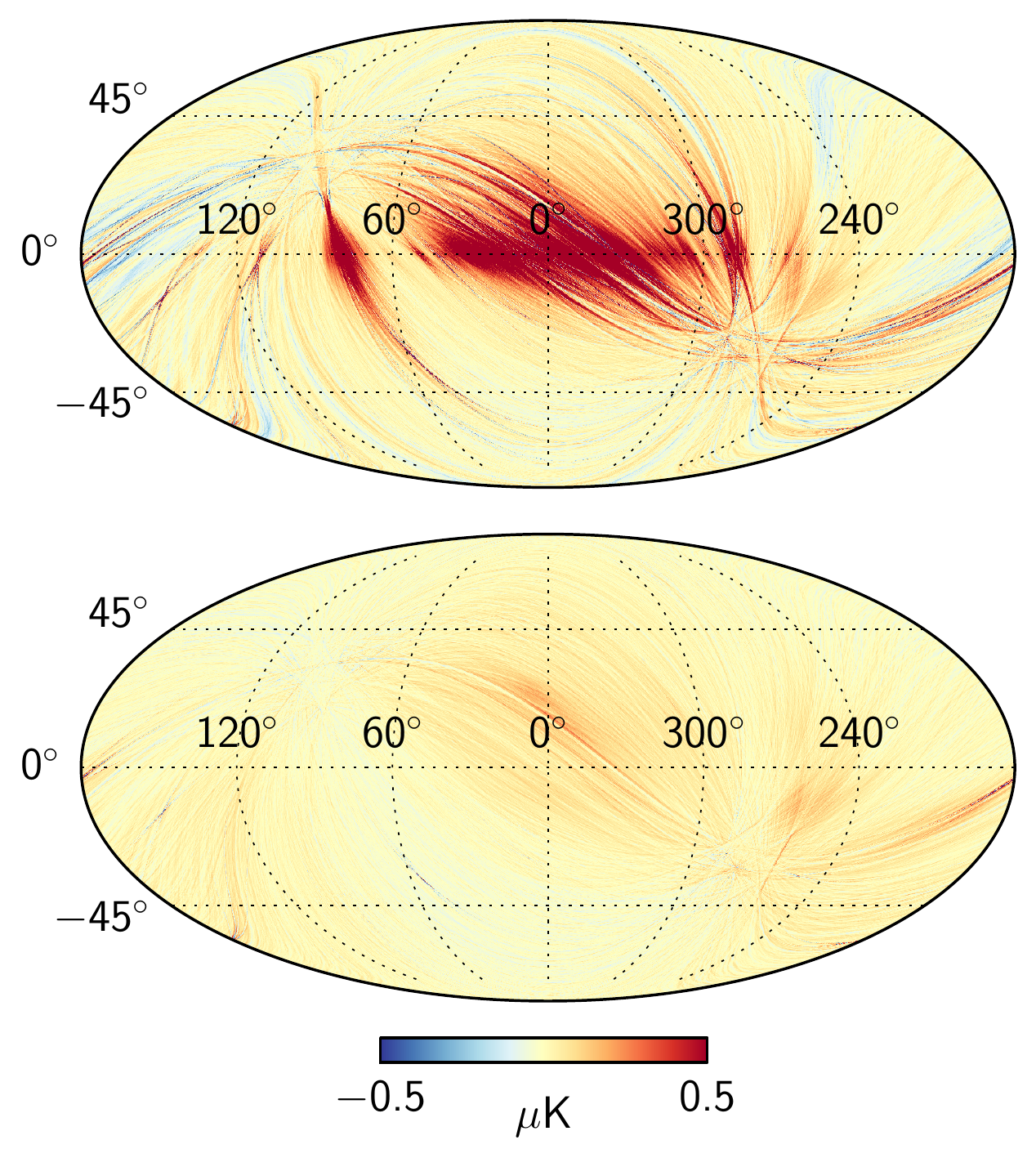}
\caption{Effect of destriping mask on 70\,GHz signal error.
We show the simulated signal error in temperature,
without (upper) and with (lower) applying a Galactic mask. 
}
\label{fig:signalerrormap}
\end{figure}

Signal error originates mostly in the Galactic plane where signal gradients as well as bandpass effects are largest.
We can greatly reduce the signal error by applying a Galactic mask in the destriping phase.
This is done by formally setting the white noise covariance to infinity for the samples under the mask,
thus ensuring consistent treatment throughout.
All samples are again included when binning the final map.
Using a mask leaves fewer crossing points between scanning rings to be used for baseline determination.
The residual noise is therefore expected to increase slightly.

The choice of the destriping mask is a trade-off between minimization of signal error and keeping enough data for accurate destriping.
We use the same masks that were used in \cite{planck2014-a06}.
The masks are illustrated in figure \ref{fig:masks}.
The effect of masking in the destriping phase on the signal error is illustrated in Fig. \ref{fig:signalerrormap}.

\subsubsection{Radiometer weighting}

Individual radiometers are weighted according to $\tens{C}_{\rm w}$
when data from different radiometers ar combined into one map according to Eqs. (\ref{baseeq}-\ref{map-binning}).
In the maximum-likelihood solution $\tens{C}_{\rm w}$ equals the white noise variance of the time-ordered data.
We call this the ``noise weighting".

The purpose of ``horn-uniform weighting" is to reduce leakage from temperature to polarization.
We replace the white noise covariance $\tens{C}_{\rm w}$ for a radiometer by the average of the covariances of the two radiometers of a horn.
The weights that go on the diagonal of $\tens{C}_{\rm w}$ are constructed as
\begin{equation}
\tens{C}_{\rm w}^{-1} = \frac{2}{\sigma_{\rm M}^2+\sigma_{\rm S}^2}
\end{equation}
where $\sigma_M$ and $\sigma_S$ are the white noise rms values for the two radiometers sharing a horn.
The same weight is applied to both radiometers.
We are neglecting variations in white noise level. $\tens{C}_{\rm w}^{-1}$ can be regarded as a radiometer-specific weighting factor. 
The values of $\tens{C}_{\rm w}^{-1}$ for both horn-uniform-weighting and pure noise weighting are given in Table \ref{tab:noise_parameters}.
The numbers are based on the $\sigma$ values listed in the same table.
Since the same matrix $\tens{C}_{\rm w}$ appears on both sides in the destriping equations, the units and overall normalization are irrelevant.
The values given in the table have been normalized so that the weights of all radiometers of the frequency channel add up 1.

At the same time we make the flags uniform, so that if a sample is flagged as unusable at one radiometer of the horn,
we flag the corresponding sample for the other radiometer as well.
This, together with the horn-uniform weighting, ensures that the polarization map is solved solely from the difference of the two timestreams. 
This has the advantage over noise weighting that
any instrumental effects that are equal for both radiometers effectively cancel out.
In particular this is expected to apply to beam effects, since the beam shapes are similar for two radiometers sharing a horn \citep{planck2014-a05}. 
The cancellation is not perfect, however, because the polarization angles of the two radiometers are not exactly orthogonal,
and because the noise priors are different.

The effect of each of these procedures on signal error and on residual noise are assessed through simulations,
which are described in Sect. \ref{sec:validation}

\subsection{White noise covariance}

When noise weighting is applied,  $\tens{C}_{\rm w}$ equals the actual white noise covariance, and the inverse of Eq. (\ref{pixelmatrix}) directly 
gives the covariance of residual white noise in a pixel.
The diagonal elements give the variance of $I,Q,U$ components, and the off-diagonal 
elements represent cross-correlations between the components.

The horn-uniform radiometer weighting complicates the picture.
If $\tens{C}_{\rm n}$ is the actual white noise variance in the time-ordered-data,
and $\tens{C}_{\rm w}$ the one used for weighting,
the residual white noise in the map domain is obtained as
\begin{equation}
\tens{C}_{\rm wn\_map} = (\tens{P}^{\rm T}\tens{C}_{\rm w}^{-1}\tens{P})^{-1} \tens{P}^{\rm T}\tens{C}_{\rm w}^{-1}\tens{C}_{\rm n}\tens{C}_{\rm w}^{-1}
\tens{P} (\tens{P}^{\rm T}\tens{C}_{\rm w}^{-1}\tens{P})^{-1} \label{wncov}.
\end{equation}
Obviously, this reduces to the inverse of the expression in  Eq. (\ref{pixelmatrix}) when $\tens{C}_{\rm w}=\tens{C}_{\rm n}$.
It can be shown that Eq. (\ref{wncov}) obtains its minimum for fixed $\tens{C}_{\rm n}$ when $\tens{C}_{\rm w}=\tens{C}_{\rm n}$,
and this minimum value is $(\tens{P}^{\rm T}\tens{C}_{\rm n}^{-1}\tens{P})^{-1}$. Noise weighting thus minimizes the residual white noise. 
We compute the white noise covariance (\ref{wncov}) for every map produced, and provide it as auxiliary information.

\subsection{Half-ring maps}

Half-ring maps are used for chacterization of total residual noise.
We split each pointing period in two, and use the first halves for the first half-ring map, 
and the second halves for the second half-ring map.  Very long pointing periods are first split into sections of one hour at maximum.
These are further split into two half-ring parts. 
We then construct maps from the two halves with the same procedure as for the full maps.

The two half-ring maps have nearly identical signal component, but independent noise contributions, 
apart from the very lowest frequency components.  We take the difference between the two maps to cancel the signal.
A correction is applied to account for the small differences in the hit count distribution between the two halves.
The ``half-ring noise map" in a given pixel is constructed as
\begin{equation}
m_{\rm h} = \frac{m_{\rm h1}-m_{\rm h2}}{w_{\rm h}}
\end{equation}
where $m_{\rm h1}$ and $m_{\rm h2}$ are the two half-ring maps,
and $w_{\rm h}$ is a weighting factor given by
\begin{equation}
w_{\rm h}= \left[ (n_{\rm h1}+n_{\rm h2}) \left(\frac1n_{\rm h1} + \frac1n_{\rm h2} \right)  \right]^{1/2}.
\end{equation}
Here $n_{\rm h1}$ and $n_{\rm h2}$ are the hit counts in the pixel under consideration.
If the hit distributions are identical, this reduces to $w_{\rm h}=2$.

The half-ring noise map consists of noise with similar properties as the residual noise in the full maps,
apart from the very longest time scales (over an hour) that are cancelled in the process.
The noise map provides a valuable tool for the characterization of residual noise in the maps,
since it is constructed from the same data as the maps themselves, and is independent of any noise model.

We expect a small error in the noise estimate based on the half-ring difference maps to come from the regions of strong temperature gradients.
In these regions the signal can ``leak'' into the half-ring difference maps. We call this effect the gradient leakage. 
It  is caused by the slightly uneven distribution of the hits within each pixel.
Although in the single-survey half-ring maps these effects are visible by eye in the Galactic region,
their effect on the noise spectrum derived from the map is negligible,
because they only affect a few pixels with strongest signal gradients.
In the full mission maps the gradient leakage is further mitigated by the much larger hit count.


\section{Validation of the mapmaking procedure}
\label{sec:validation}

\subsection{Simulations}
\label{sec:Simulations}

We have performed a series of simulations to assess the effect of various parameter choices on map accuracy.
Our goal here is not to build an error budget, but to justify the parameter choices done in the mapmaking process.
The impact of various systematic effects on the \Planck\ data products
is assessed in \cite{planck2014-a04}.

The simulations include noise with realistic noise parameters,
and CMB+foreground with realistic beam shapes.
In all cases we used realistic LFI radiometer pointings and flags,
and constructed $I,Q,U$ maps at {\tt HEALPix} resolution $N_\textrm{side}$=1024.

\subsubsection{Noise simulations}

{\tt Madam} includes an internal noise generator that we used for noise simulations.
The internal generator has an advantage over external codes, because it avoids the step of writing the generated timeline on disk,
thus reducing the computational cost of the simulation.

The code generates white and $1/f$ noise according to a given noise spectrum, 
inserts it in place of the time-ordered data, and processes it into a map in the same way as actual data.
The resulting map is the ``residual noise map".

The residual noise is dominated by white noise.
It therefore makes sense to look at the correlated residual noise component in isolation.
The residual noise map can be split into two independent components
\begin{equation}
\vec{m}_{\rm n} = \vec{m}_{\rm wn}+\vec{m}_{\rm CRN}.
\end{equation}
Here $\vec{m}_{\rm wn}$ represents the binned white noise map, constructed as
\begin{equation}
  \vec{m}_{wn} = (\tens{P}^{\rm T}\tens{C}_{\rm n}^{-1}\tens{P})^{-1}\tens{P}^{\rm T}\tens{C}_{\rm a}^{-1}\vec{y}_{wn}
\end{equation}
where $\vec{y}_{\rm wn}$ is the separately generated white noise timeline.
The binned white noise map does not depend on the destriping procedure, apart from the radiometer weighting.

We then find the ``correlated residual noise'' (CRN) map  by subtracting $\vec{m}_{wn}$ from the residual noise map.
The CRN component includes the correlated noise residual
that remains after destriping, and the error in baseline determination. 
It can be shown that as long as the destriping resolution is equal to the map resolution or exceeds it,
the two components are statistically independent, i.e.,
 \begin{equation}
 \langle  \vec{m}_{\rm wn}^{\rm T}\vec{m}_{\rm CRN}\rangle =0.
 \end{equation}
This makes the rms of the CRN map a convenient figure of merit.
It shows the effects of changes in the mapmaking procedure more clearly than the full noise.
The total residual noise rms is obtained as the quadrature sum of the CRN and white noise values. 

In Monte Carlo simulations we generate the white noise stream independently from the correlated $1/f$ noise stream.
This makes it possible to construct the binned white noise map and to subtract it from the total noise map to obtain the CRN map.
With flight data this distinction is not possible.

When generating noise we used a time-dependent noise model, where the time span of the mission was split into 5 sections,
each of which had a different noise spectrum.  At the time of constructing the simulations this was the most accurate model available.
When destriping the data we used a single noise prior constructed from a constant set of noise parameters.
This is equivalent to the procedure used for actual mapmaking.
To reduce scatter, we have generated 10 noise realizations for each combination of mapmaking parameters.
The values shown are obtained as quadrature sum of the individual rms values.

\subsubsection{Signal simulations}
\label{subsec:signal_simulations}

We perform signal simulations where we generate time-ordered data starting from a sky model.
We used as input the FFP7 sky model \citep{planck2014-a14} which was the version available at the time of our simulations.
Differences with respect to the release version FFP8 are small, and unlikely to affect the results.
Each radiometer has its own input sky,
integrated with the radiometer's frequency response.
We use the tools of the {\tt LevelS} package ({\tt conviqt},  {\tt multimod}) \citep{reinecke2006}, to convolve the sky with the beam,
and to scan a time-ordered data stream from the input sky according to the real radiometer pointing.
We include the main and intermediate beam components, but not the far sidelobe component,
since we assume that the latter is removed in the calibration process.
We apply flags, and run the data through the mapmaking procedure.

The ``signal error rms" is computed as the rms of the difference between destriped and binned signal maps constructed from same data stream.
It has become customary to call this quantity signal error, though its interpretation as pure error is questionable.
For instance,
destriping may remove some of the systematic polarization signal that arises from bandpass mismatch, but this
shows up as an increased signal error.
The difficulty here is that the binned map itself is subject to bandpass and beam mismatch,
and there is not a unique ``true" map to which we could compare the destriped maps.
Signal error is rather a measure of how much the destriping process modifies the signal, compared to naive binning.


Signal error rms is not a useful figure of merit when we consider the effect of radiometer weighting,
because the chosen weighting scheme affects the binned map as well.
We performed another series of signal simulations where we included only unpolarized CMB in the input sky.
All signal in the output polarization is thus leakage from temperature.
Analysis of the polarization maps gives information on how the chosen radiometer weighting scheme impacts the leakage.

\subsection{Results}

\subsubsection{Baseline length and mask}

\begin{figure}
\centering
\includegraphics[width=1\columnwidth]{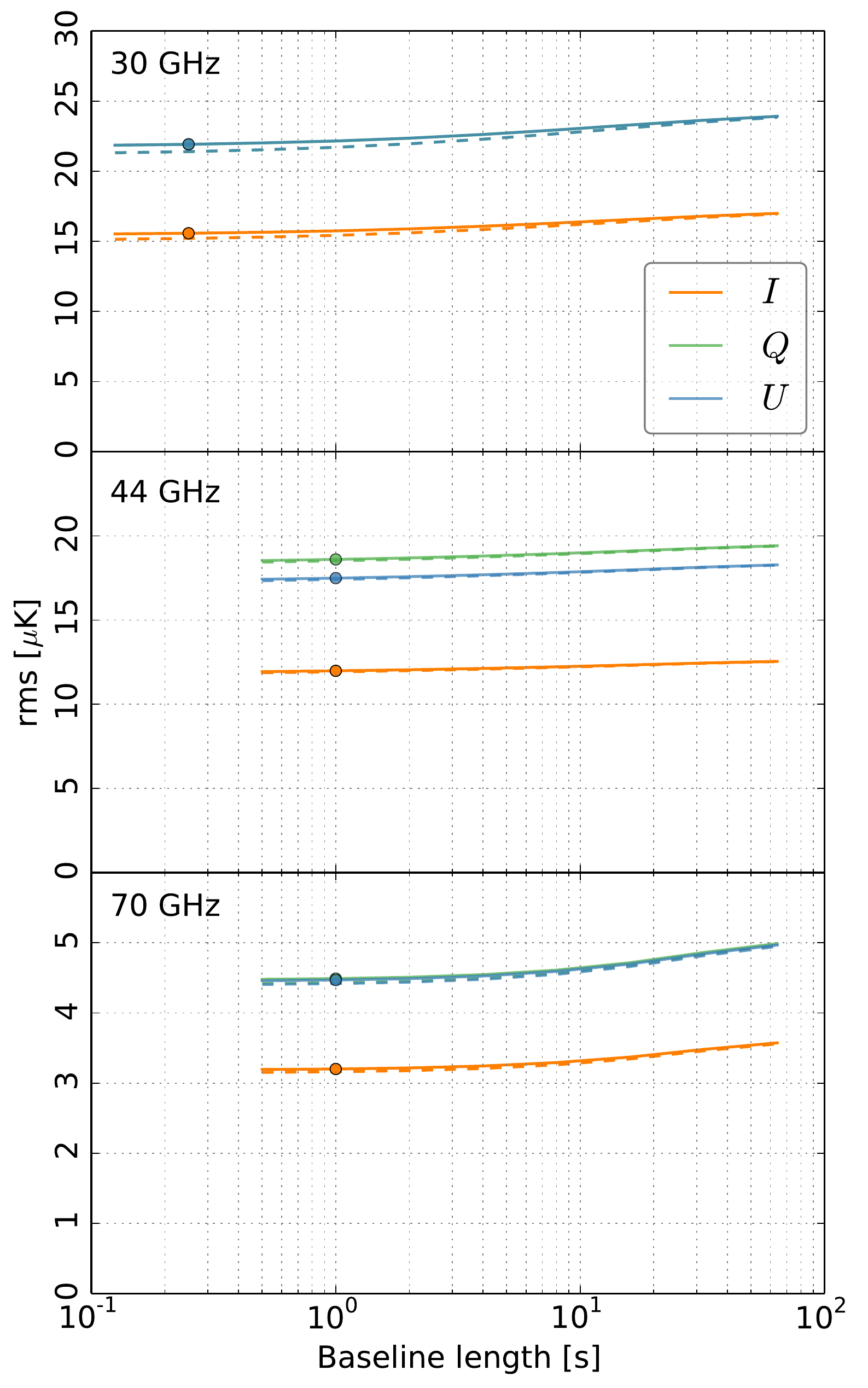}
\caption{Correlated residual noise (CRN) as a function of baseline length for all LFI frequencies.
CRN is computed as the rms of the difference between the destriped map and the binned white noise map.
We plot CRN for $I$ (orange), $Q$ (green), and $U$ (blue) as a function of baseline length.
Solid and dashed  lines show the result with and without the destriping mask, respectively.
The circles indicate the baseline value used in actual mapmaking.}
\label{fig:noise_base}
\end{figure}

\begin{figure}
\centering
\includegraphics[width=1\columnwidth]{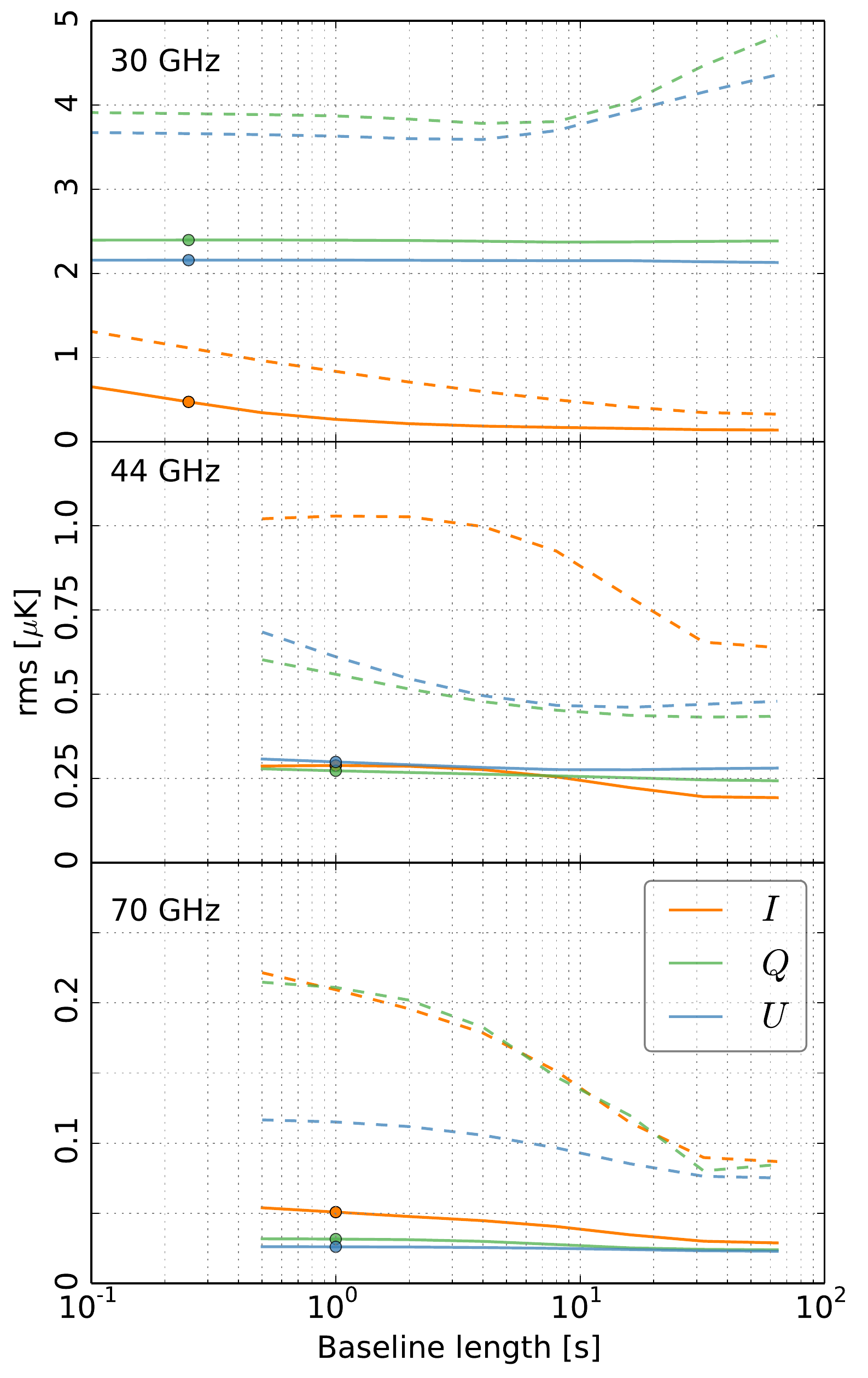}
\caption{Signal error as a function of baseline length for all LFI frequencies,
for $I$ (orange), $Q$ (green), and $U$ (blue).
Signal error is computed as the rms of the difference between destriped and binned maps.
Solid and dashed  lines show the result with and without the destriping mask, respectively. }
\label{fig:sig_base}
\end{figure}

\begin{figure}
\centering
\includegraphics[width=1\columnwidth]{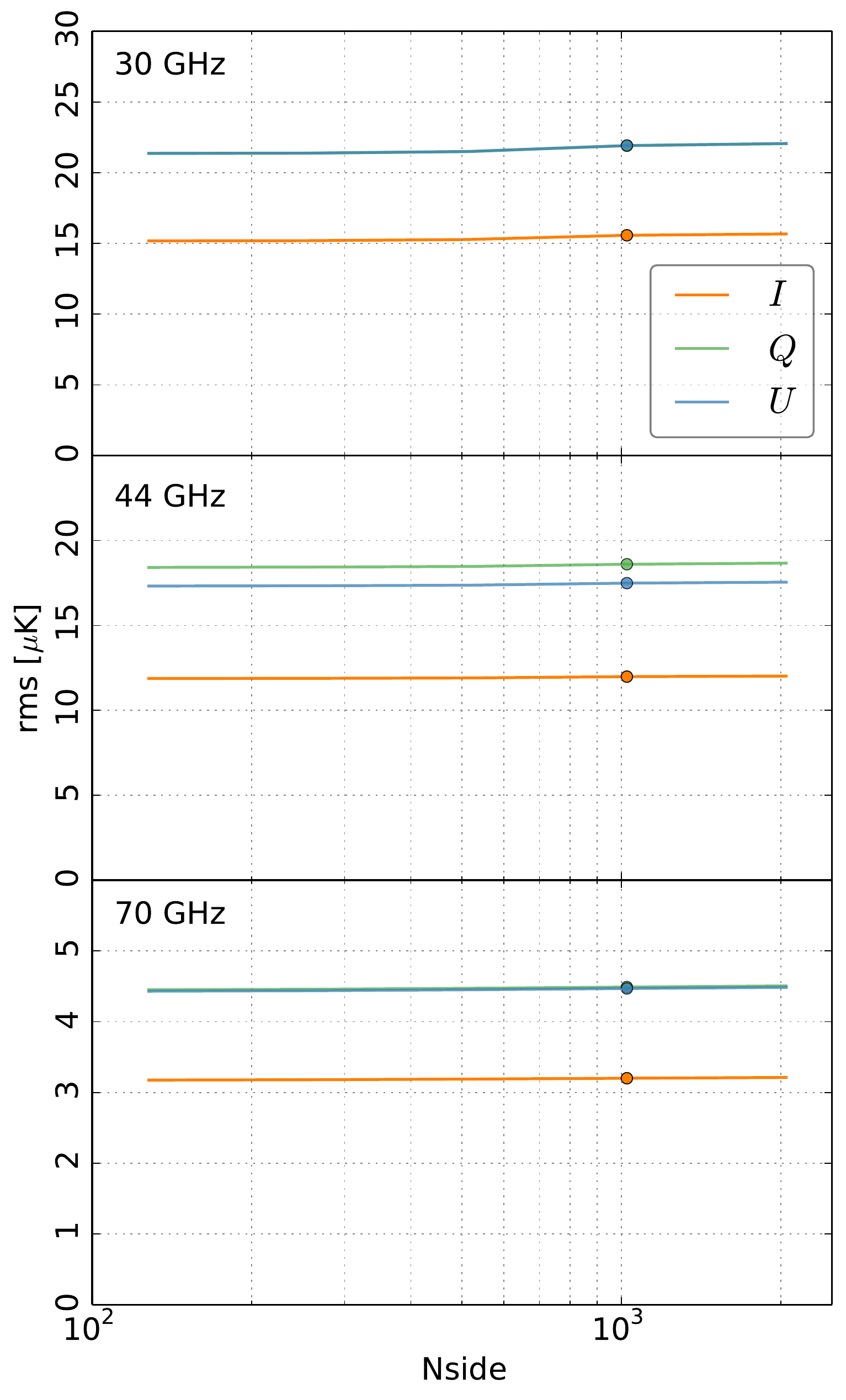}
\caption{Correlated residual noise (CRN) as a function of destriping resolution for all LFI frequencies.
CRN is computed as the rms of the difference between the destriped map and the binned white noise map.
We plot CRN for $I$ (orange), $Q$ (green), and $U$ (blue) as a function of destriping resolution.
The circles indicate the  resolution ($N_\textrm{side}$=1024) used in actual mapmaking.}
\label{fig:noise_nside}
\end{figure}

\begin{figure}
\centering
\includegraphics[width=1\columnwidth]{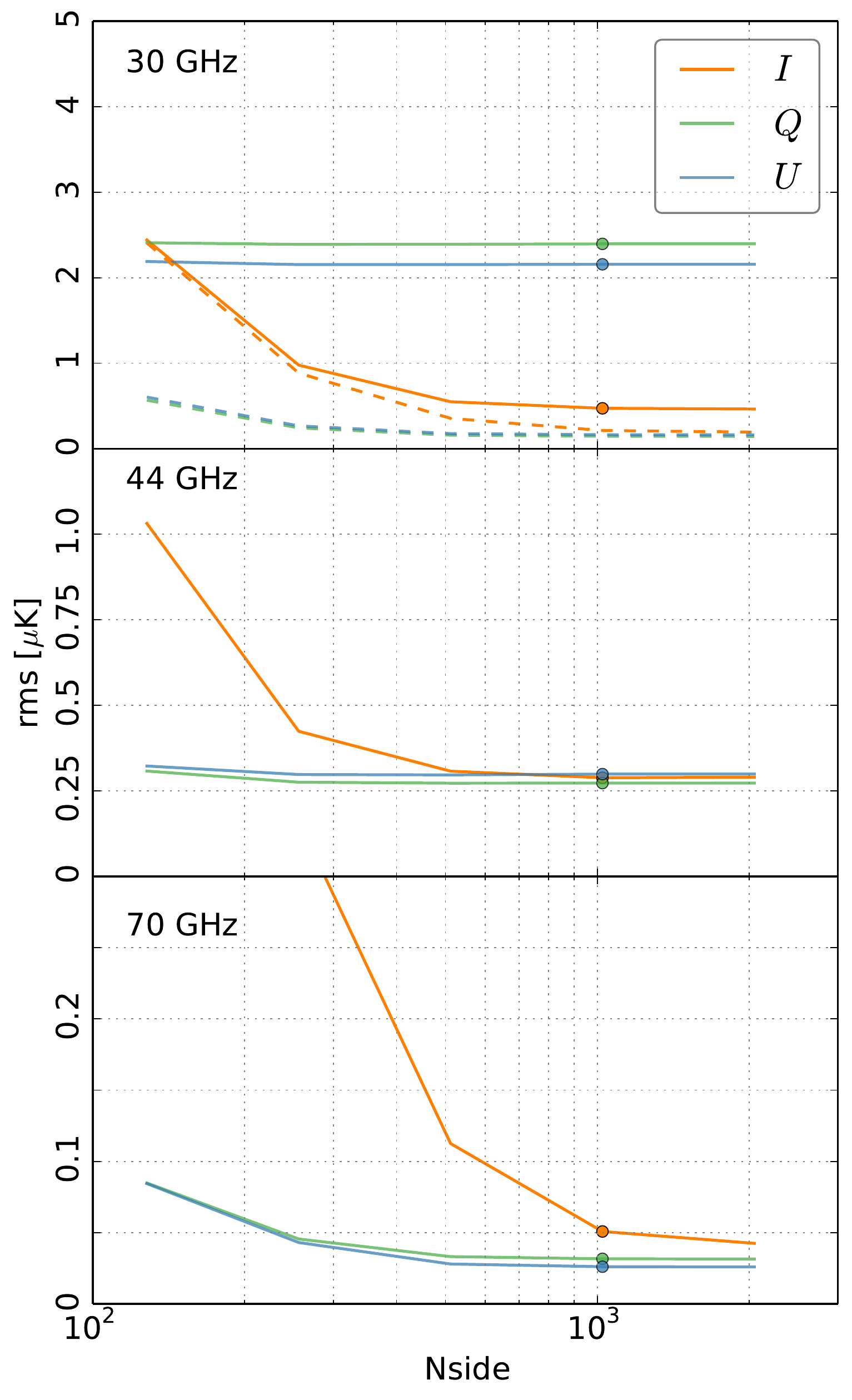}
\caption{Signal error as a function of destriping resolution for all LFI frequencies,
for $I$ (orange), $Q$ (green), and $U$ (blue).
Signal error is computed as the rms of the difference between destriped and binned maps. 
The dashed lines in the 30\,GHz case show results from a simulation without bandpass mismatch.}
\label{fig:sig_nside}
\end{figure}

In the first series of simulations we varied the baseline length  and examined its effect on residual noise and signal error.
The results are plotted in Figs. \ref{fig:noise_base} and \ref{fig:sig_base}.
We show the CRN and signal error RMS for temperature and $Q,U$ polarization.
Because the amplitudes of residual noise and signal error are very different, we plot them in different panels.
The total residual noise is a sum of the CRN component, the binned white noise component, and the signal error.
Since the three components are independent, the square of total residual noise  rms is obtained as a quadrature sum of the individual rms values.
The rms of the binned white noise component, again averaged over 10 noise realizations, are found in Table \ref{tab:noise_detweight}.

As expected, residual noise decreases with decreasing baseline width. The values used in actual mapmaking are indicated by circles.
The 0.25 s baseline length used at 30\,GHz appears to be a bit of overkill, since the residual noise level 
changes only very little below one second (0.2 $\mu$K).

At 30\,GHz and 70\,GHz, $Q$ and $U$ have nearly identical noise levels. This is a consequence of the focal plane design.
LFI horns are arranged in pairs that have their direction of polarization sensitivity at 45$^\circ$ angles to each other,
so that when one horn is measuring $Q$, the other is measuring $U$, and both components are recovered with same accuracy.
An exception from this is the 44\,GHz channel, where horn LFI-24 does not have a counterpart.
This leads to a difference in noise levels between $Q$ and $U$ at 44\,GHz.

The level of signal error is a result of complicated interplay between destriping, phenomena causing signal error, and instrument design.
Different phenomena dominate at different frequencies.
Signal variations within a pixel and beam shape mismatch contribute more strongly to the signal error in temperature than in polarization,
because the beam patterns of two radiometers of a horn are similar, so that the effect partly cancels in polarization.
The same does not apply to bandpass mismatch.

We ran the same series of  simulations with and without destriping mask.
As discussed in Sect. \ref{sec:systematics}, the mask is expected to reduce the signal error arising from high signal gradients.
The masks used are depicted in Fig. \ref{fig:masks}. 
Figure \ref{fig:signalerrormap} illustrates the effect of destriping mask on signal error.
We plot a map of signal error in 70\,GHz temperature,
first without a destriping mask, and below it with mask.  
The mask greatly reduces the error.

The mask has the effect of slightly increasing the residual noise, as there is less data to solve the baselines from;
however, the effect is small, well below $1\, \mu$K, as can be seen from Fig. \ref{fig:noise_base}.
The effect is the opposite direction for signal error. The signal error increases towards shorter baselines.
The overall level of the signal error is still well below that of the residual noise.
The destriping mask has a dramatic effect on the signal error, decreasing it typically by a factor of 2.

We can conclude that our chosen baseline lengths are short enough for nearly optimal noise removal.
At the same time the signal error is kept under control by the application of a mask.

\subsubsection{Destriping resolution}

In another series of simulations we varied the destriping resolution.
The final maps were constructed at a fixed resolution $N_\textrm{side}$=1024 in all cases.
The baseline length was fixed to its nominal value, and a destriping mask was applied.
The results are shown in Figs. \ref{fig:noise_nside} and \ref{fig:sig_nside}.

A higher resolution increases the residual noise slightly, since there are more unknowns to solve for the same amount of data;
however, the effect is again below 1 $\mu$K.
In signal error the effect is in the opposite direction.
In the case of temperature maps, increasing the resolution reduces the signal error drastically,
but the effect levels out near our chosen destriping resolution.
In polarization, the effect is less dramatic.

The 30\,GHz channel has the peculiar feature that the signal error is significantly higher in polarization than in temperature.
It is also nearly completely insensitive to baseline length and destriping resolution.
The insensitivity to baseline length suggests that the effect is coupled to a foreground component distributed over
a large fraction of the sky. The likely explanation is leakage of synchrotron emission signal from temperature
to polarization through bandpass mismatch.
To verify this, we run yet another series of simulations where
we artificially turned off the bandpass mismatch
by using the same input sky model for all radiometers of 30\,GHz.
The results are shown with dashed lines in the uppermost panel of Fig. \ref{fig:sig_nside}.
The signal error is greatly reduced, confirming that bandpass mismatch is the dominant factor in the signal error.

\begin{table}
\begin{center}
\begin{tabular}{l  rr rr  rr}
\hline\hline
30\,GHz \vphantom{\Large I} & \multispan2  $I$  & \multispan2 $Q$   & \multispan2 $U$  \\
\vphantom{\Large I} & NW & HUW & NW & HUW & NW &HUW  \\
  \noalign{\vskip 2pt\hrule\vskip 2pt}
white & 56.76 & 56.78 & 80.28 & 80.42 & 80.20 & 80.41  \\
CRN & 15.56 & 15.57 & 21.75 & 21.91 & 21.73 & 21.92 \\
full &  58.85 & 58.87 & 83.17 & 83.34 & 83.09 & 83.32 \\
  \noalign{\vskip 2pt\hrule\vskip2pt}
44\,GHz \vphantom{\Large I} & \multispan2 $I$   &\multispan2  $Q$   & \multispan2 $U$ \\
\vphantom{\Large I} & NW & HUW & NW & HUW & NW &HUW  \\
  \noalign{\vskip 2pt\hrule\vskip 2pt}
white & 65.42 & 65.81 & 99.88 & 100.44 & 93.80 & 94.37 \\
CRN & 12.52 & 11.99 & 19.36 & 18.61 & 18.28 & 17.49  \\
full & 66.60 & 66.90 & 101.74 & 102.15 & 95.57 & 95.99 \\
  \noalign{\vskip 2pt\hrule\vskip 2pt}
70\,GHz \vphantom{\Large I} & \multispan2 $I$  & \multispan2 $Q$ & \multispan2 $U$ \\
\vphantom{\Large I} & NW & HUW & NW & HUW & NW &HUW  \\
  \noalign{\vskip 2pt\hrule\vskip 2pt}
white & 57.82 & 57.95 & 82.00 & 82.32 & 81.59 & 81.94 \\
CRN & 3.22 & 3.20 & 4.50 & 4.49 & 4.48 & 4.47 \\
full & 57.91 & 58.04 & 82.12 & 82.44 & 81.71 & 82.06 \\
\hline
\end{tabular}
\end{center}
\caption{Impact of detector weighting scheme on residual noise.
We show the rms (in $\mu$K) of binned white noise (``white"), correlated residual noise (``CRN"),
and their sum (``full"). All values are averages over 10 noise realizations.
NW refers to noise weighting, and
HUW to the horn-uniform weighting scheme.
Noise weighting minimises the total residual noise.
In actual mapmaking we used instead horn-uniform weighting
to reduce effects from beam shape mismatch.}
\label{tab:noise_detweight}
\end{table}

\begin{table}
\begin{center}
\begin{tabular}{l rr  rr  rr}
\hline\hline
30\,GHz \vphantom{\Large I} & \multispan2 $Q$ &  \multispan2 $U$ \\
\vphantom{\Large I} &  NW & HUW & NW &HUW  \\
  \noalign{\vskip 3pt\hrule\vskip 5pt}
ideal                     & 0.1175 & 6.6e-15 & 0.1385  & 6.6e-15  \\
non-orthogonality & 0.1158 & 0.0080  & 0.1377  & 0.0067  \\
realistic pointing   &  0.1163 & 0.0124   &  0.1381 & 0.0119 \\
realistic beam       & 0.1671 & {\bf 0.0890}   & 0.1863   & {\bf 0.0893}  \\
\noalign{\vskip 2pt\hrule\vskip 2pt}
44\,GHz \vphantom{\Large I} & \multispan2  $Q$ & \multispan2 $U$ \\
\vphantom{\Large I} & NW & HUW & NW &HUW  \\
\hline
ideal                     & 0.5073  & 6.5e-15 & 0.5817 & 6.3e-15  \\
non-orthogonality & 0.5239 & 0.0166   & 0.6016 & 0.0187  \\
realistic pointing   & 0.5313  & 0.0663  & 0.6081   & 0.0632  \\
realistic beam       & 0.7007 & {\bf 0.2946}    & 0.8189   & {\bf 0.3246}   \\
  \noalign{\vskip 2pt\hrule\vskip 2pt}
70\,GHz \vphantom{\Large I} & \multispan2 $Q$ &  \multispan2 $U$ \\
\vphantom{\Large I} & NW & HUW & NW & HUW  \\
\hline
ideal                     & 0.0731  & 4.7e-15 &  0.0832 & 4.7e-15  \\
non-orthogonality & 0.0740  & 0.0025  & 0.0841 & 0.0023  \\
realistic pointing   &  0.0758 & 0.0144   & 0.0858    & 0.0144  \\
realistic beam       &  0.1900 & {\bf 0.1658}   & 0.2177   & {\bf 0.1834}   \\
\hline

\end{tabular}
\end{center}
\caption{Impact of detector weighting scheme on temperature leakage.
Noise weighting (NW) is compared against horn-uniform weighting (HUW).
We show the rms (in $\mu$K) of the $Q$ and $U$ maps from a CMB simulation with unpolarised inputs.
All signal in $Q,U$ maps is leakage from temperature through beam and pointing non-idealities.
The four rows correspond to simulations with increasing complexity:
idealised simulation (see main text); non-orthogonal polarization angles;
realistic pointing; and realistic beam model.
The case that most closely corresponds to actual mapmaking is shown in boldface.
}
\label{tab:leakage_detweight}
\end{table}

\subsubsection{Horn-uniform weighting}

The purpose of the horn-uniform weighting scheme is to reduce
leakage from temperature to polarization through
systematics that are correlated
between radiometers sharing a horn.
In particular this concerns leakage through beamshape mismatch.

Changing the detector weighting changes both the binned and the destriped map. 
Therefore CRN alone is no longer a suitable figure of merit, but we have to look at the total residual noise.

Results from noise simulations are shown in Table \ref{tab:noise_detweight}.
For completeness we show also the CRN values.
We compare the horn-uniform weighting scheme with weighting based on the radiometer's white noise level.
The latter is expected to minimise the total residual noise.
This is confirmed by our simulations. 
In all cases studied the noise weighting scheme yields a slightly smaller total noise residual than the horn-uniform weighting,
but the differences are small.
The white noise rms (see Table \ref{tab:noise_parameters}) typically vary by 10\,\%, but this translates to a $<1$\, \% effect in residual noise.

To study the effect of radiometer weighting scheme on $T$-$P$ leakage,
we performed a series of signal simulations where we only include the CMB temperature anisotropies in inputs.
The resulting $Q/U$ maps consist purely of temperature leakage.
Since foregrounds are not present, bandpass mismatch plays no role,
but beam shape mismatch and non-idealities in pointing are the only sources of leakage.
We use as a simple measure of error in these maps the rms of the $Q$ and $U$ polarization maps.
The results are shown in Table \ref{tab:leakage_detweight}.

We compare noise weighting and horn-uniform weighting in four simulations of increasing complexity.
The first one is an idealized simulation, where we have eliminated any non-idealities that could cause leakage from temperature to polarization.
We use identical elliptic beams and the same pointing for both radiometers of a horn.  The polarization responses ($\psi_{\rm pol}$ angle)
are set at exactly $90^\circ$ from each other, while they in reality differ from orthogonality by $\pm0.5^\circ$  at maximum \citep{planck2014-a05}.
With these assumptions, the horn-uniform weighting is expected to fully cancel the temperature leakage.
This is confirmed by our simulation.

In the second simulation we set the polarization angles to their real values,
and in the third we add to that realistic pointing, which is slightly different for the M and S radiometers.
Under these circumstances, the benefit of horn-uniform weighting as compared to noise weighting is still dramatic.
In the last simulation we keep the realistic pointing and polarization angles, and replace the elliptic beam by the realistic main beam model.
In this case the difference between the two weighting schemes reduces significantly, 
but is still a factor of 2--3 at 30\,GHz\,GHz and 44\,GHz.

The level of temperature leakage reflects the beam shape mismatch at each frequency. 
Although the beams of a 44\,GHz horn are similar in ellipticity, they have different FWHM at the 3\,\% level, 
as contrasted to $<1\,\%$ at 30\,GHz and 70\,GHz \citep{planck2014-a05}.
For this reason the leakage is strongest at 44\,GHz.

With foregrounds present, the temperature leakage is dominated by bandpass mismatch.
We do not attempt to correct for it at the mapmaking level.  Bandpass mismatch is easier to model and account for at later 
stages of data processing than beam mismatch.  Bandpass correction is discussed in \cite{planck2014-a03}.


\section{Low resolution products}

\label{sec:lowres_products}

Low-resolution products are an integral part of the low-$\ell$ likelihood. 
To fully exploit the information contained in the largest structures of the CMB sky, a full statistical description 
of the residual noise present in the maps is required. This information is provided in the form of pixel-pixel noise covariance 
matrices (NCVM). However, due to resource limitations they are impossible to employ at the native map resolution. 
Therefore a low-resolution data set is needed for the low-$\ell$ analysis. 
This data set consist of low-resolution maps and corresponding noise covariance matrices. At present, the low-resolution 
data set can be efficiently used only at resolution $N_\textrm{side} = 16$ (or lower). All the low-resolution products 
are produced at this target resolution.

We will begin by reviewing how the low-resolution maps were produced for the 2015 release.
Then we will discuss the production of  the 2015 LFI low-resolution noise covariance matrices. The noise covariance matrix 
section deepens the study performed for \cite{planck2013-p02}.

\subsection{ Low-resolution maps}
\label{sec:lowres_maps}

Number of different approaches for producing the low-resolution maps exist. 
Previous work on production of low-resolution maps have been published in 
\cite{keskitalo2009}. 
An ideal method would simultaneously minimise the pixelization effects and residual noise.
Realistic methods involve a trade-off between these two goals.

We use the high-resolution maps described in Section \ref{sec:mapmaking} as input for the low-resolution map production. 
The chosen downgrading scheme is nearly identical to the one used in the previous 2013 release \citep{planck2013-p02}, 
the only difference being the addition of regularization noise to the final products. 

The signal in a given high-resolution pixel is weighted with the inverse of its noise variance
when binning the signal into larger pixels,
and hence the downgrading scheme is called the ``noise-weighted scheme". 
The weights are given by Eq. (\ref{pixelmatrix}).  The resulting map is equivalent to one that is 
directly binned onto the target resolution from the TOI destriped at the high resolution. 
Specifically we apply to a high-resolution map the operation
\begin{equation}
   \vec m_{\mathrm{l}} = \left( \tens{P}^{\mathrm{T}}_{\mathrm{l}}\tens{C}_{\rm w}^{-1}\tens{P}_{\mathrm{l}} \right)^{-1} \tens{X}  
   \left( \tens{P}^{\mathrm{T}}_{\mathrm{h}}\tens{C}_{\rm w}^{-1}\tens{P}_{\mathrm{h}} \right) \vec m_{\mathrm{h}} 
   \equiv \tens{D} \vec m_{\mathrm{h}},
\end{equation}
where
\begin{equation}
   \tens{X}_{qp} = \begin{cases}
   1,\: \: \: \mathrm{if}\: p\: {\rm is\: a\: subpixel\: of\: } q    \\
   0,\: \: \: \mathrm{otherwise}
   \end{cases}
\end{equation}
is the sum of high-resolution pixels to low-resolution pixels. Here subscripts ``h" and ``l" refer to high and low resolution version 
of the pointing matrix. The same matrix $\tens{X}$ also downgrades the pointing matrix, 
$\tens{P}_{\mathrm{l}} = \tens{P}_{\mathrm{h}}\tens{X}^{\mathrm{T}}$.
After downgrading, the temperature component is smoothed with a Gaussian window function with FWHM $= 440\arcmin$,
 to prevent  aliasing due to high frequency power in the map.

Our implementation of the downgrading scheme first downgrades the maps to an intermediate 
resolution of  $N_{\textrm{side, mid}} = 32$. The Stokes $I$ part of the map is expanded in spherical harmonics, treated with 
the smoothing beam, and the final map is then synthesised at the target resolution. 
The last downgrading step for Stokes $Q$ and $U$ maps is performed by naive averaging of higher resolution pixels,
to minimize signal distortion.

\subsection{Noise covariance matrices}

The statistical description of the residual noise present in a low-resolution map is given in the form of a pixel-pixel 
noise covariance matrix.  For generalised destriping the formalism was developed in \cite{keihanen2005} 
and \cite{keskitalo2009}.
Using the formalism introduced in Sect. \ref{sec:mapmaking}, the noise noise covariance matrix $\tens{N}$ is given by
\begin{equation}
   \tens{N} = \left[ \tens{P}^{\mathrm{T}}\left(\tens{C}_{\rm w} + 
   \tens{F}\tens{C}_{\rm a}\tens{F}^{\mathrm{T}}\right)^{-1}\tens{P}  \right]^{-1},
\end{equation}
which can be written in a dimensionally reduced form as
\begin{equation}
   \tens{N}^{-1} = \tens{P}^{\mathrm{T}}\tens{C}_{\rm w}^{-1}\tens{P} 
   +  \tens{P}^{\mathrm{T}}\tens{C}_{\rm w}^{-1}\tens{F}\left(\tens{F}^{\mathrm{T}}\tens{C}_{\rm w}^{-1}\tens{F}
   + \tens{C}_{\rm a}^{-1}\right)^{-1}\tens{F}^{\mathrm{T}}\tens{C}_{\rm w}^{-1}\tens{P}.
   \label{eq:inv_N}
\end{equation}
Here $\tens{P}$, is the pointing matrix, $\tens{F}$ spreads the noise baselines to the TOI,
$\tens{C}_{\rm a}$ is the noise prior, and $\tens{C}_{\rm w}$ is white noise covariance.

For the current release, we use an implementation by {\tt{Madam/TOAST}},  a Time Ordered Astrophysics Scalable Tools 
({\tt{TOAST}}; \cite{toast}) port of {\tt{Madam}}. The {\tt{TOAST}} interface was chosen on the basis of added flexibility. 
 {\tt{Madam/TOAST}} is capable of handling arbitrarily complex noise models. It can also handle bigger chunks of data, i.e.,
  full mission versus single survey in the previously used implementation, leading to reduced wall-clock time in computing. 
  Some approximations and differences compared to high-resolution map production are involved, and their effects will be discussed
   in Sect. \ref{sec:ncvm_simu}. These are the differences in baseline length and destriping resolution, as well as simplifying approximations.

The above formulation in Eq. (\ref{eq:inv_N}) describes the noise correlations of a map destriped and
 binned at the same resolution, according to Eqs. (\ref{baseeq}-\ref{map-binning}). 
For an exact description we should construct the matrices at resolution $N_{\textrm{side}} = 1024$, 
and subsequently downgrade to the target resolution. This is, however, computationally unfeasible. 
Hence we calculate the noise covariance
matrices at a practical intermediate resolution, which is not necessarily the target resolution, 
and subsequently downgrade to target resolution. 
For consistency, the noise covariance matrices must go through the same smoothing and regularization
steps as is applied to the 
low-resolution maps. 

The covariance matrix does not model the effects of using a destriping mask and horn-uniform weighting,
which were introduced to reduce systematics.
The impact of these approximation on the accuracy of the NCVM products is assessed through simulations in Sect. \ref{sec:ncvm_simu}.

\begin{figure}
\centering
\includegraphics[width=1\columnwidth]{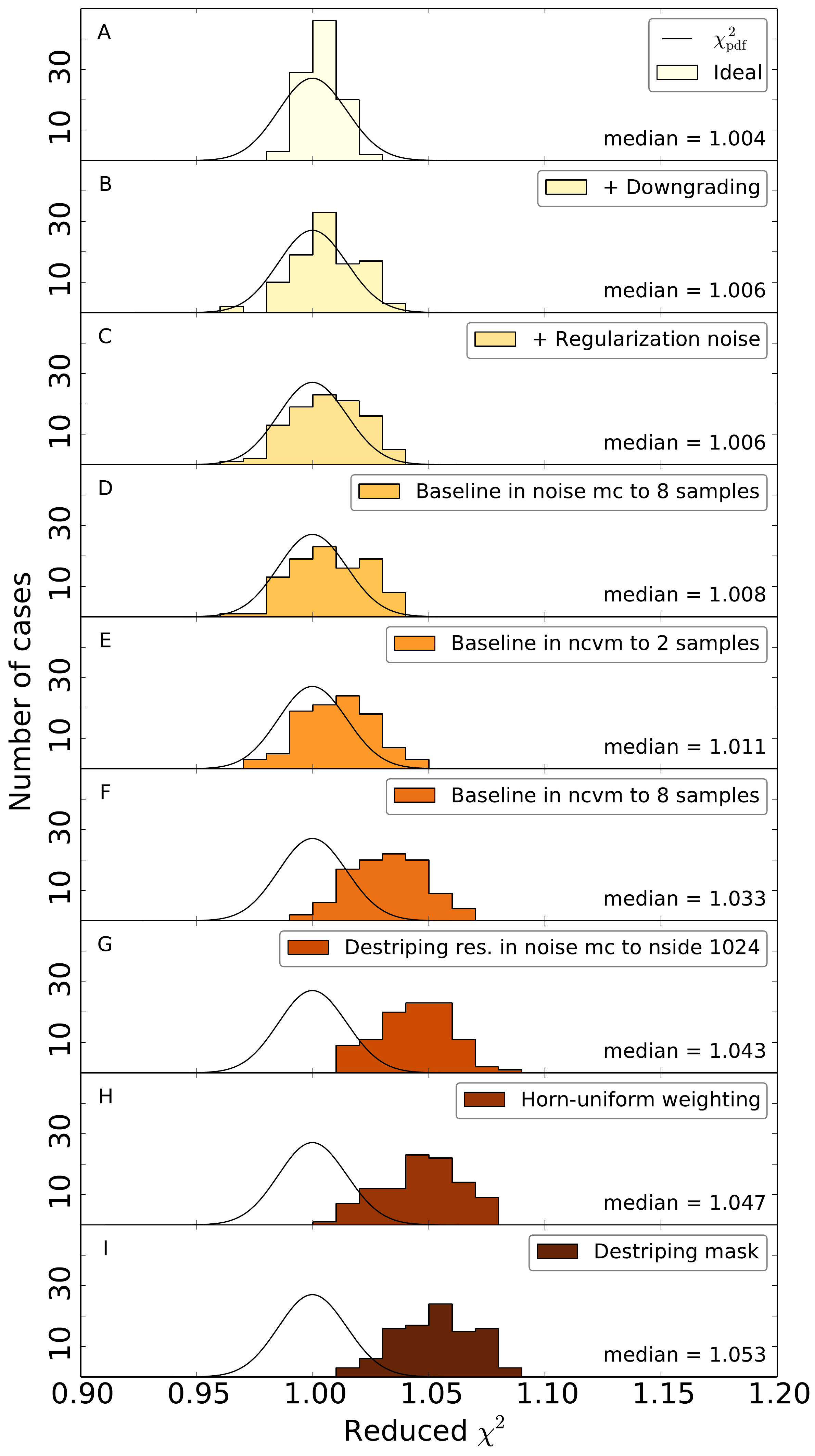}
\caption{Reduced $\chi^2$ statistics for simulated full mission 30\,GHz data.
 Initial resolution of the noise covariance matrix is $N_{\textrm{side}}=32$.
  The level of idealization decreases from top to bottom.}
\label{fig:ncvm_ideal}
\end{figure}

\begin{figure}
\centering
\includegraphics[width=1\columnwidth]{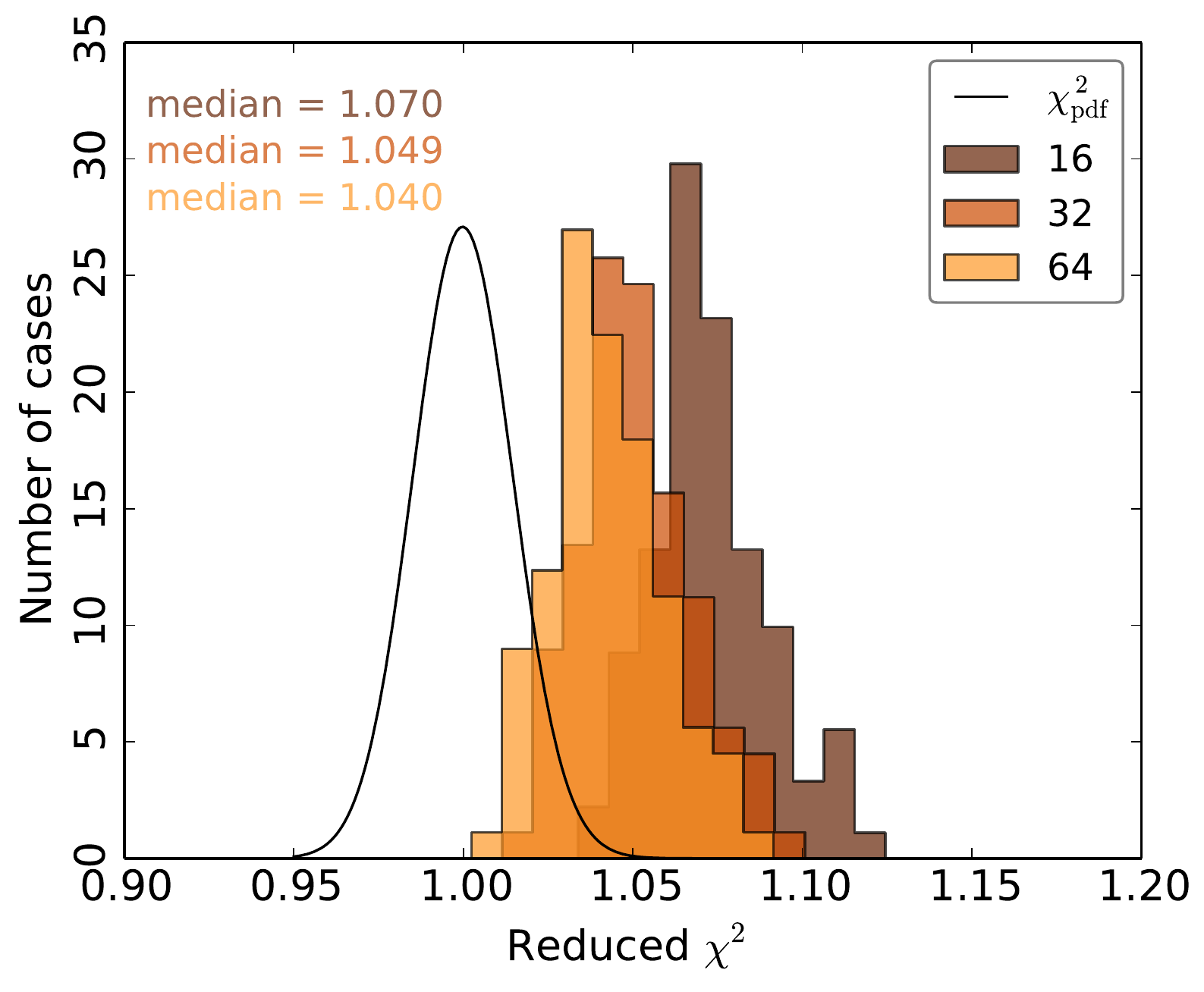}
\caption{Reduced $\chi^2$ statistics for simulated full mission 30\,GHz data. 
Initial resolution of the noise covariance matrix is varied from $N_\textrm{side} = 16$ to 64.}
\label{fig:chi2_nside16}
\end{figure}

Applying equation (\ref{eq:inv_N})  in practice will require inversion of  a symmetric $3 N_\mathrm{pix} \times 3 N_\mathrm{pix}$ matrix.  The inversion is performed via the eigen-decomposition of the matrix. The monopole of the $I$ map cannot be resolved by the map-maker, 
and the matrix becomes singular. This ill-determined mode is left out of the analysis. 

The chosen downgrading scheme leads to a singular covariance matrix, since after smoothing  
the number of pixels in the map is larger than the number of non-zero spherical harmonics,
resulting in a significant number of zero eigenvalues.
We regularise the products by adding a small amount of white noise to the maps,
and a corresponding diagonal covariance to the matrices. 
The level of regularization noise is chosen to be  2 $\mu$K RMS for $I$, and 0.02 $\mu$K RMS for $Q$ and $U$ 
at $N_\textrm{side} = 16$ resolution.

As seen from equation (\ref{eq:inv_N}), the noise covariance computation requires two inputs: 
the detector pointings and the noise model derived from the flight data.

  \begin{table*}
  \begingroup
  \newdimen\tblskip \tblskip=5pt
  \caption{Frequency-specific information related to the noise covariance matrix production. From left to right: the statistics of difference between 
  DPC and FFP8 full mission hitmaps for all LFI frequency channels; the statistics of DPC full mission hits maps;
  and the baseline lengths 
  used in the noise covariance matrix production, both in seconds and as a number of samples.}
  \label{table:hitdiff}
  \nointerlineskip
  \vskip -3mm
  \footnotesize
  \setbox\tablebox=\vbox{
  \newdimen\digitwidth
  \setbox0=\hbox{\rm 0}
  \digitwidth=\wd0
  \catcode`*=\active
  \def*{\kern\digitwidth}
  \newdimen\signwidth
  \setbox0=\hbox{+}
  \signwidth=\wd0
  \catcode`!=\active
  \def!{\kern\signwidth}
  \halign{\hbox {#}\tabskip=2em&
         \hfil#\hfil& \hfil#\hfil&\hfil#\hfil& \hfil#\hfil& \hfil#\hfil&  \hfil#\hfil&
      \hfil#\hfil  &  \hfil#\hfil &  \hfil#\hfil \tabskip=0pt\cr
     \noalign{\doubleline}
  \omit&
  \multispan3\hfil D{\sc ifference hitmap statistics} \hfil
  &\multispan3\hfil H{\sc itmap statistics} \hfil 
  &\multispan2\hfil B{\sc aseline length}\hfil
  \cr
  \noalign{\vskip -4pt}
  \omit&\multispan3\hrulefill &\multispan3\hrulefill &\multispan2\hrulefill \cr
Channel & Min  & Max & St. Dev. & Mean  & Min & Max & [s] & Samples \cr
  \noalign{\vskip 3pt\hrule\vskip 5pt}
 30\,GHz & $-18$ & 18 & 0.71 & 1180 & 172  & 92100  & 0.2461 & 8 \cr
44\,GHz & $-22$ & 28 & 1.02 & 2530 & 574  &  134000 & 0.2578& 12 \cr
70\,GHz & $-34$ & 44 & 1.88 & 8560 & 2130 & 266000 & 0.2539 & 20 \cr
  \noalign{\vskip 5pt\hrule\vskip 3pt}}}
  \endPlancktable
  \endgroup
  \end{table*}

\begin{figure}
\centering
\includegraphics[width=8.86cm]{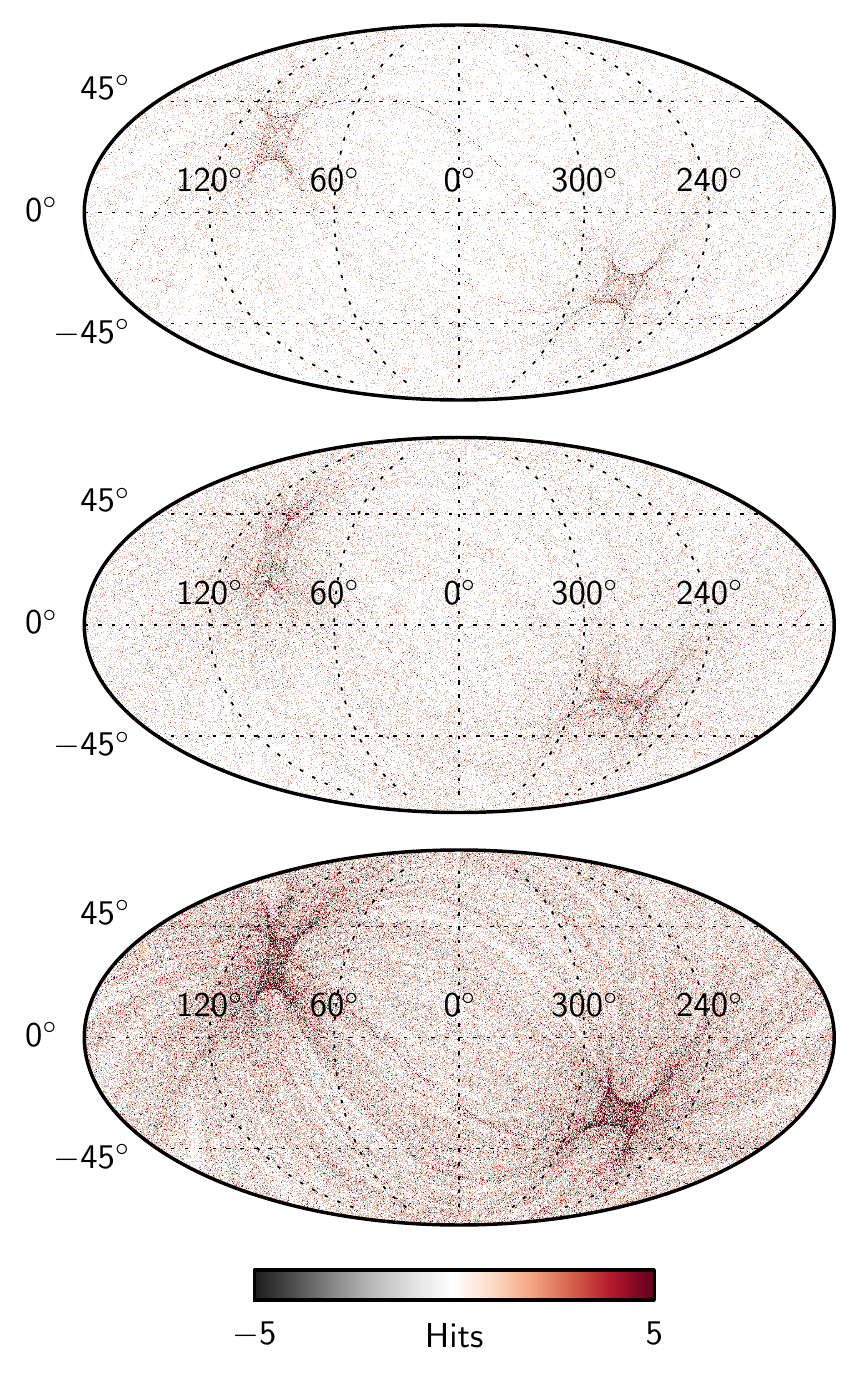}
\caption{The difference of DPC and FFP8 full mission hitmaps for all LFI frequency channels. 
The maps are shown in galactic coordinates. {\em{Top:}} 30\,GHz. {\em{Middle:}} 44\,GHz. {\em{Bottom:}} 70\,GHz.}
\label{fig:hitdiff}
\end{figure}

For the noise covariance matrix production we use the most representative noise models available, 
namely the FFP8 noise estimates \citep{planck2014-a14}. 
The noise model comprises of daily 1/$f$ model parameters.

The NCVM production uses the FFP8 pointing solution of {\tt{TOAST}} as an input. It reproduces the data processing 
centre (DPC) pointing 
to sub arc second accuracy except for a few isolated time periods where the flagging in TOAST differs from the DPC one. 
The difference between DPC and FFP8 hitmaps for all LFI frequency channels are illustrated in Fig. \ref{fig:hitdiff}, and the statistics 
of hits map differences are summarised in Table \ref{table:hitdiff}. For survey-by-survey comparisons see \cite{planck2014-a14};
this level of discrepancy is negligible for the results. The difference in hit counts is greatest around the ecliptic poles, 
where we have the most hits per pixel.

For noise covariance computations the output of {\tt{Madam/TOAST}} consists of inverse noise covariance matrices defined 
by Eq. (\ref{eq:inv_N}), specifically one inverse matrix per radiometer for a given time period. For all LFI frequency channels 
we use 0.25\,s baselines in the matrix production; the exact numbers are given in Table \ref{table:hitdiff}. 
The matrices are computed at the highest feasible resolution of $N_{\textrm{side}} = 64$. The individual inverse matrices are 
merged together to form the actual inverse NCVMs. These inverse matrices are then inverted using the eigen-decomposition 
of a matrix, and subsequently downgraded to the target-resolution using the same downgrading scheme as applied for the maps.
The 70\,GHz full mission noise covariance matrix computation took approximately 6 hours of wall-clock time and 23 000 CPU hours.

\subsection{Validation}
\label{sec:ncvm_simu}

We have performed a series of simulations to assess the effect of different parameters and 
approximations to the quality of noise
covariance matrices. 

We performed noise MC simulations with the FFP8 pipeline \citep{planck2014-a14}, 
but by varying the mapmaking parameters. 
For all combinations of simulation parameters we have generated 100 realizations of noise-only maps. 
We altered the baseline length, the destriping resolution, the weighting scheme, or the inclusion of destriping mask.
In the noise  covariance matrix production we altered the baseline length and destriping resolution. 
We chose to perform the simulations for the 30\,GHz channel, which requires the least computational resources. 
Figures \ref{fig:ncvm_ideal} and \ref{fig:chi2_nside16} summarise the results from our simulations. 

We calculate for each simulated noise map $\vec{m}$ the reduced $\chi^2$ statistics, 
defined as 
\begin{equation}
\chi^2 = \frac{\vec m^{\mathrm{T}} \tens{N}^{-1}\vec m}{N_{\rm dof}},
\end{equation}
where $\tens{N}$ is the noise covariance matrix, and the number of degrees of freedom $N_{\rm dof}=3N_{\textrm{pix}}$. 
In the presence of ill-conditioned eigenmodes, the degrees of freedom are reduced accordingly. 
For unregularized smoothed noise covariance matrices the effective number of degrees of freedom equals 
the number of non-zero eigenvalues. 
 The distribution of reduced $\chi^2$ will peak at 1 for a noise covariance matrix that models 
perfectly the properties of the simulated noise maps.

In Fig. \ref{fig:ncvm_ideal} the degree of idealization reduces from top to bottom. We add one effect at a time,
so that the last panel represents a fully realistic simulation.
We start from an idealized case (panel A), where both the noise maps and noise covariance matrices have been calculated 
at resolution $N_{\textrm{side}}=32$, with one sample baselines (0.03125\,s). In panel B we downgrade both maps and 
matrix to the resolution of $N_{\textrm{side}}=16$, and in panel C we regularize the maps and matrices. 
We observe that the effects of downgrading and regularization on the $\chi^2$ result are marginal.

The key parameter in noise covariance matrix computation is the baseline length. In earlier studies we have demonstrated
 that shorter baseline length in the noise covariance matrix production models the residual noise better \citep{planck2013-p02}. 
 Given the enhanced computing resources it was possible to bring the baseline length to one sample.
 Panels D - F summarize the results. 
 In panel D we increase the baseline length in MC simulations from one sample to 8 samples (0.25\,s),
 which was the value used in actual mapmaking.
Panels E-F show the effect of increasing the baseline length
 in matrix computation
 from one sample to first 2 samples, and then to 8 samples (0.25\,s).
 
 We find that reducing the baseline length in NCVM computation below 0.25 sec does 
 indeed improve the results, but at the same time the computer memory 
 requirements increase rapidly. 
 For the 2015 release we have used 0.25 sec baselines for all LFI frequency channels, 
 but we propose the adoption of even shorter baselines in future releases,
resources permitting.

Panel G shows the effect of changing the destriping resolution in MC simulations from
$N_{\textrm{side}} = 32$ to the realistic value $N_{\textrm{side}} = 1024$.

The noise covariance computation makes two further deviations from the high-resolution mapmaking: the horns are not uniformly 
weighted (panel H) and it does not take into account the destriping mask (panel I). 
Panel I finally represents a fully realistic simulation.
The effects in panels G-I are shown to be much smaller than the effects of the baseline length 
and the destriping resolution.

We examine the impact of resolution further in Fig. \ref{fig:chi2_nside16}. 
As motivated in previous section the matrices should be calculated at the highest computationally feasible resolution. We calculated a few noise covariance matrices with varying initial resolution.
Each matrix was downgraded to resolution  $N_{\textrm{side}} = 16$ and compared against a downgraded set of FFP8 noise maps (100 realisations).
 As expected, $N_{\textrm{side}} = 64$ gives the best agreement. Increasing the initial resolution 
 beyond $N_{\textrm{side}} = 64$ is likely to improve results further, but the expected improvement would be a very small for 
 a huge computational effort. Increasing the initial resolution beyond $N_{\textrm{side}} = 64$ to $N_{\textrm{side}} = 128$  will increase the matrix size 
 by a factor of 16 to 2.5 TB.


\section{Results}

\subsection{High-resolution maps}

\begin{figure}
\centering
\includegraphics[width=1\columnwidth]{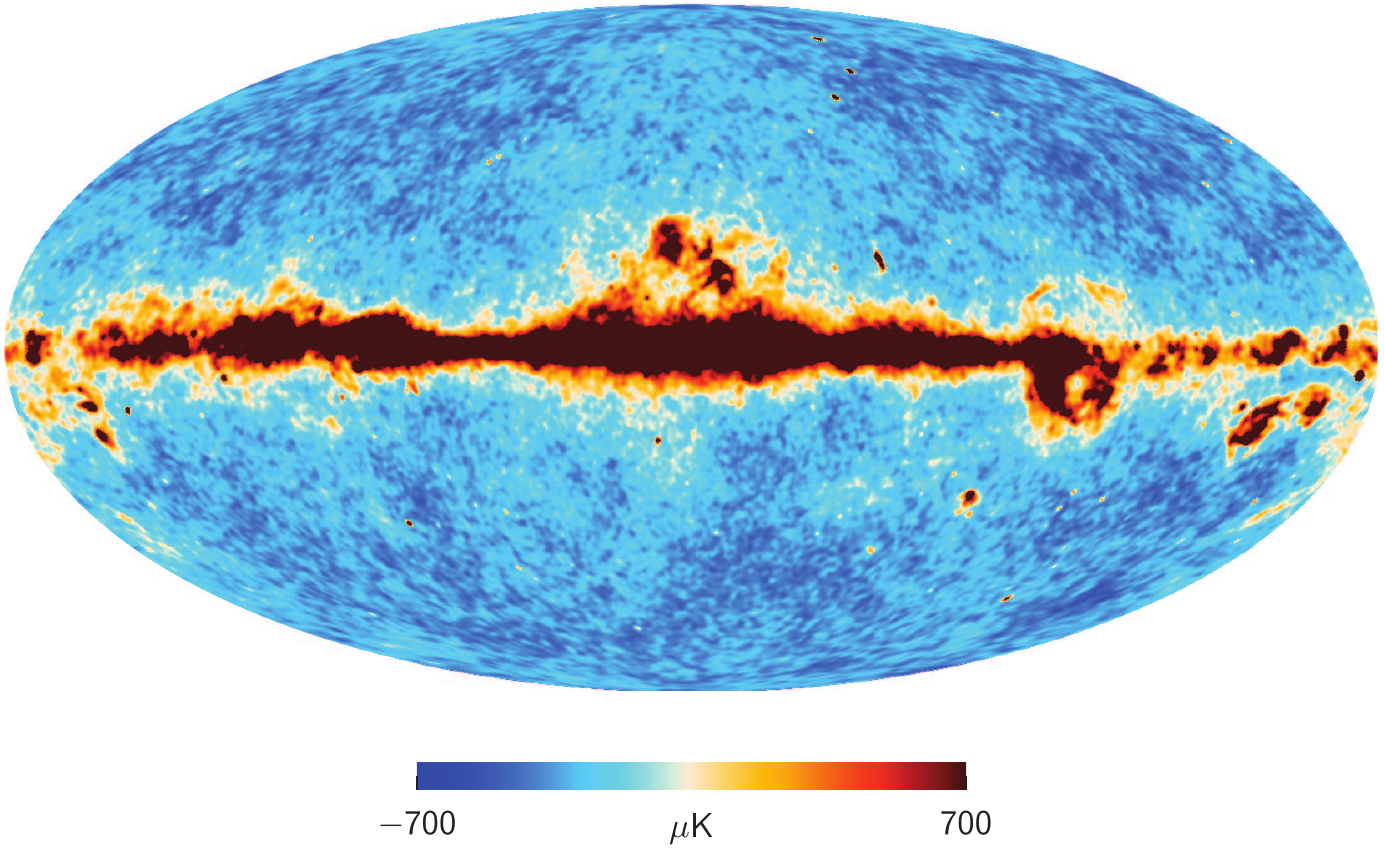}
\includegraphics[width=1\columnwidth]{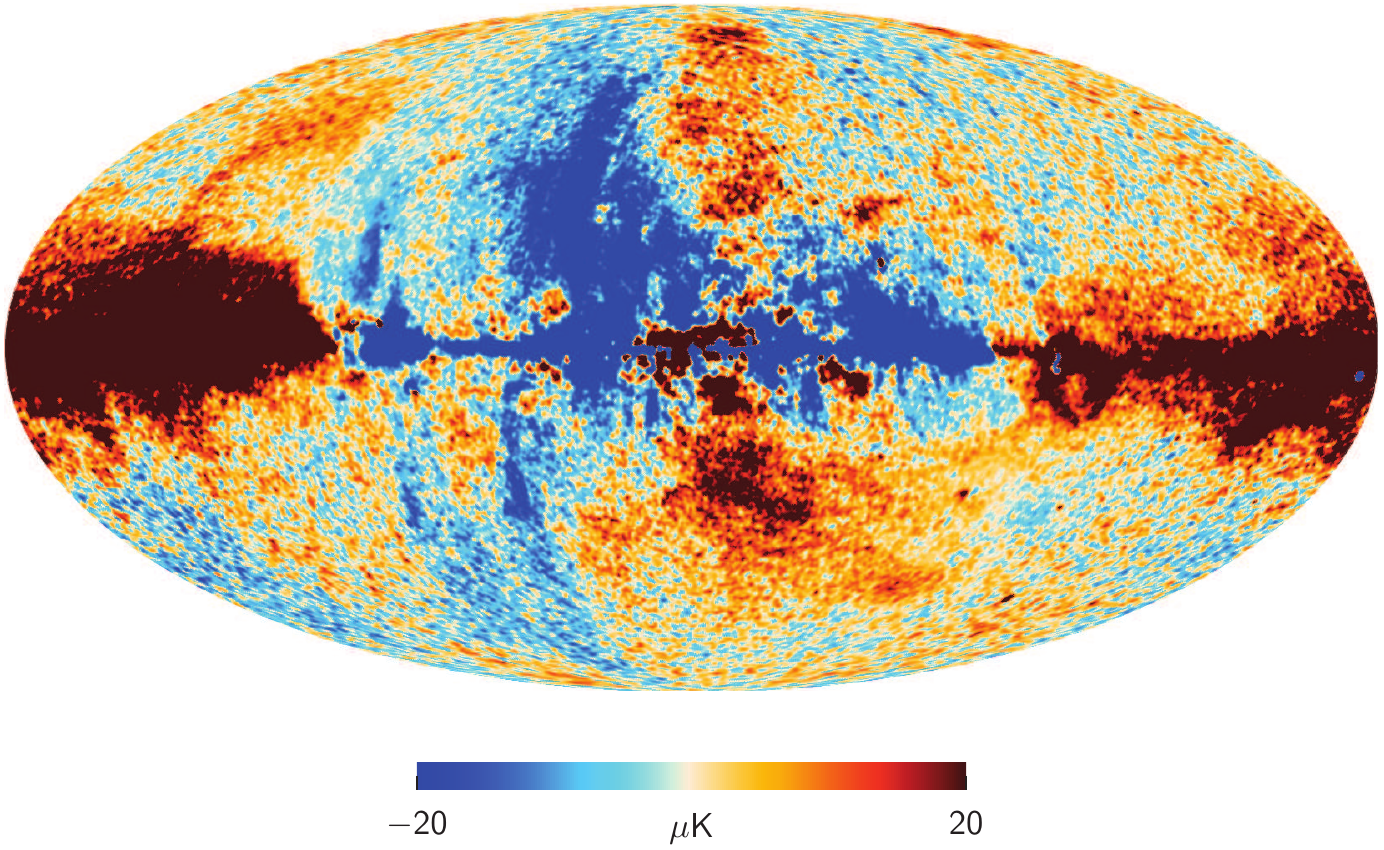}
\includegraphics[width=1\columnwidth]{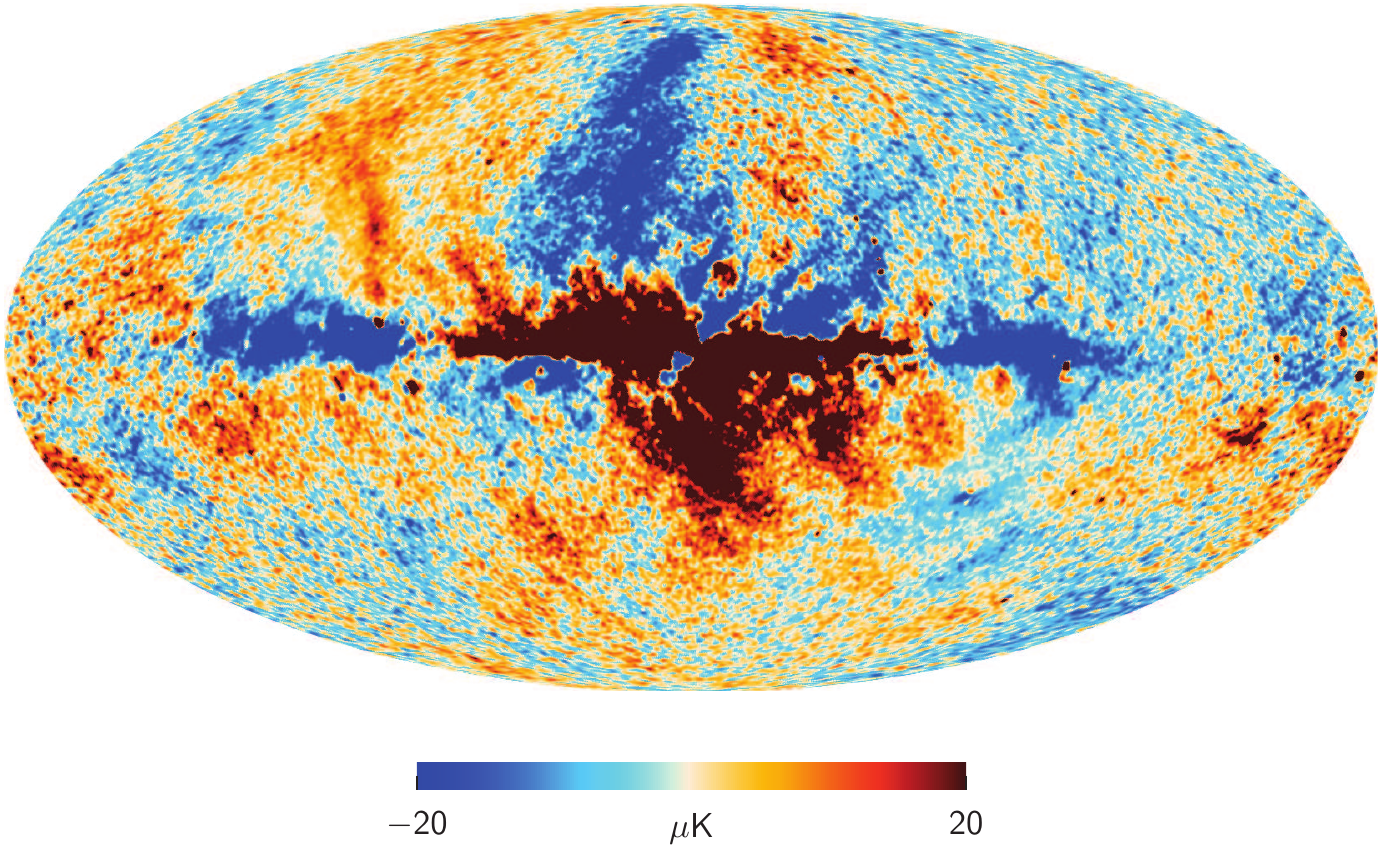}
\caption{LFI 30\,GHz channel maps. From top to bottom: temperature; $Q$, and $U$ polarization. 
The temperature map has resolution $N_\textrm{nside}$=1024.
The polarization maps have been downgraded to resolution $N_\textrm{side}$=256 and smoothed with 
a FWHM=1$^\circ$ Gaussian beam. 
The polarization maps are not corrected for bandpass mismatch leakage.}
\label{fig:freqmaps30}
\end{figure}

\begin{figure}
\centering
\includegraphics[width=1\columnwidth]{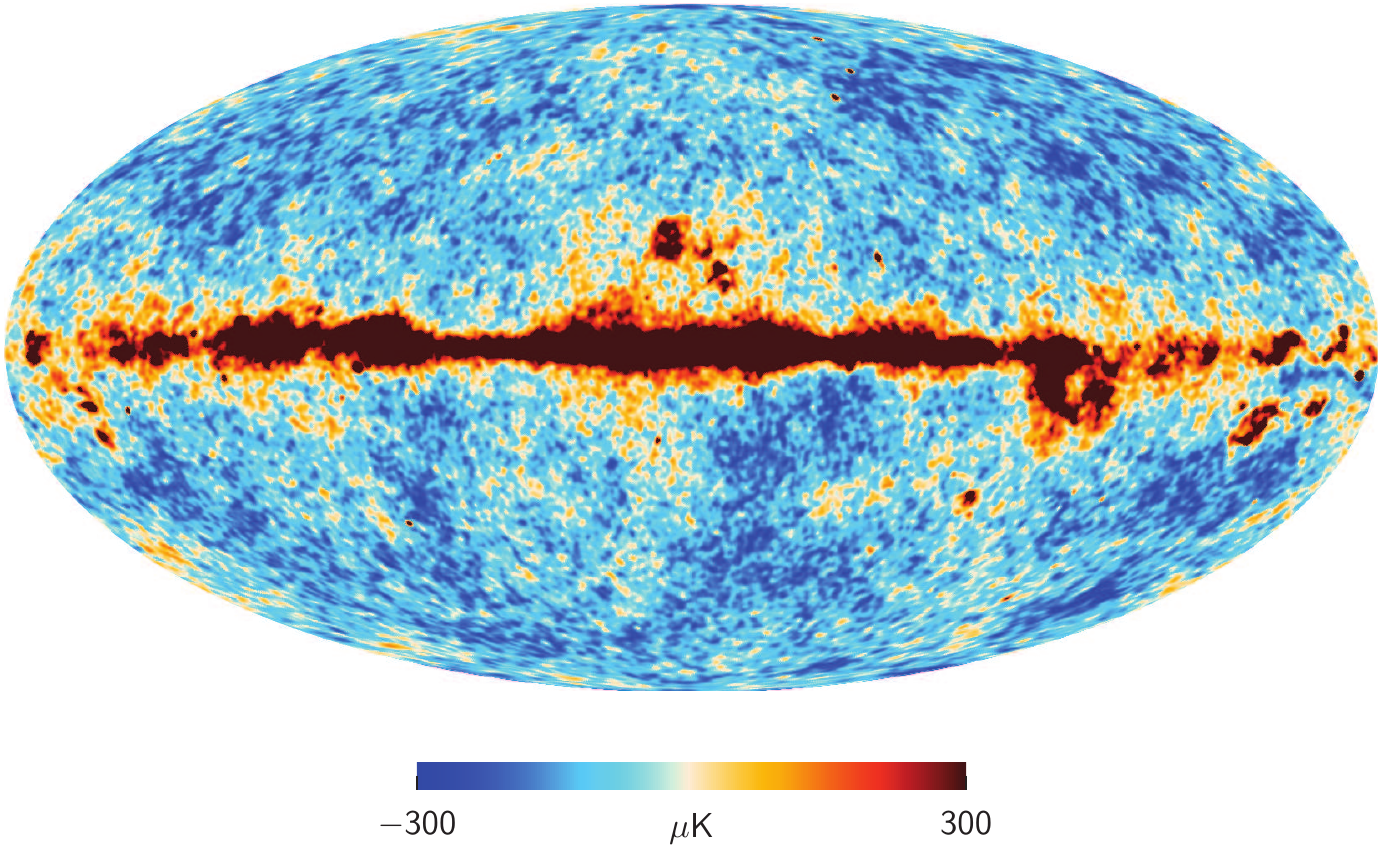}
\includegraphics[width=1\columnwidth]{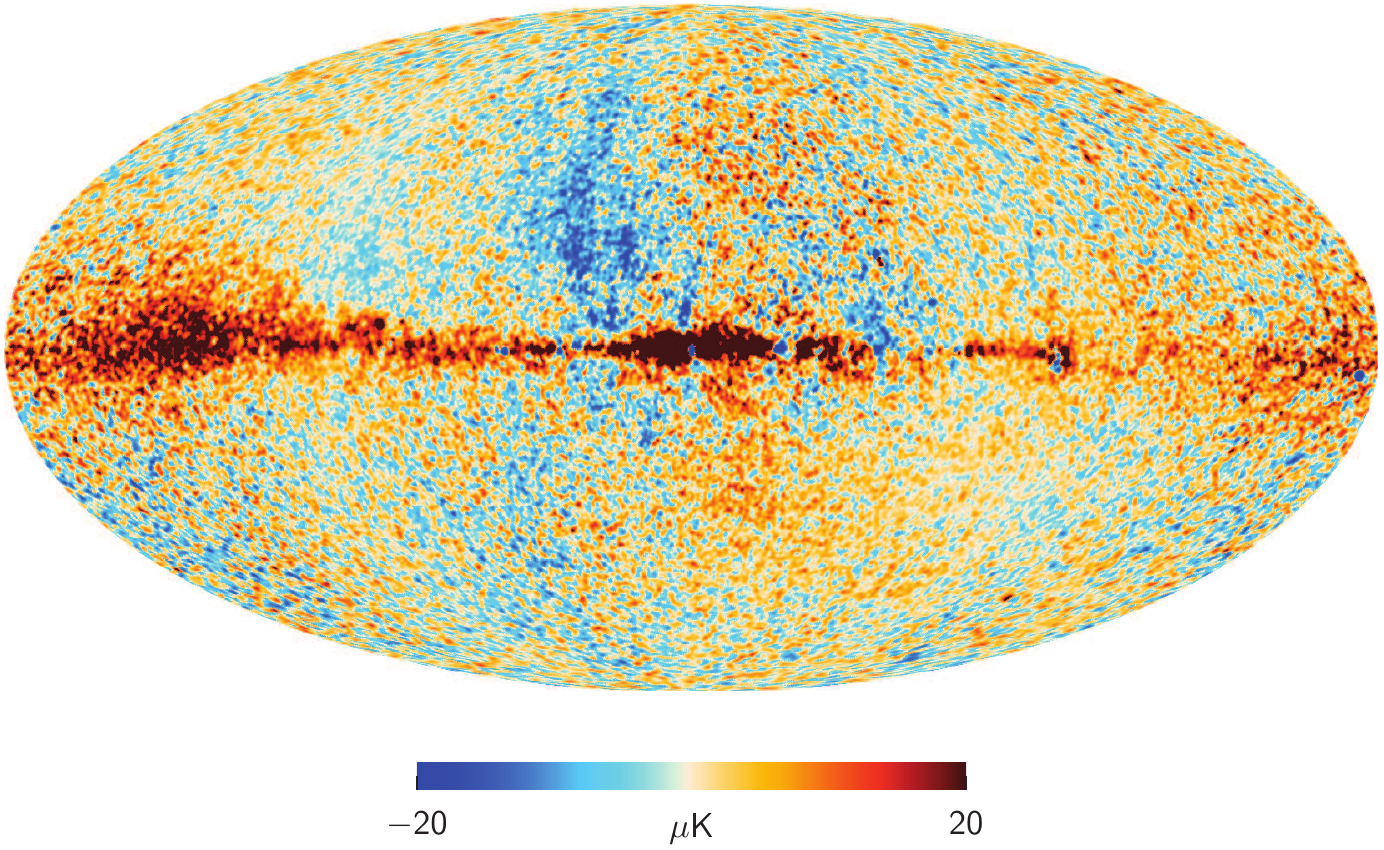}
\includegraphics[width=1\columnwidth]{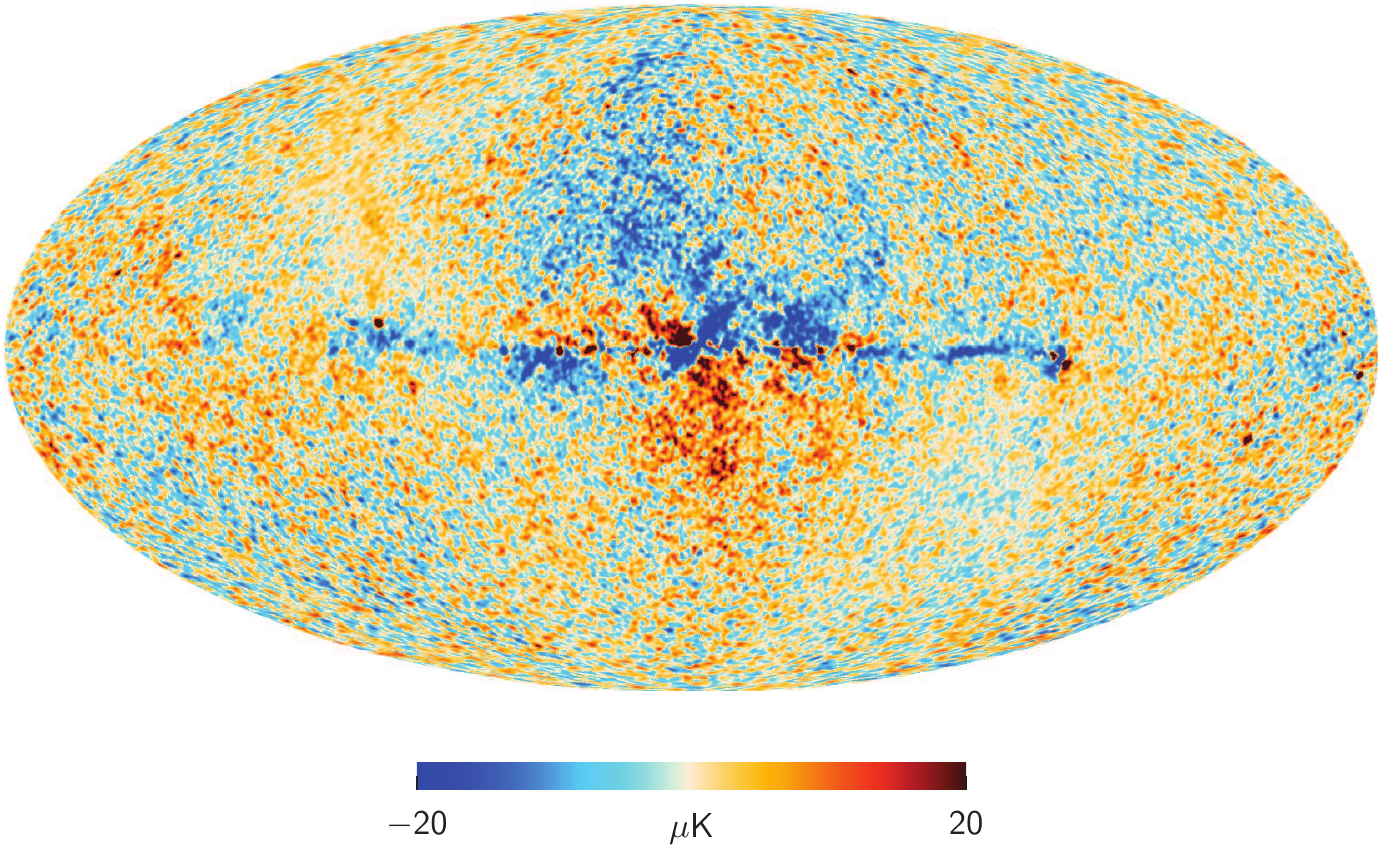}
\caption{LFI 44\,GHz channel maps. 
Fom top to bottom: temperature, $Q$, and $U$ polarization. }
\label{fig:freqmaps44}
\end{figure}

\begin{figure}
\centering
\includegraphics[width=1\columnwidth]{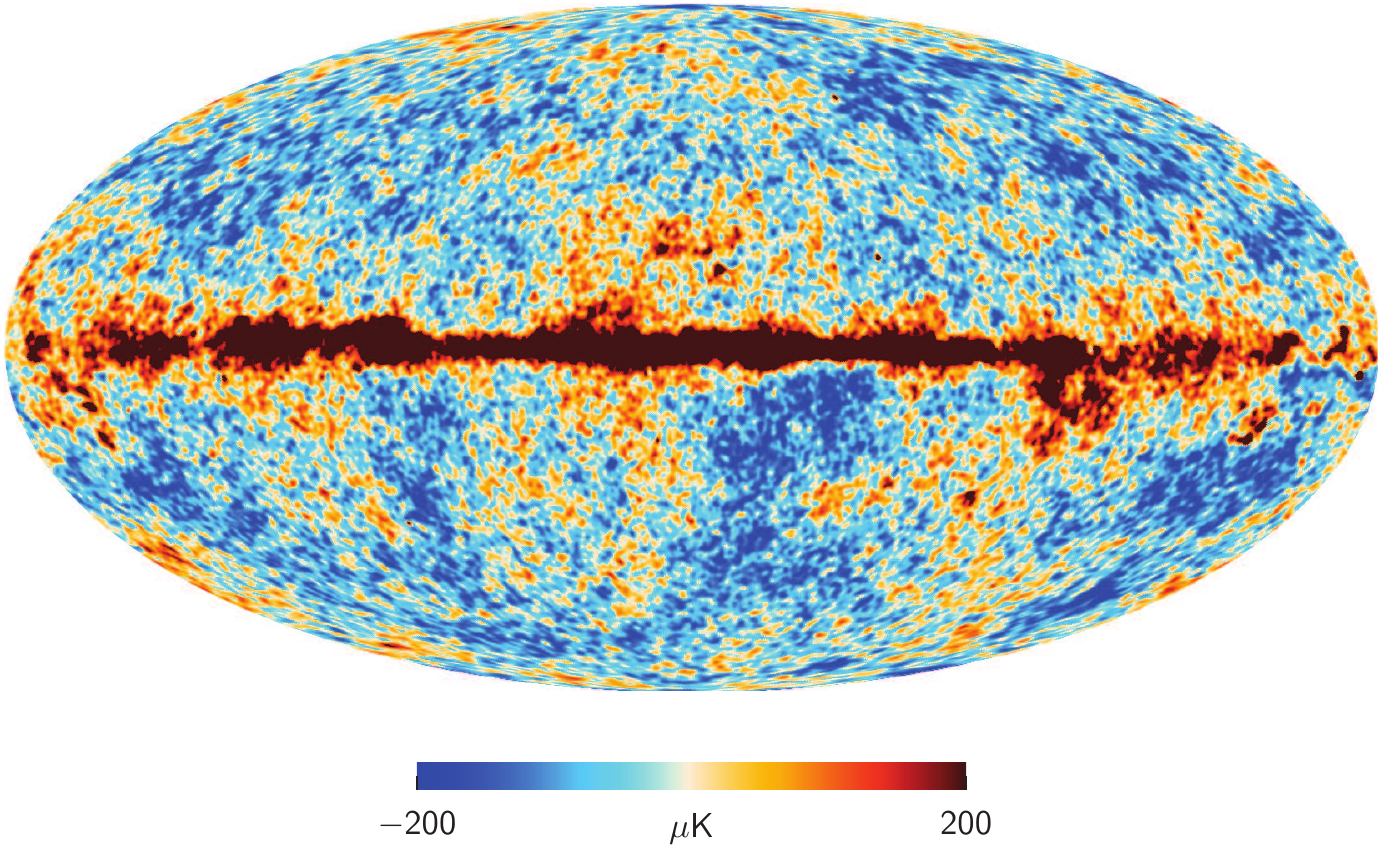}
\includegraphics[width=1\columnwidth]{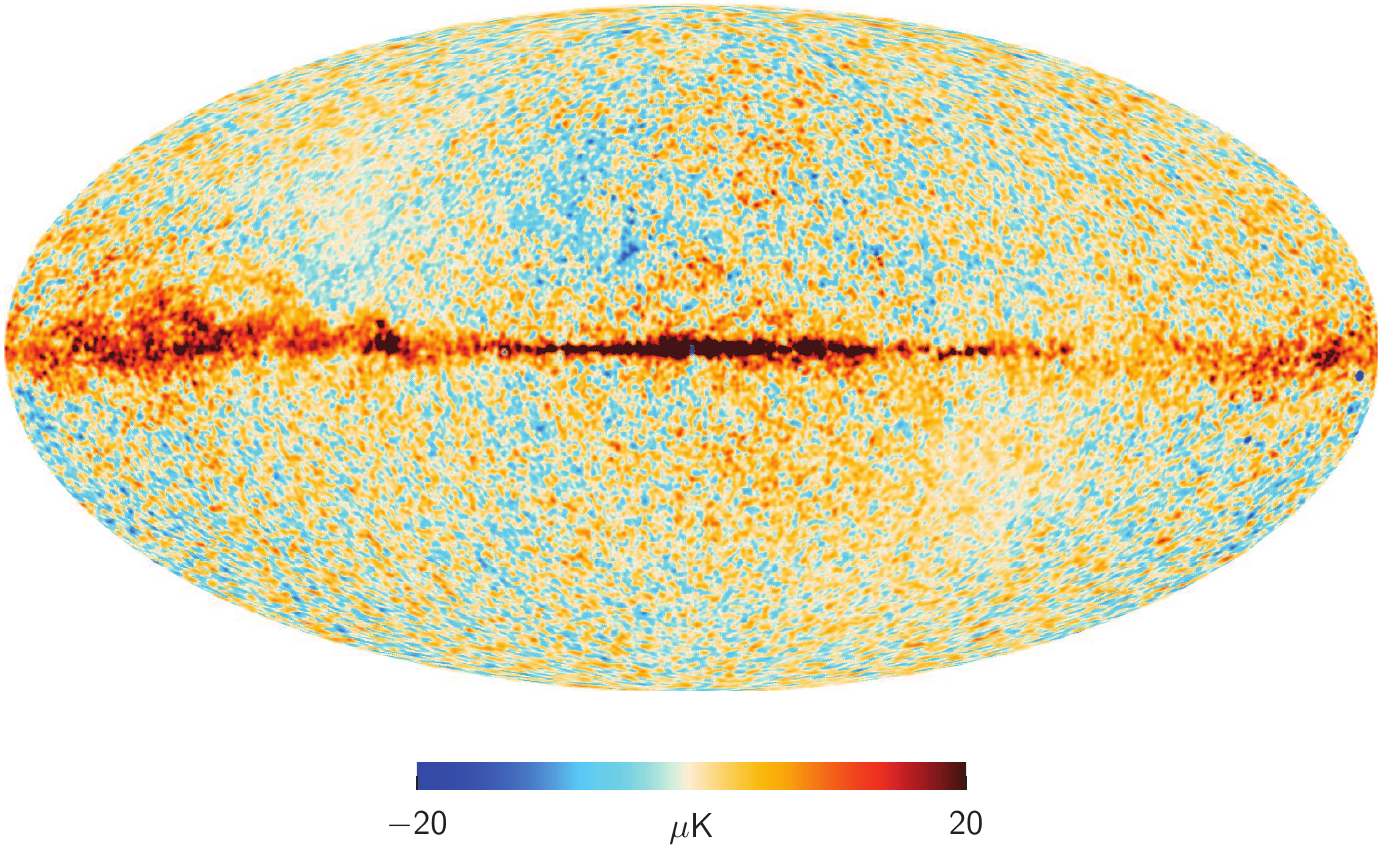}
\includegraphics[width=1\columnwidth]{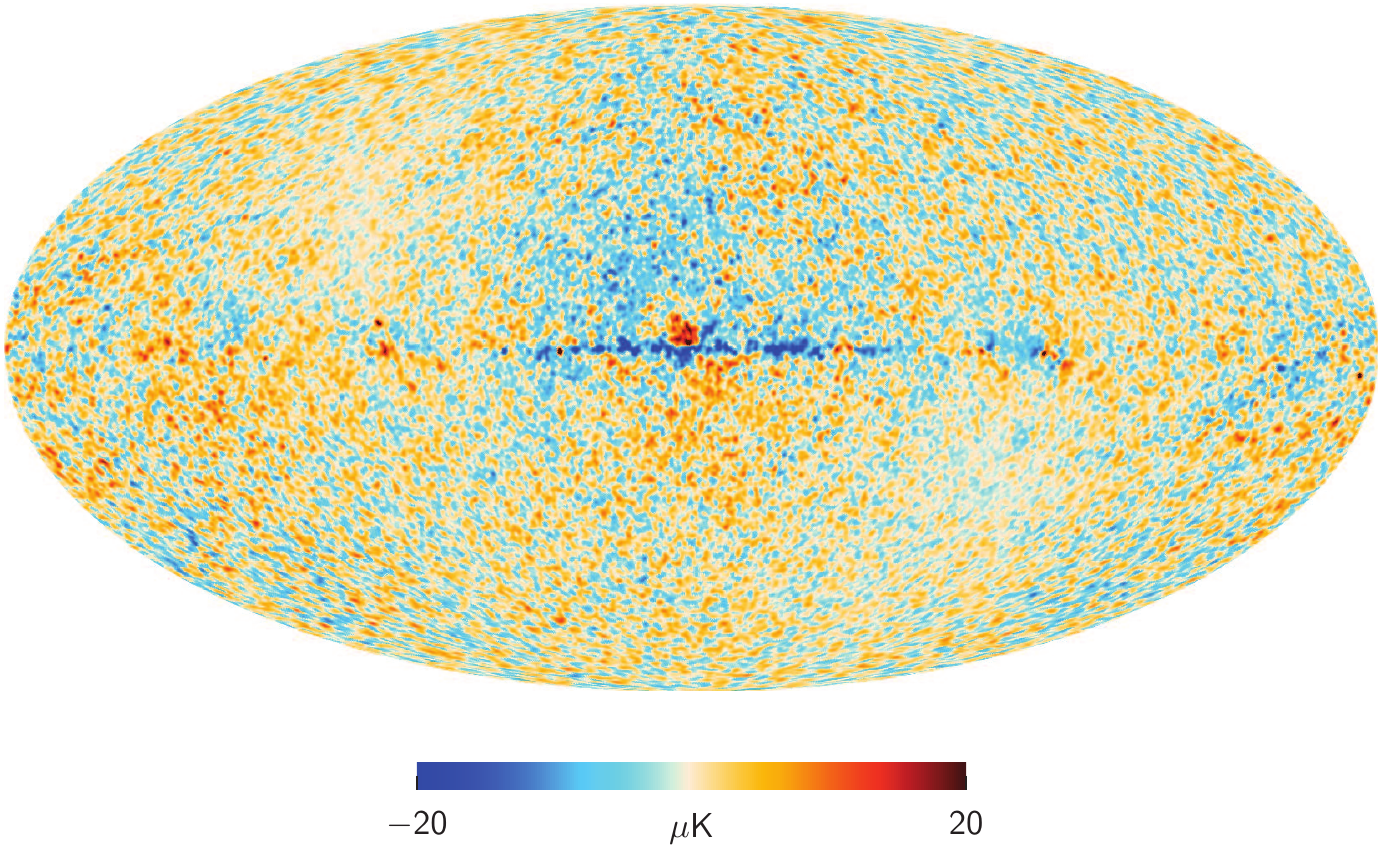}
\caption{LFI 70\,GHz channel maps.
From top to bottom: temperature, $Q$, and $U$ polarization.  }
\label{fig:freqmaps70}
\end{figure}

The 2015 release includes a large number of map products.
A complete list is given \cite{planck2014-a03}.
Figures \ref{fig:freqmaps30}--\ref{fig:freqmaps70} show the main products,
which are the frequency  maps for the complete four-year data set. 
Shown in these figures are the temperature map, and $Q$ and $U$ polarization.
For illustration purposes the polarization maps have been smoothed
with a symmetric Gaussian beam with FWHM=$1\deg$.

The released maps are in Galactic coordinates.
The division of the polarization signal into $Q$ and $U$ components is dependent on the chosen coordinate frame.
The $Q$ map represents a component where the polarization direction is aligned with the local meridian (positive $Q$) 
or with the local latitude (negative $Q$).
The $U$ map represents a component where the polarization direction is at a 45$^\circ$ angle with respect to the meridian.
The polarization maps shown here are subject to bandpass leakage \citep{planck2014-a03},
which must be taken into account in the subsequent analysis.

We do not correct for beam shape in the mapmaking process.
The shape of a point source in a map directly reflects the beam shape. In particular,
the power that is lost into the far sidelobes \citep{planck2014-a05}, is missing from the point source as well,
and must be taken into account when determining source fluxes.

As auxiliary information we provide for each high-frequency map an estimate of the white noise covariance.
The matrix gives the noise correlation between Stokes components within each pixel.  
We provide also a corresponding hit count map for each map.

\subsection{Sky coverage}

Among other map products are single-survey maps that include data from one 6 month survey,
and single-year maps than include two surveys each.
These do not give full sky coverage.
As discussed in Sect.
\ref{sec:missing_pixels}, construction of ($I,Q,U$) Stokes components requires that a pixel is scanned by a pair of horns,
with complementary polarization sensitivity.  Pixels that are scanned by a single horn or radiometer, can be used for destriping,
but not for the final map.

The sky coverage fraction of the frequency maps are given in Table {\ref{tab:sky_coverage}}.
We present three numbers for each map.
In the first column (``map") we give the sky coverage of the final map.
This is the fraction of sky that has been scanned by two radiometer pairs so that we have been able to solve the full $I,Q,U$ triplet.

The second column (``scan") gives the fraction of sky that is scanned by at least one radiometer, and is thus
available for the solution of the destriping equation. 
The focal plane is designed in such a way that in most cases a pair of horns follows the same scanning path.
Each pixel is thus covered by four radiometers,
and the difference between the two columns is small.
An exception to this is the 44\,GHz frequency channel, where horn LFI-24 does not have a counterpart.
A significant fraction of the sky in single surveys is covered by horn LFI-24 alone.
As a consequence, the sky fraction available for destriping is significantly higher than that available for the final map.
In the third column (``destriping") we give the sky fraction that remains after the galactic mask is applied.
This is the fraction eventually used for destriping.

\begin{figure*}[htbp]
        \centering
        \includegraphics[width=18cm]{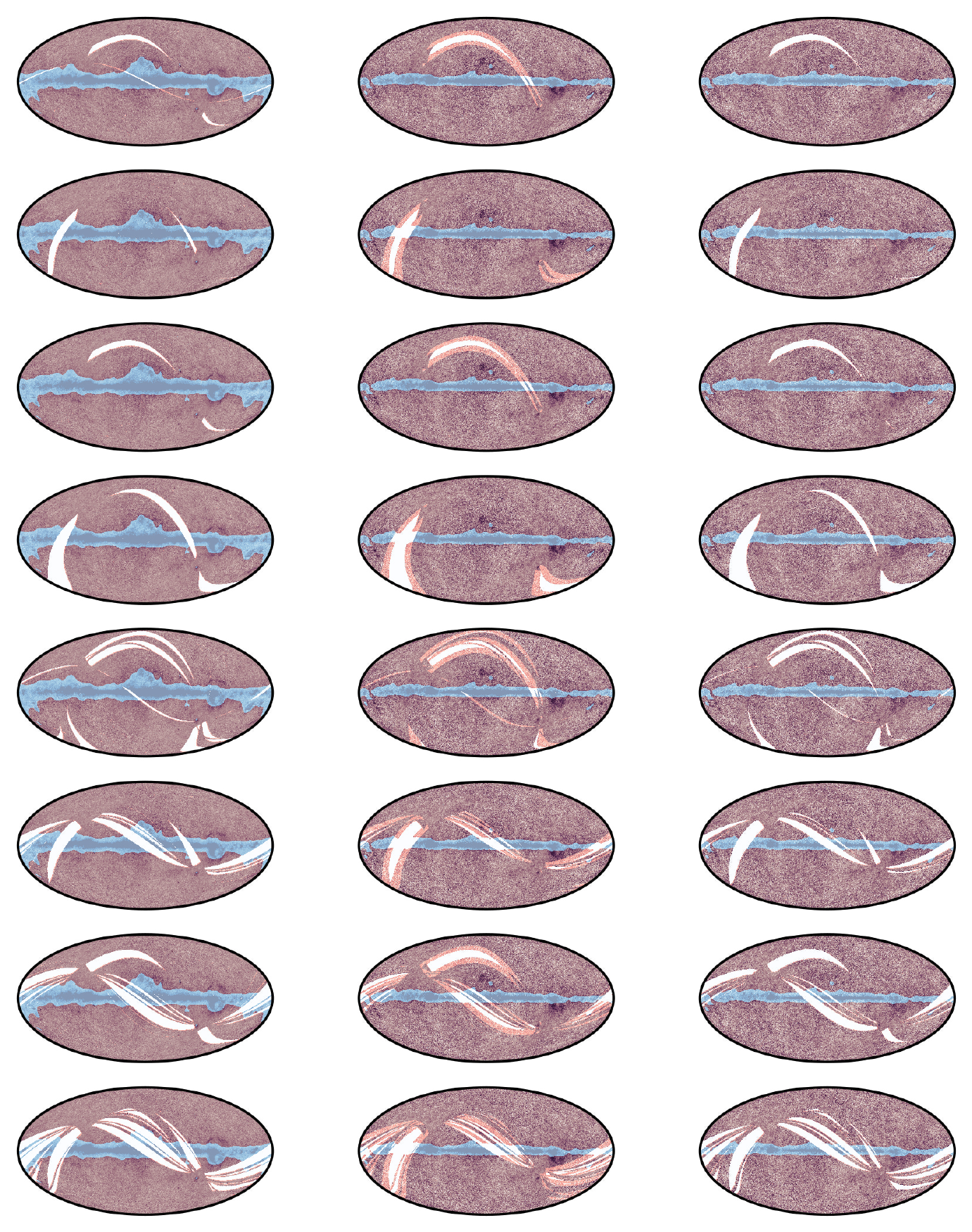}
                \caption{Sky-coverage of single-survey temperature maps for 30, 44, and 70\,GHz. 
                {\em Left:} 30\,GHz. {\em Middle:} 44\,GHz. {\em Right:} 70\,GHz. 
        From top to bottom, Surveys 1 to 8. The combination of red and purple regions is used for destriping.
        The combination on purple and blue regions is included in the final map.
        Unobserved regions are shown in white.}
        \label{fig:surveymaps}

\end{figure*}

Single-survey temperature maps for each full frequency are shown in Fig. {\ref{fig:surveymaps}}.
We indicate by different colour scales the regions used for the final map or for destriping.
The combination of blue and purple regions is included in the final map. The size of this region is given in the ``map'' column 
of Table \ref{tab:sky_coverage}.
The Galactic region, shown in blue, is masked out in the destriping phase, and not used for solving the baselines.
The red colour, mostly visible at 44\,GHz, indicates the region that is scanned by a single horn, and is used for destriping,
but is not included in the final map.  The purple and red regions together correspond to the column headed ``destriping'' in Table  \ref{tab:sky_coverage}. 
The region shown in white does not contain useful data.

  \begin{table*}
  \begingroup
  \newdimen\tblskip \tblskip=5pt
  \caption{Sky coverage (\%) of full, single-year, and single-survey frequency maps.
  We give the fraction of the sky covered by the final map (``map"),
  the fraction that was scanned by at least one horn (``scan") and is available for destriping,
  and the fraction that remains after applying a galactic mask and that was used for destriping (``destriping").
  The coverage of the destriping mask can be read from the first row of  the ``destriping" column.)
  }
  \label{tab:sky_coverage}
  \nointerlineskip
  \vskip -3mm
  \footnotesize
  \setbox\tablebox=\vbox{
  \newdimen\digitwidth
  \setbox0=\hbox{\rm 0}
  \digitwidth=\wd0
  \catcode`*=\active
  \def*{\kern\digitwidth}
  \newdimen\signwidth
  \setbox0=\hbox{+}
  \signwidth=\wd0
  \catcode`!=\active
  \def!{\kern\signwidth}
  \halign{\hbox \tabskip=12pt{#}&
      \hfil#  & \hfil#\ &\hfil# & \hfil#\hfil& \hfil#\hfil&  
      \hfil#\hfil& \hfil#\hfil& \hfil#\hfil& \hfil#\hfil& \hfil#\hfil\tabskip=0pt\cr
  \noalign{\doubleline}
 \omit & \omit & \multispan3 \hfill 30\,GHz\hfill &   \multispan3 \hfill 44\,GHz\hfill &   \multispan3 \hfill 70\,GHz\hfill  \cr
  \noalign{\vskip -4pt}
  \omit & \omit & \multispan3 \hrulefill &   \multispan3 \hrulefill &   \multispan3 \hrulefill \cr
  \noalign{\vskip 0pt}
Survey & PID range & map & scan & destriping & map & scan & destriping & map & scan & destriping \cr
  \noalign{\vskip 3pt\hrule\vskip 5pt} \cr
  \noalign{\vskip 4pt}
full      & 3-44072         & 100.0   & 100.0    &  78.74    &100.0 & 100.0 &89.37    &100.0 & 100.0 &  89.67\cr
  \noalign{\vskip 4pt}
year 1 & 3--10957         & 99.999 & 100.0   &  78.74    & 99.998 & 100.0 & 89.37     & 100.0  & 100.0 & 89.67 \cr
year 2 & 10958--21482 & 99.98 & 100.0   &  78.74    & 99.998 & 100.0 &  89.37    & 100.0  & 100.0 &  89.67 \cr
year 3 & 21483--32761 & 99.72 & 99.74 &  78.66    & 99.91 & 100.0 &  89.37    & 99.65  & 99.65 & 89.38  \cr
year 4 & 32762--44072 & 95.68 & 95.76 &  76.61    & 96.87 & 98.68  & 88.35    & 97.38 & 97.43 &  87.63 \cr
  \noalign{\vskip 4pt}
survey 1 & 3--5483             & 97.20  & 98.11 &  76.88    & 93.93  & 98.03  &  87.41   & 97.94  & 97.95 & 87.62 \cr
survey 2 & 5484--10957     & 97.48  & 97.58 & 77.40      & 93.31  &  98.13 &  87.89    & 97.47   & 97.49 &  87.62 \cr
survey 3 & 10958--16454   & 97.62  & 97.71  &  76.49     & 93.65  &  98.23 &  87.61   & 97.61  & 97.63 &  87.30  \cr
survey 4 & 16455--21482   & 91.88  & 91.98  &  72.37     & 89.53  &  94.92 &  84.85  & 92.40  & 92.43 & 82.89   \cr
survey 5 & 21483--27404   & 90.89  & 91.02  &  70.27     & 88.43  &  96.70 &  86.08    & 92.44  & 92.50 &  82.33  \cr
survey 6 & 27405--32761   & 87.79  & 88.06  &  72.35     & 86.10  & 94.56  &  85.29    & 89.95  & 90.18 &  82.04  \cr
survey 7 & 32762--38574   & 85.40  & 85.60 &  68.29      & 83.70  & 92.49  &  82.80    & 88.43   & 88.52 & 79.65  \cr
survey 8 & 38575--44072   & 80.01  & 80.26  &  67.54     & 78.92  & 89.61  &  81.15    & 83.83  & 84.09 & 77.05   \cr
   \noalign{\vskip 5pt\hrule\vskip 3pt}}} 
 \endPlancktable
  \endgroup
  \end{table*}

\subsection{Low-resolution products}

The 2015 release includes low-resolution maps and matrices both for the full mission and for a data selection covering surveys 
1, 3, and 5--8 \citep{planck2014-a03} at resolution of  $N_{\textrm{side}} = 16$. They have been processed with the noise-weighted 
downgrading scheme, and subsequently 
the Stokes $I$ component has been smoothed with a Gaussian beam of FWHM $ = 440 \arcmin$. 
Finally the products have been regularised by adding 2 $\mu$K RMS for $I$, and 0.02 $\mu$K RMS
 for $Q$ and $U$ of white noise at $N_\textrm{side} = 16$ resolution. 

\begin{figure*}[htbp]
        \centering
         \includegraphics[width=18cm]{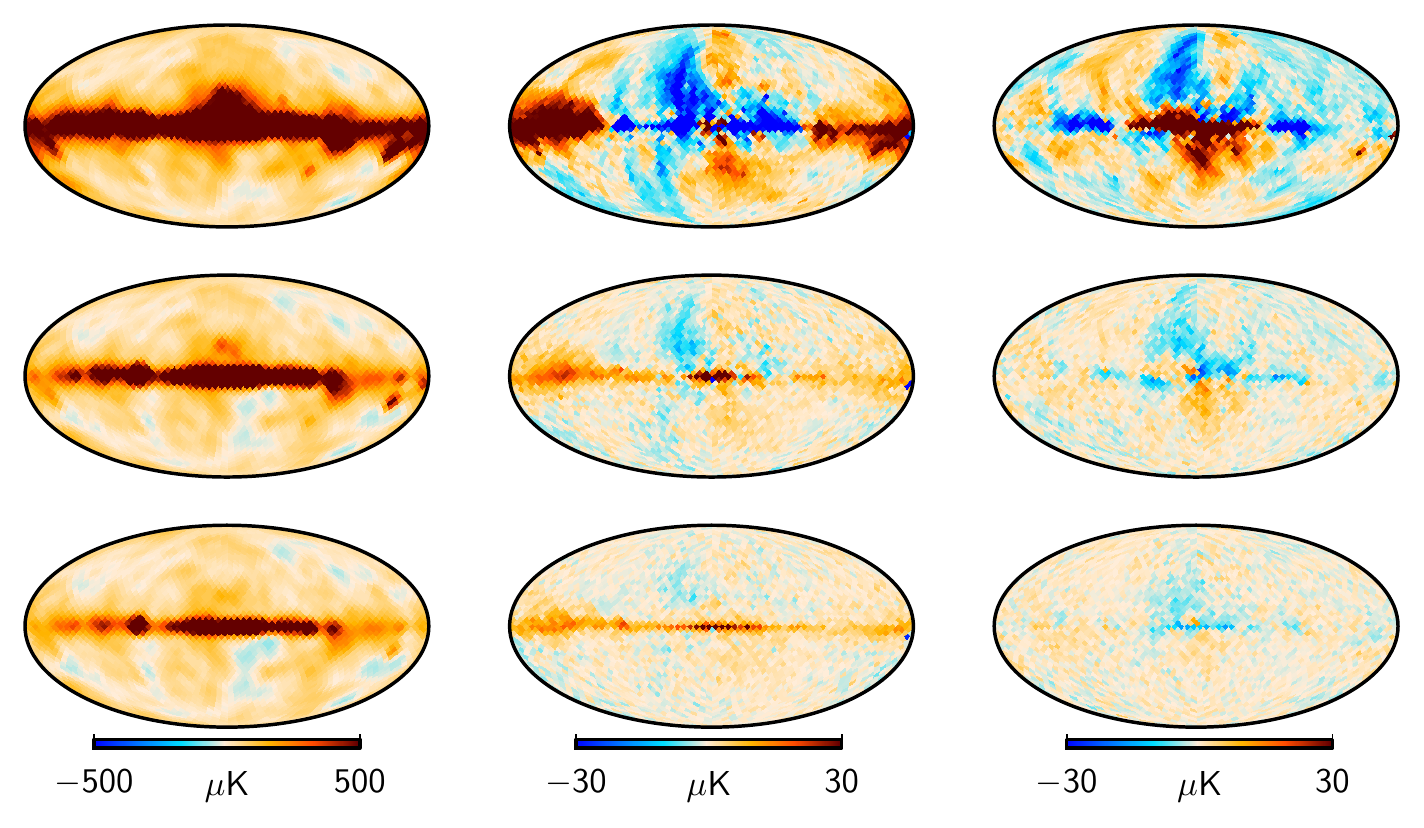}
    \caption{LFI full mission low resolution maps, at $N_\textrm{side}=16$.  
    From left to right: intensity $I$; polarization $Q$ component; and polarization $U$ component.
     \,\,{\it Top}: 30\,GHz. \,\,{\it Middle}: 44\,GHz  \,\,{\it Bottom}: 70\,GHz. 
     Units are in $\mu$K$_{\rm CMB}$.}
        \label{fig:low_maps}
\end{figure*}

\begin{figure*}
\centering
\includegraphics[width=18cm]{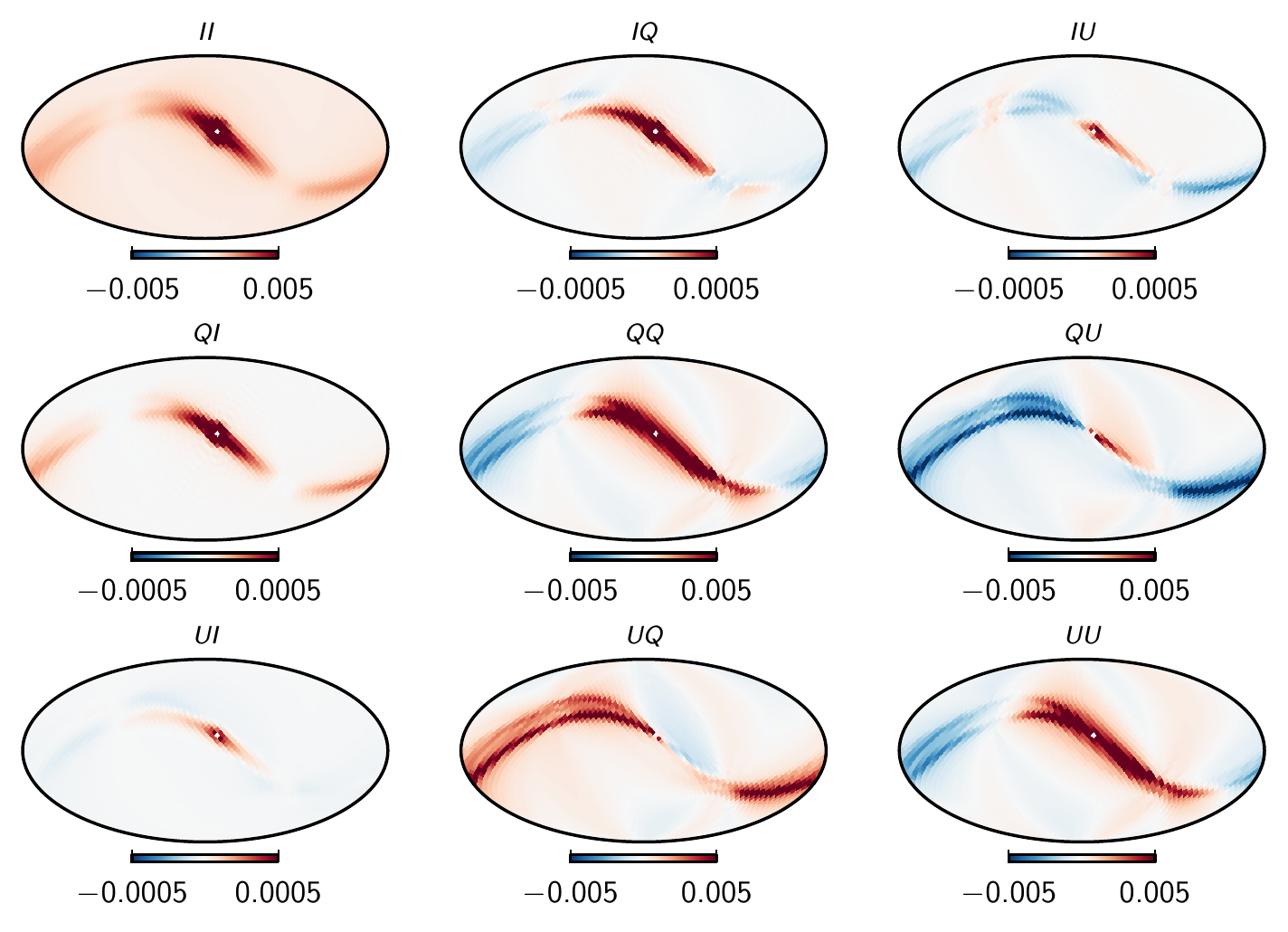}
\caption{A single column of the 70\,GHz full mission noise covariance matrix. The reference pixel is pixel 1248 in the {\tt{HEALPix}} 
nested pixelization scheme. The noise covariance matrix has been normalised to show correlation coefficients. The reference pixel 
is set to zero to highlight more of the finer details of the correlation structure.}
\label{fig:ncvm_col1248}
\end{figure*}

Figure \ref{fig:low_maps} illustrates the full mission low-resolution Stokes $I$, $Q$ and $U$ maps for 
each \Planck\ LFI frequency channel.
Figure \ref{fig:ncvm_col1248} illustrates a single column of the 70\,GHz full mission noise covariance matrix. 
The column corresponds 
to a pixel number 1248 in the  {\tt{HEALPix}} nested pixelization scheme for $N_\mathrm{side} = 16$ resolution. 
The pixel values in the plot 
represent correlation coefficients, since each pixel value $\langle m_p m_q \rangle$ has been normalised by 
$\sqrt{\langle m_p^2\rangle \langle m_q^2\rangle}$. 
In this normalization the reference pixel automatically gets unit value, and is later set to zero to bring out finer details of the 
noise covariance matrix.

\subsection{Characterization of noise}

\subsubsection{Residual noise at high multipoles}

Residual noise at high multipoles is dominated by non-uniform white noise,
which is uncorrelated between pixels.
We provide a description of the white noise component in the form of the white noise covariance,
computed according to Eq. (\ref{wncov}). 
The covariance matrix consists of a symmetric $3\times3$ matrix for every map pixel.
The matrix gives the noise correlation between Stokes components, 
$I,Q$, and $U$, within each pixel. 
White noise gives a lower limit for the full residual noise level.

\begin{table*}
  \begingroup
  \newdimen\tblskip \tblskip=5pt
  \caption{White noise level for frequency maps, computed from the white noise covariance matrix.
 We show the average white noise covariance per pixel for Stokes components,
 and  white noise spectrum level derived from the same numbers.
 All numbers are in $\mu$K$^2$ units.}
  \label{tab:whitenoise}
  \nointerlineskip
  \vskip -3mm
  \footnotesize
  \setbox\tablebox=\vbox{
  \newdimen\digitwidth
  \setbox0=\hbox{\rm 0}
  \digitwidth=\wd0
  \catcode`*=\active
  \def*{\kern\digitwidth}
  \newdimen\signwidth
  \setbox0=\hbox{+}
  \signwidth=\wd0
  \catcode`!=\active
  \def!{\kern\signwidth}
  \halign{\hbox {#}\tabskip=2em&
      \hfil#\hfil& \hfil#\hfil&\hfil#\hfil& \hfil#\hfil& \hfil#\hfil&  \hfil#\hfil&
      \hfil#\hfil  & \hfil#\hfil\tabskip=0pt\cr
  \noalign{\doubleline}
  \noalign{\vskip 0pt}
Channel & $II$ & $QQ$ & $UU$ & $IQ$  & $IU$ & $QU$ & $C_{TT}$ & $C_{EE}$, $C_{BB}$ \cr
  \noalign{\vskip 3pt\hrule\vskip 5pt}
30\,GHz &  3224.4 & 6467.8   & 6465.1 & $-19.2$ &  $-71.0$ &  $ -41.2$   &  0.003220 & 0.006458 \cr
44\,GHz &  4331.0 & 10088.0 & 8906.9 & $-2.4$  &   $24.3$ & $-136.9$   &  0.004325 & 0.009485 \cr
70\,GHz &  3358.5 & 6775.7   & 6713.7 & $28.1$  & $-69.0$ &  $-83.3$   &  0.003354 & 0.006736 \cr
  \noalign{\vskip 5pt\hrule\vskip 3pt}}}
  \endPlancktable
  \endgroup
  \end{table*}
  
Table \ref{tab:whitenoise} shows the average white noise covariance per pixel in the frequency maps.
We show the average value of $II, IQ, IU, QQ, QU, UU$ noise covariance,
as average over all pixels.  We also give the expected white noise level in the angular power spectrum,
calculated as
\begin{equation}
C_{T\ell}^{\rm wn} = \langle II \rangle \frac{4\pi}{N_{\rm pix}}
\end{equation}
and
\begin{equation}
C_{E\ell}^{\rm wn} = C_{B\ell }^{\rm wn} = \frac12 (\langle QQ\rangle +\langle UU\rangle) \frac{4\pi}{N_{\rm pix}}
\end{equation}
where $N_{\rm pix}$ is the number of pixels in the map (12\,582\,912)
and brackets indicate an average over pixels.

The 44\,GHz channel shows an imbalance between $Q$ and $U$ components.
This is due to horn LFI-24, which does not have a counterpart with complementary polarization sensitivity.
The same effect can be seen in the CRN plots in Sect. \ref{sec:validation}.

\begin{figure}
\centering
\includegraphics[width=1\columnwidth]{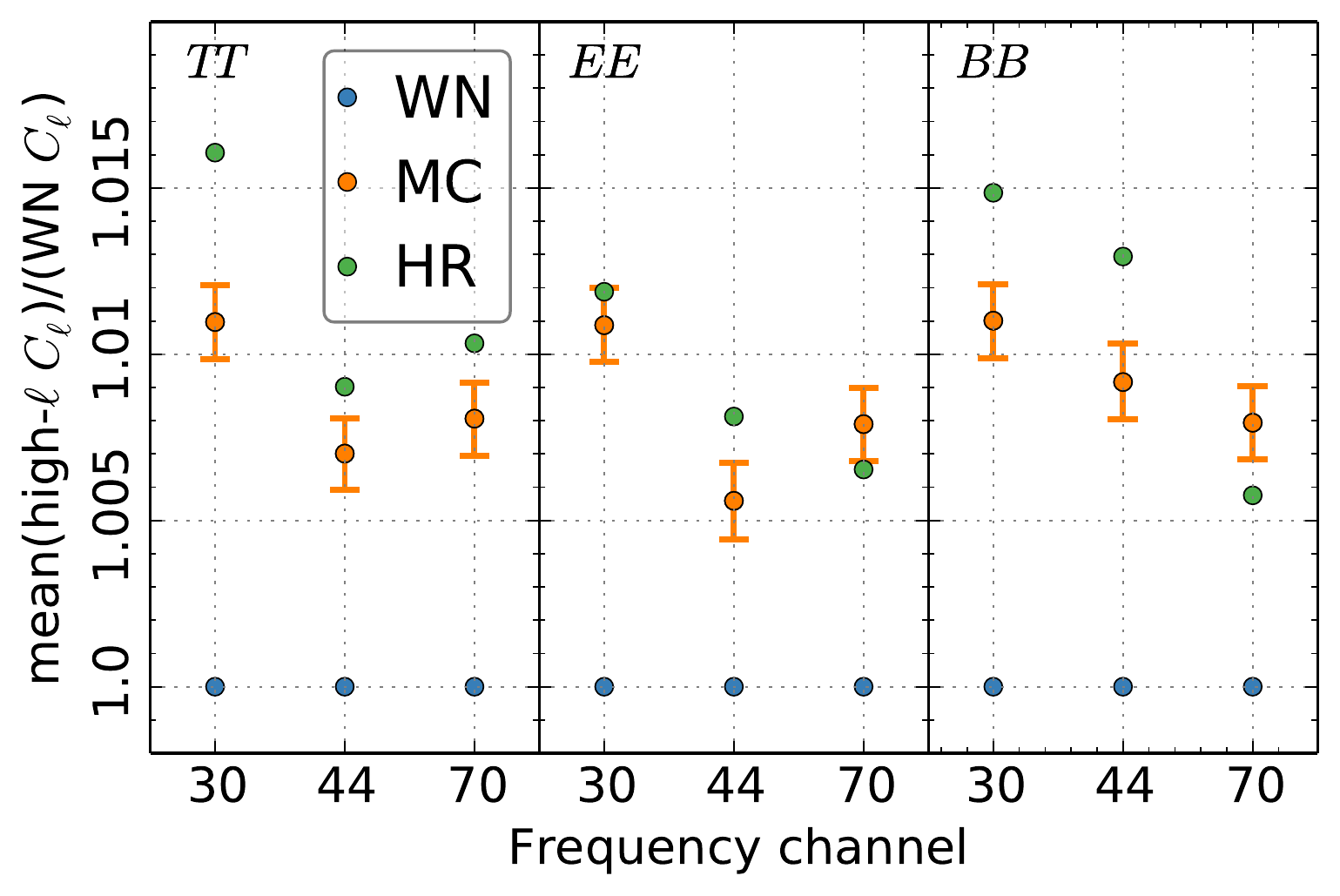}
\caption{Comparison of noise estimates at high multipoles,
as a fraction white noise level (blue).  The green dots show the estimate from half-ring maps.
The orange dots show the mean of 10\,000 Monte Carlo simulations. The error bars include the statistical variation only.
The values shown are mean values of the noise spectrum in the range $\ell=1150-1800$. 
The absolute white noise levels in $\mu$K$^2$ are reported in the last two columns of Table \ref{tab:whitenoise}.
 }
\label{fig:high_ell}
\end{figure}

We use half-ring maps to derive an estimate for the level of correlated residual noise at high multipoles.
We find that even after destriping there remains correlated residual noise at the highest multipoles,
and it affects the auto-spectra at the 1\% level.
In Fig. \ref{fig:high_ell} we plot the estimated noise spectrum in the high-multipole regime,
as a fraction of the white noise level reported in Table \ref{tab:whitenoise}.
We show the mean of the spectrum over multipoles $\ell=1150-1800$.
We show in the same figure also the mean of 10\,000 FFP8 noise Monte Carlo realizations,
all derived from same noise spectrum.
The error bars on the MC estimated indicate the statistical $1\sigma$ variation of the realizations,
not the uncertainty of the noise model.

\subsubsection{Residual noise at low multipoles}

We have three complementary methods for chracterization of correlated residual noise at low multipoles: 
the noise covariance matrix; MC simulations;
and the half-ring noise map.
We now compare the noise estimates from the three methods against each other.

\begin{figure}
\centering
\includegraphics[width=1\columnwidth]{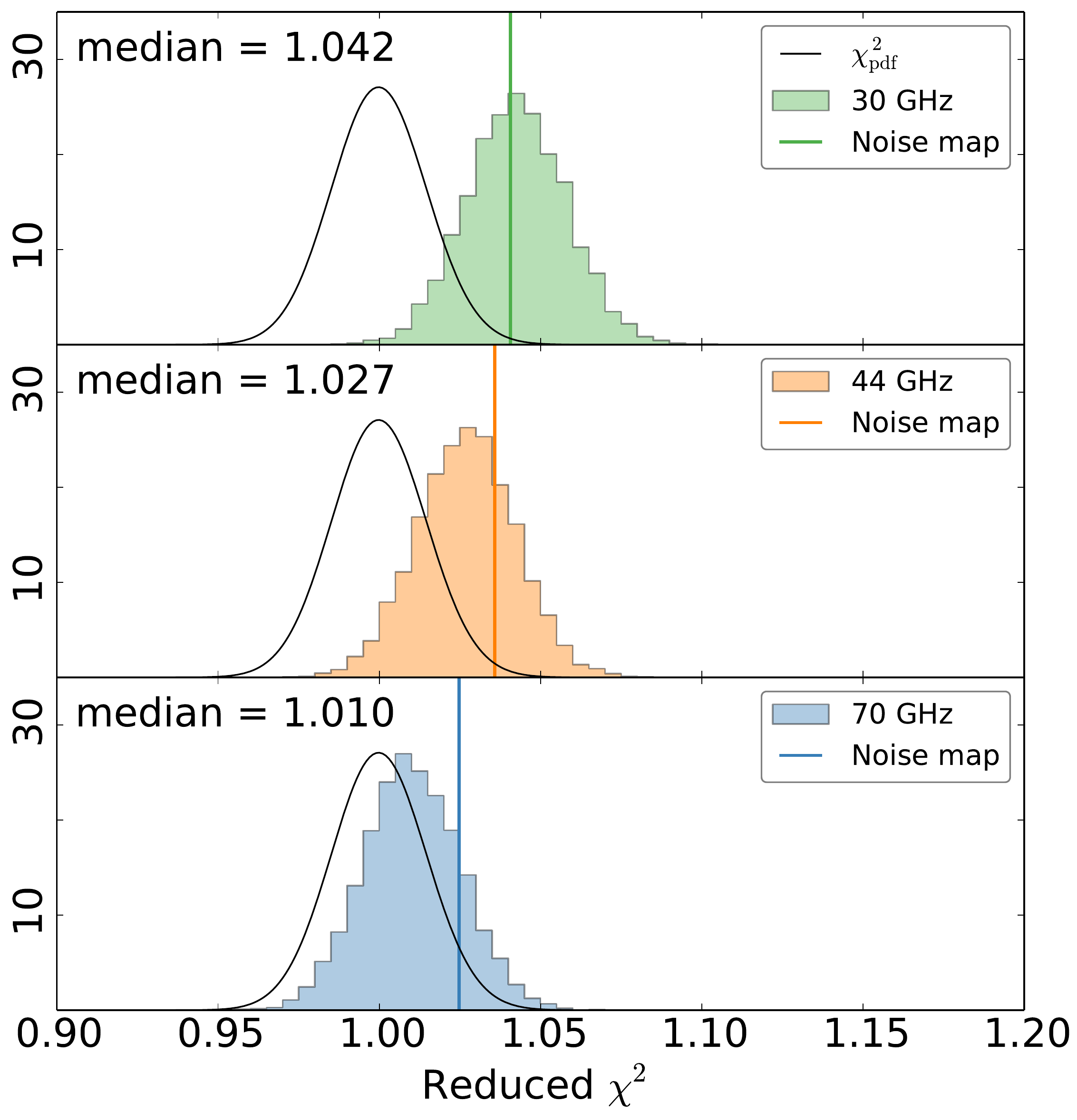}
\caption{Reduced $\chi^2$ statistics for 2015 full mission noise covariance matrices. $\chi^2$  tests have been carried out with 
10\,000 FFP8 noise realizations. Vertical lines mark the reduced $\chi^2$ value for half-ring noise maps.}
\label{fig:chi2_DDX11D_full}
\end{figure}

Figure \ref{fig:chi2_DDX11D_full} shows the reduced $\chi^2$ statistics for full mission noise covariance matrices. 
The $\chi^2$  tests have been carried out with 
10 000 FFP8 noise realizations with realistic mapmaking parameters.
The vertical lines indicate the reduced $\chi^2$ value for the half-ring noise maps. 
We observe that at the 70\,GHz frequency channel, which is the most important for the cosmological parameter estimation,
the noise covariance matrix models the residual noise the best. 

\begin{figure*}
\includegraphics[width=18cm]{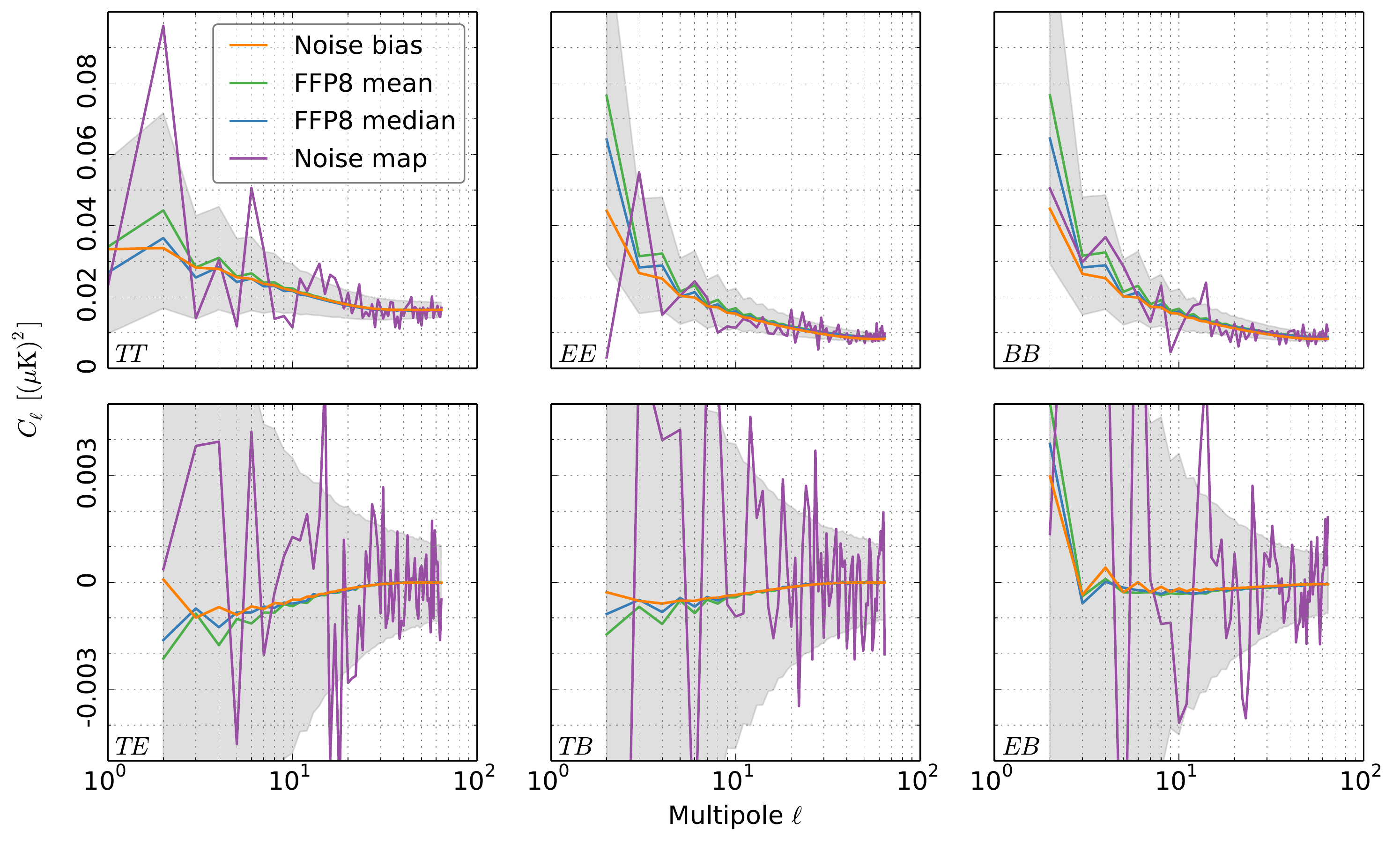}
\caption{Noise bias levels at 30\,GHz. Noise bias from 2015 full mission noise covariance matrix
 is plotted in orange, the noise MC statistics 
from 10\,000 FFP8 simulations are plotted in green (for the mean), blue (median) and grey ($\pm$1 sigma region), 
as well as the noise estimate of the half ring noise map in purple.}
\label{fig:noise_bias30}
\end{figure*}

\begin{figure*}
\includegraphics[width=18cm]{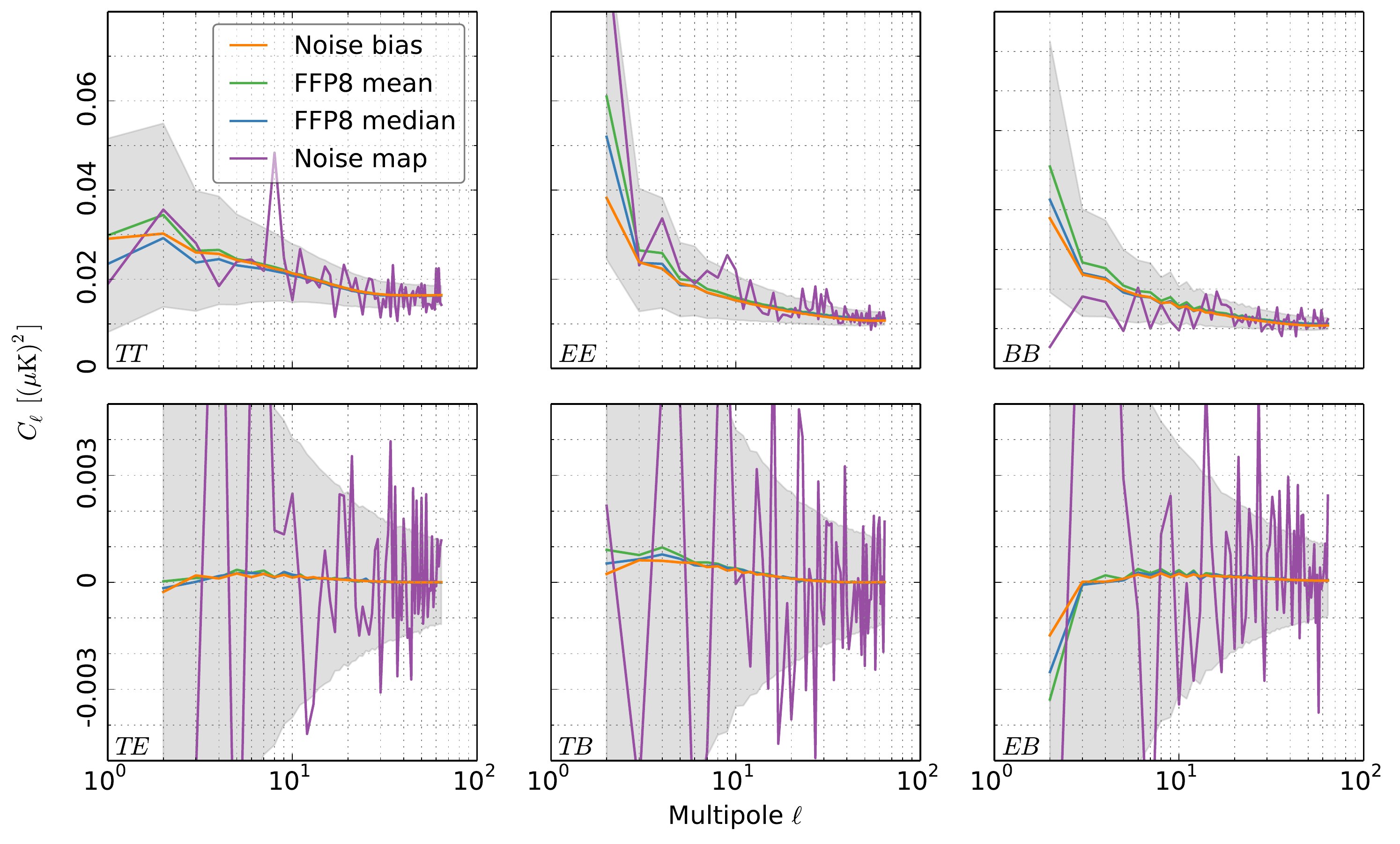}
\caption{Noise bias levels at 44\,GHz.}
\label{fig:noise_bias44}
\end{figure*}

\begin{figure*}
\includegraphics[width=18cm]{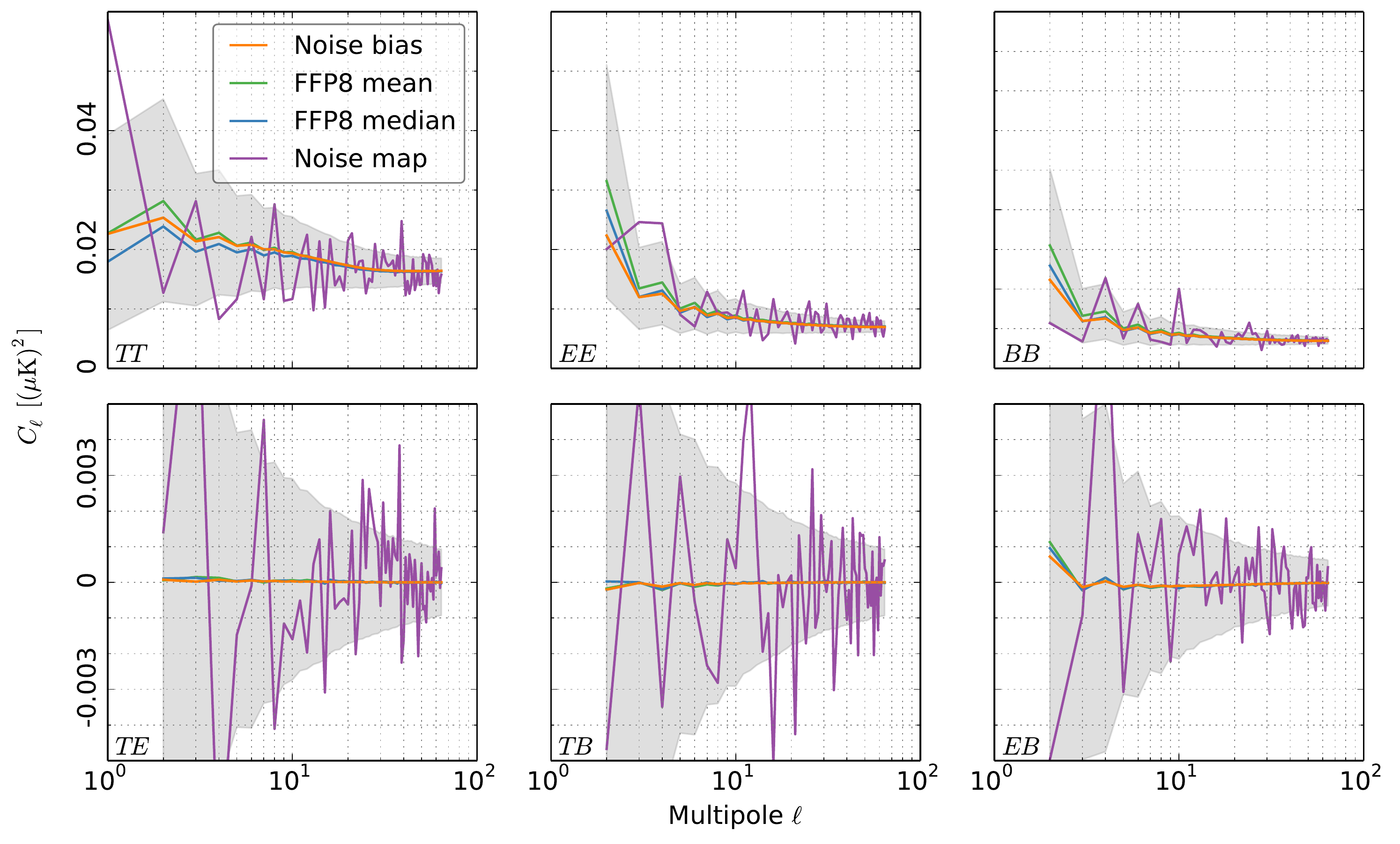}
\caption{Noise bias levels at 70\,GHz.}
\label{fig:noise_bias70}
\end{figure*}

Figures \ref{fig:noise_bias30} -- \ref{fig:noise_bias70} show noise bias estimates for full frequency maps at low multipoles.  
We show the six noise spectra ($TT$, $EE$, $BB$, $TE$, $TB$, $EB$) constructed from the NCVMs, and corresponding 
estimates from FFP8 noise Monte Carlo simulations. We show the average and median over 10\,000 noise realizations. 
We plot in the same figure also the spectrum of the half-ring noise map. As this represents one realization only, it has much 
stronger variations than the other estimates.

\section{Conclusions}

In this paper we have described the mapmaking procedure used for the 2015 data release from \Planck\ LFI,
and characterised the noise in the map products.
For the first time, we have shown polarization $Q$ and $U$ maps at LFI frequencies.

High-resolution maps were produced with the {\tt Madam} mapmaking code. 
The  algorithm is based on a destriping technique,
which is extended to short baseline lengths with the help of a noise prior.
We validated the procedure through noise and signal simulations.
We studied the effect of two main parameters that control the mapmaking procedure, namely  the baseline length, 
and the destriping resolution.
We showed that the values chosen for actual pipeline are appropriate for the data set and allow for good noise removal,
while keeping the signal error at low level. 

We have paid special attention to the reduction of unwanted systematic effects.
We applied a Galactic mask to reduce signal error arising from bandpass mismatch and 
strong signal gradients.  
We assessed the impact through simulations, and showed that the signal error is indeed 
reduced significantly, while the cost in 
residual noise is low.

The horn-uniform weighting scheme was introduced to reduce leakage of temperature signal to 
polarization through beamshape mismatch.
The effect at 30\,GHz and 44\,GHz was significant, less so at 70\,GHz.
 In all cases, however, the cost in residual noise level was small.

The polarization map products presented here are subject to bandpass leakage.
The leakage correction applied is discussed in \cite{planck2014-a03}.

The maps are not corrected for beam shape, or for the power to the far sidelobe part of the beam.
In particular, images of point sources reflect the radiometer beam shapes.
The exact beam shapes must be accounted in the estimation of source fluxes from LFI maps.

We have also described the production of low-resolution maps and noise covariance matrices. 
The low-resolution maps are produced by downgrading the high-resolution maps to the target 
resolution of $N_\textrm{side} = 16$ with 
the noise-weighted scheme, and the $I$ component is subsequently smoothed with a Gaussian beam. 
The noise covariance matrices are 
initially calculated at the resolution of $N_\textrm{side} = 64$ with {\tt{Madam/TOAST}} software, 
and subsequently the same downgrading 
scheme is applied to them as for the maps. Finally both the low-resolution maps and 
matrices are regularised by adding a small amount of white noise to all products.

The matrices were validated against noise Monte Carlo simulations.
The match between the matrices and the simulations was best at 70\,GHz, 
which is the most important channel for cosmology analysis.
We performed a series of simulations to quantify the effect of each approximation involved 
in the noise covariance computation.
The most important effects turned out to be the high destriping resolution, and the limited baseline length.

In addition to the noise covariance matrices, we characterised the residual noise in maps through 
noise Monte Carlo simulations 
and the noise maps constructed from half-ring maps.  
The noise estimates from the complementary methods show good agreement.

\begin{acknowledgements}

The Planck Collaboration acknowledges the support of: ESA; CNES, and
CNRS/INSU-IN2P3-INP (France); ASI, CNR, and INAF (Italy); NASA and DoE
(USA); STFC and UKSA (UK); CSIC, MINECO, JA and RES (Spain); Tekes, AoF,
and CSC (Finland); DLR and MPG (Germany); CSA (Canada); DTU Space
(Denmark); SER/SSO (Switzerland); RCN (Norway); SFI (Ireland);
FCT/MCTES (Portugal); ERC and PRACE (EU). A description of the Planck
Collaboration and a list of its members, indicating which technical
or scientific activities they have been involved in, can be found at
\url{http://www.cosmos.esa.int/web/planck/planck-collaboration}.
Some of the results of this research have been achieved
  using the PRACE-3IP project (FP7 RI-312763) resource Sisu
  based in Finland at CSC.
  We thank CSC -- IT Center for Science Ltd (Finland) for computational resources.  
  This research used resources of the National Energy Research Scientific Computing Center, a DOE Office of Science User Facility supported by the Office of Science of the U.S. Department of Energy under Contract No. DE-AC02-05CH11231.
 This
  work has made use of the Planck satellite simulation package
  (Level-S), which is assembled by the Max Planck Institute for
  Astrophysics Planck Analysis Centre (MPAC) \cite{reinecke2006}. Some of the results in this paper have been
  derived using the HEALPix package \cite{gorski2005}.

\end{acknowledgements}

\bibliographystyle{aat}
\bibliography{Planck_bib,LFI_DPC_bib}

\end{document}